\documentclass[twocolumn]{aastex63}
\usepackage{graphicx}

\shorttitle{Proper motions of MW satellites}
\shortauthors{McConnachie \& Venn}

\begin{document}
\title{Revised and new proper motions for confirmed and candidate Milky
  Way dwarf galaxies}

\author{Alan W. McConnachie}
\correspondingauthor{Alan W. McConnachie}
\email{alan.mcconnachie@nrc-cnrc.gc.ca}
\affil{NRC Herzberg Astronomy and Astrophysics, 5071 West Saanich
  Road, Victoria, B.C., Canada, V9E 2E7}
\author{Kim A. Venn} 
\affil{Physics \& Astronomy Department, University of Victoria, 3800 Finnerty Rd, Victoria, B.C., Canada, V8P 5C2}

\begin{abstract}
 A new derivation of systemic proper motions of Milky Way satellites
  is presented, and applied to 59 confirmed or candidate dwarf galaxy
  satellites using Gaia Data Release 2. This constitutes all known
  Milky Way dwarf galaxies (and likely candidates) as of May 2020 except
  the Magellanic Clouds, the Canis Major and Hydra 1 stellar
  overdensities, and the tidally disrupting Bootes III and Sagittarius
  dwarf galaxies. We derive systemic proper motions for the first time  for Indus 1, DES J0225+0304, Cetus 2, Pictor 2 and Leo T, but note that the latter three rely on photometry that is of poorer quality than for the rest of the sample. We cannot resolve a signal for Bootes 4, Cetus 3, Indus 2, Pegasus 3, or Virgo 1.  Our method is inspired by the maximum likelihood
  approach of \cite{pace2019} and examines simultaneously the spatial,
  color-magnitude, and proper motion distribution of sources. Systemic proper motions are
  derived without the need to identify confirmed radial velocity
  members, although the proper motions of these stars, where
  available, are incorporated into the analysis through a prior
  on the model. The associated
  uncertainties on the systemic proper motions are on average a factor
  of $\sim 1.4$ smaller than existing literature values.  Analysis of the implied membership distribution of the
  satellites suggests we accurately identify member stars with a
  contamination rate less than 1 in 20. 
\end{abstract}


\section{Introduction}

With respect to the formation and evolution of nearby dwarf galaxies, few areas of research are more satisfying than the study of the orbital dynamics of Milky Way satellites. Almost every other aspect of the evolution of  dwarf galaxies appears to be affected by a  multitude of unknown (or only partially-understood) physical processes. For many areas of interest, such as the star formation histories of dwarfs or their chemical evolution, these physical processes ordinarily combine to create a level of complexity per unit stellar mass that can appear to be out of whack compared to the meagre luminosities of dwarf galaxies. But for their orbits at least, only gravity really matters.

The Second Data Release from Gaia (DR2; \citealt{gaia2018b}) - to be followed in the near future with Early Data Release 3 (EDR3) - is a particularly impressive resource for near field cosmology. From a user perspective, it has transformed what is ultimately an incredibly complex and sensitive measurement - that of the absolute proper motion of intrinsically rather faint stars - into a measurement that is now readily available for more than 1 billion objects in the sky. Combined with precision radial velocities and their positions, complete knowledge of the space motions of the Milky Way satellites allows both a better estimation of the mass profile of the Galaxy (e.g., \citealt{eadie2019, posti2019, watkins2019, cautun2020, fritz2020, li2020}), and an estimation of the orbital histories of the satellites (e.g., \citealt{helmi2018b, fritz2018, simon2018, erkal2020, patel2020}).

The past two years has seen several influential papers demonstrating the utility of Gaia DR2 for dwarf galaxy proper motions, starting with \cite{helmi2018b}. This paper studied the brighter dwarfs, for which there are generally hundreds or more member stars in Gaia DR2, and determined systemic proper motions through an iterative procedure whereby a first estimate of the proper motion is refined by comparison to the data. Subsequent work included these brighter satellites but also started to include the fainter satellites, where there are far fewer member stars expected in Gaia DR2. Initially, these studies estimated systemic proper motions by taking a weighted mean of ``known'' members, that is stars that were considered high likelihood members based upon spectroscopy. To be targeted for spectroscopy in the first place implies they also had consistent positions on the sky and color-magnitude information, and importantly they were known to have consistent radial velocities (as well as potentially other spectroscopic information such as metallicities; \citealt{simon2018, simon2019, fritz2018, fritz2019,kallivayalil2018, longeard2018, longeard2020a}). A variant of this technique was introduced by \cite{massari2018}, that measures systemic proper motions incorporating information from stars that have not necessarily been spectroscopically ``pre-confirmed'' as members. A first estimate of the proper motion was made from either radial velocity members, or from candidate blue horizontal branch stars, and an iterative procedure refined these estimates based on all plausible members in the vicinity (in proper-motion space, color-magnitude space, and on-sky location).

Particularly relevant to the current study is the work of \cite{pace2019}. These authors were able to estimate the proper motions of thirteen dwarf galaxies without any dependence on ``known'' members. Rather, they selected stars in the appropriate regions of the colour-magnitude diagram (CMD), and created a mixture model for the data comprising both satellite members and contamination from the Milky Way. They considered spatial and proper motion information, and sought to find the systemic proper motion for the satellite that maximises the likelihood of the dataset. This involved the creation of models of the Milky Way contamination, and did not use any radial velocity information. This technique was very successful, and they showed good consistency with the radial velocity-based methods (see also confirmation of their technique for several objects by \citealt{simon2019}). Their technique has since been applied to the recent discovery of Bliss 1 and Centaurus 1 (\citealt{mau2019, mau2020}). 

Here, we present a new algorithm for the determination of satellite proper motions that is based on the method of \cite{pace2019}. Unlike those authors, we develop completely empirical models for the foreground that do not require us to marginalize over additional unknown parameters, and we incorporate radial velocity information if it exists. These modifications allow us to determine proper motions for some systems for which \cite{pace2019} were unable to derive solutions, and we argue that the remaining satellites in our sample for which we do not have solutions are due to a lack of any member stars with reliable data in Gaia DR2. Despite the subject of this paper, our ultimate goal is not focused towards determining the systemic proper motions of dwarfs, but to the related problem of correctly identifying member stars for detailed follow-up, which will be the subject of future contributions. As such, we examine the contamination and completeness of our member selections as a way to determine the robustness of our algorithm, and we present evidence that suggests we are able to identify member stars (without requiring radial velocities) with a contamination rate of $\lesssim 5\%$. In Section~\ref{sec:data}, we present our galaxy sample and the relevant data. Section~\ref{sec:method} describes our methodology, and Section~\ref{sec:results} presents our results, summarized in Section~\ref{sec:summ}.

\section{Target galaxies and data preparation}\label{sec:data}

\subsection{Dwarf galaxy candidates}

\begin{table*}{\scriptsize
\begin{center}
 \begin{tabular*}{\textwidth}{l|rrrrrrrrr}
Galaxy & RA (degs) & Dec (degs) & $(m - M)_0$ & $r_h$ (arcmins)& $e = 1 - b/a$ & $\theta ^\circ$ & $v_h$ (km\,s$^{-1}$)&$\sigma_v$ (km\,s$^{-1}$) & $<$[Fe/H]$>$\\
\hline\\
Antlia2          & 143.8867 & -36.7672 & $20.6^{+0.11}_{-0.11}$ & $76.2^{+7.2}_{-7.2}$ & $0.38^{+0.08}_{-0.08}$ & $156.0^{+6.0}_{-6.0}$ & $290.7^{+0.5}_{-0.5}$ & $5.71^{+1.08}_{-1.08}$ & $-1.36^{+0.04}_{-0.04}$ \\
Aquarius2        & 338.4812 & -9.3275 & $20.16^{+0.07}_{-0.07}$ & $5.1^{+0.8}_{-0.8}$ & $0.39^{+0.09}_{-0.09}$ & $121.0^{+9.0}_{-9.0}$ & $-71.1^{+2.5}_{-2.5}$ & $5.4^{+3.4}_{-0.9}$ & $-2.3^{+0.5}_{-0.5}$ \\
Bootes1        & 210.025 & 14.5 & $19.11^{+0.08}_{-0.08}$ & $11.26^{+0.27}_{-0.27}$ & $0.25^{+0.02}_{-0.02}$ & $7.0^{+3.0}_{-3.0}$ & $99.0^{+2.1}_{-2.1}$ & $2.4^{+0.9}_{-0.5}$ & $-2.55^{+0.11}_{-0.11}$ \\
Bootes2          & 209.5 & 12.85 & $18.1^{+0.06}_{-0.06}$ & $3.05^{+0.45}_{-0.45}$ & $0.24^{+0.12}_{-0.12}$ & $-70.0^{+27.0}_{-27.0}$ & $-117.0^{+5.2}_{-5.2}$ & $10.5^{+7.4}_{-7.4}$ & $-1.79^{+0.05}_{-0.05}$ \\
{\it Bootes4}          & 233.6892 & 43.7261 & $21.6^{+0.2}_{-0.2}$ & $7.6^{+0.8}_{-0.8}$ & $0.64^{+0.05}_{-0.05}$ & $3.0^{+4.0}_{-4.0}$ & --- & --- & --- \\
CanesVenatici1 & 202.0146 & 33.5558 & $21.69^{+0.1}_{-0.1}$ & $8.9^{+0.4}_{-0.4}$ & $0.39^{+0.03}_{-0.03}$ & $70.0^{+4.0}_{-4.0}$ & $30.9^{+0.6}_{-0.6}$ & $7.6^{+0.4}_{-0.4}$ & $-1.98^{+0.01}_{-0.01}$ \\
CanesVenatici2   & 194.2917 & 34.3208 & $21.02^{+0.06}_{-0.06}$ & $1.51^{+0.23}_{-0.23}$ & $0.46^{+0.11}_{-0.11}$ & $10.0^{+11.0}_{-11.0}$ & $-128.9^{+1.2}_{-1.2}$ & $4.6^{+1.0}_{-1.0}$ & $-2.21^{+0.05}_{-0.05}$ \\
Carina           & 100.4029 & -50.9661 & $20.11^{+0.13}_{-0.13}$ & $11.43^{+0.12}_{-0.12}$ & $0.37^{+0.01}_{-0.01}$ & $60.0^{+1.0}_{-1.0}$ & $222.9^{+0.1}_{-0.1}$ & $6.6^{+1.2}_{-1.2}$ & $-1.72^{+0.01}_{-0.01}$ \\
Carina2          & 114.1067 & -57.9992 & $17.79^{+0.05}_{-0.05}$ & $8.69^{+0.75}_{-0.75}$ & $0.34^{+0.07}_{-0.07}$ & $170.0^{+9.0}_{-9.0}$ & $477.2^{+1.2}_{-1.2}$ & $3.4^{+1.2}_{-0.8}$ & $-2.44^{+0.09}_{-0.09}$ \\
Carina3          & 114.63 & -57.8997 & $17.22^{+0.1}_{-0.1}$ & $3.75^{+1.0}_{-1.0}$ & $0.55^{+0.18}_{-0.18}$ & $150.0^{+14.0}_{-14.0}$ & $284.6^{+3.4}_{-3.1}$ & $5.6^{+4.3}_{-2.1}$ & $-1.8^{+0.2}_{-0.2}$ \\
Centaurus1       & 189.585 & -40.902 & $20.33^{+0.1}_{-0.1}$ & $2.88^{+0.5}_{-0.4}$ & $0.4^{+0.1}_{-0.1}$ & $20.0^{+11.0}_{-11.0}$ & --- & --- & $-1.8^{+999.0}_{-999.0}$ \\
Cetus2          & 19.47 & -17.42 & $17.1^{+0.1}_{-0.1}$ & $1.9^{+1.0}_{-0.5}$ & --- & --- & --- & --- & $-1.28^{+0.07}_{-0.07}$ \\
{\it Cetus3}           & 31.3308 & -4.27 & $22.0^{+0.2}_{-0.1}$ & $1.23^{+0.42}_{-0.19}$ & $0.76^{+0.06}_{-0.08}$ & $101.0^{+5.0}_{-6.0}$ & --- & --- & --- \\
Columba1        & 82.86 & -28.03 & $21.3^{+0.22}_{-0.22}$ & $1.9^{+0.5}_{-0.4}$ & --- & --- & $153.7^{+5.0}_{-4.8}$ & --- & --- \\
ComaBerenices    & 186.7458 & 23.9042 & $18.2^{+0.2}_{-0.2}$ & $5.63^{+0.3}_{-0.3}$ & $0.37^{+0.05}_{-0.05}$ & $-58.0^{+4.0}_{-4.0}$ & $98.1^{+0.9}_{-0.9}$ & $4.6^{+0.8}_{-0.8}$ & $-2.6^{+0.05}_{-0.05}$ \\
Crater2          & 177.31 & -18.4131 & $20.35^{+0.02}_{-0.02}$ & $31.2^{+2.5}_{-2.5}$ & --- & --- & $87.5^{+0.4}_{-0.4}$ & $2.7^{+0.3}_{-0.3}$ & $-1.98^{+0.1}_{-0.1}$ \\
DESJ0225+0304    & 36.4267 & 3.0695 & $16.88^{+0.06}_{-0.05}$ & $2.68^{+1.33}_{-0.7}$ & $0.61^{+0.14}_{-0.23}$ & $31.25^{+11.48}_{-13.39}$ & --- & --- & $-1.26^{+0.03}_{-0.03}$ \\
Draco            & 260.0517 & 57.9153 & $19.4^{+0.17}_{-0.17}$ & $9.93^{+0.09}_{-0.09}$ & $0.3^{+0.01}_{-0.01}$ & $87.0^{+1.0}_{-1.0}$ & $-291.0^{+0.1}_{-0.1}$ & $9.1^{+1.2}_{-1.2}$ & $-1.93^{+0.01}_{-0.01}$ \\
Draco2          & 238.1983 & 64.5653 & $16.67^{+0.05}_{-0.05}$ & $3.0^{+0.7}_{-0.5}$ & $0.23^{+0.15}_{-0.15}$ & $76.0^{+22.0}_{-32.0}$ & $-342.5^{+1.1}_{-1.2}$ & --- & $-2.7^{+0.1}_{-0.1}$ \\
Eridanus2        & 56.0879 & -43.5333 & $22.9^{+0.2}_{-0.2}$ & $1.81^{+0.17}_{-0.17}$ & $0.37^{+0.06}_{-0.06}$ & $82.0^{+7.0}_{-7.0}$ & $75.6^{+3.3}_{-3.3}$ & $6.9^{+1.2}_{-0.9}$ & $-2.38^{+0.13}_{-0.13}$ \\
Eridanus3       & 35.6896 & -52.2836 & $19.7^{+0.2}_{-0.2}$ & $0.29^{+0.23}_{-0.23}$ & $0.32^{+0.13}_{-0.13}$ & $62.0^{+11.0}_{-11.0}$ & --- & --- & --- \\
Fornax           & 39.9971 & -34.4492 & $20.84^{+0.18}_{-0.18}$ & $18.4^{+0.2}_{-0.2}$ & $0.28^{+0.01}_{-0.01}$ & $46.0^{+1.0}_{-1.0}$ & $55.3^{+0.1}_{-0.1}$ & $11.7^{+0.9}_{-0.9}$ & $-0.99^{+0.01}_{-0.01}$ \\
Grus1           & 344.1767 & -50.1633 & $20.4^{+0.2}_{-0.2}$ & $2.08^{+0.87}_{-0.87}$ & $0.54^{+0.26}_{-0.26}$ & $11.0^{+32.0}_{-32.0}$ & $-140.5^{+2.4}_{-1.6}$ & $9.8$ & $-1.42^{+0.55}_{-0.42}$ \\
Grus2           & 331.02 & -46.44 & $18.62^{+0.21}_{-0.21}$ & $6.0^{+0.9}_{-0.5}$ & --- & --- & $-110.0^{+0.5}_{-0.5}$ & --- & $-2.51^{+0.11}_{-0.11}$ \\
Hercules         & 247.7583 & 12.7917 & $20.6^{+0.2}_{-0.2}$ & $5.99^{+0.58}_{-0.58}$ & $0.69^{+0.04}_{-0.04}$ & $-73.0^{+2.0}_{-2.0}$ & $45.2^{+1.1}_{-1.1}$ & $3.7^{+0.9}_{-0.9}$ & $-2.41^{+0.04}_{-0.04}$ \\
Horologium1     & 43.8821 & -54.1189 & $19.5^{+0.2}_{-0.2}$ & $1.54^{+0.34}_{-0.34}$ & $0.31^{+0.16}_{-0.16}$ & $50.0^{+26.0}_{-26.0}$ & $112.8^{+2.5}_{-2.6}$ & $4.9^{+2.8}_{-0.9}$ & $-2.76^{+0.1}_{-0.1}$ \\
Horologium2     & 49.1338 & -50.0181 & $19.46^{+0.2}_{-0.2}$ & $2.83^{+1.31}_{-1.31}$ & $0.86^{+0.19}_{-0.19}$ & $130.0^{+16.0}_{-16.0}$ & $168.7^{+12.9}_{-12.6}$ & --- & $-2.1$ \\
Hydra2           & 185.4254 & -31.9853 & $20.64^{+0.16}_{-0.16}$ & $1.5^{+0.32}_{-0.32}$ & $0.17^{+0.13}_{-0.13}$ & $29.0^{+25.0}_{-25.0}$ & $303.1^{+1.4}_{-1.4}$ & --- & $-2.02^{+0.08}_{-0.08}$ \\
Hydrus1          & 37.3892 & -79.3089 & $17.2^{+0.04}_{-0.04}$ & $7.42^{+0.62}_{-0.54}$ & $0.21^{+0.15}_{-0.07}$ & $97.0^{+14.0}_{-14.0}$ & $80.4^{+0.6}_{-0.6}$ & $2.69^{+0.51}_{-0.43}$ & $-2.52^{+0.09}_{-0.09}$ \\
Indus1          & 317.2046 & -51.1656 & $20.0^{+0.2}_{-0.2}$ & $0.87^{+0.45}_{-0.45}$ & $0.72^{+0.29}_{-0.29}$ & $5.0^{+20.0}_{-20.0}$ & --- & --- & --- \\
{\it Indus2}          & 309.72 & -46.16 & $21.65^{+0.16}_{-0.16}$ & $2.9^{+1.1}_{-1.0}$ & --- & --- & --- & --- & --- \\
Leo1             & 152.1171 & 12.3064 & $22.02^{+0.13}_{-0.13}$ & $3.3^{+0.03}_{-0.03}$ & $0.3^{+0.01}_{-0.01}$ & $78.0^{+1.0}_{-1.0}$ & $282.5^{+0.1}_{-0.1}$ & $9.2^{+1.4}_{-1.4}$ & $-1.43^{+0.01}_{-0.01}$ \\
Leo2             & 168.37 & 22.1517 & $21.84^{+0.13}_{-0.13}$ & $2.48^{+0.03}_{-0.03}$ & $0.07^{+0.02}_{-0.02}$ & $43.0^{+8.0}_{-8.0}$ & $78.0^{+0.1}_{-0.1}$ & $6.6^{+0.7}_{-0.7}$ & $-1.62^{+0.01}_{-0.01}$ \\
Leo4             & 173.2375 & -0.5333 & $20.94^{+0.09}_{-0.09}$ & $2.61^{+0.31}_{-0.31}$ & $0.19^{+0.09}_{-0.09}$ & $-29.0^{+27.0}_{-27.0}$ & $132.3^{+1.4}_{-1.4}$ & $3.3^{+1.7}_{-1.7}$ & $-2.54^{+0.07}_{-0.07}$ \\
Leo5             & 172.79 & 2.22 & $21.46^{+0.16}_{-0.16}$ & $1.0^{+0.22}_{-0.22}$ & $0.35^{+0.07}_{-0.07}$ & $-65.0^{+21.0}_{-21.0}$ & $173.3^{+3.1}_{-3.1}$ & $3.7^{+2.3}_{-1.4}$ & $-2.0^{+0.2}_{-0.2}$ \\
LeoT             & 143.7225 & 17.0514 & $23.1^{+0.1}_{-0.1}$ & $1.25^{+0.14}_{-0.14}$ & $0.23^{+0.09}_{-0.09}$ & $-107.0^{+16.0}_{-16.0}$ & $38.1^{+2.0}_{-2.0}$ & $7.5^{+1.6}_{-1.6}$ & $-2.02^{+0.05}_{-0.05}$ \\
{\it Pegasus3}        & 336.0942 & 5.42 & $21.56^{+0.2}_{-0.2}$ & $1.3^{+0.5}_{-0.4}$ & $0.46^{+0.18}_{-0.26}$ & $133.0^{+17.0}_{-17.0}$ & $-222.9^{+2.6}_{-2.6}$ & $5.4^{+3.0}_{-2.5}$ & $-2.1$ \\
Phoenix          & 27.7762 & -44.4447 & $23.06^{+0.12}_{-0.12}$ & $2.3^{+0.07}_{-0.07}$ & $0.3^{+0.03}_{-0.03}$ & $8.0^{+4.0}_{-4.0}$ & $-21.2^{+1.0}_{-1.0}$ & $9.3^{+0.7}_{-0.7}$ & $-1.49^{+0.04}_{-0.04}$ \\
Phoenix2        & 354.9975 & -54.4061 & $19.6^{+0.2}_{-0.2}$ & $1.61^{+0.27}_{-0.27}$ & $0.61^{+0.15}_{-0.15}$ & $-19.0^{+14.0}_{-14.0}$ & $32.4^{+3.7}_{-3.8}$ & $11.0^{+9.4}_{-5.3}$ & --- \\
Pictor1       & 70.9475 & -50.2831 & $20.3^{+0.2}_{-0.2}$ & $0.66^{+0.32}_{-0.32}$ & $0.24^{+0.19}_{-0.19}$ & $72.0^{+10.0}_{-10.0}$ & --- & --- & --- \\
Pictor2          & 101.18 & -59.8969 & $18.3^{+0.12}_{-0.15}$ & $3.8^{+1.5}_{-1.0}$ & $0.13^{+0.22}_{-0.13}$ & $14.0^{+60.0}_{-66.0}$ & --- & --- & $-1.8^{+0.3}_{-0.3}$ \\
Pisces2          & 344.6292 & 5.9525 & $21.31^{+0.17}_{-0.17}$ & $1.22^{+0.2}_{-0.2}$ & $0.4^{+0.1}_{-0.1}$ & $99.0^{+13.0}_{-13.0}$ & $-226.5^{+2.7}_{-2.7}$ & $5.4^{+3.6}_{-2.4}$ & $-2.45^{+0.07}_{-0.07}$ \\
Reticulum2      & 53.9254 & -54.0492 & $17.4^{+0.2}_{-0.2}$ & $5.59^{+0.21}_{-0.21}$ & $0.56^{+0.03}_{-0.03}$ & $69.0^{+2.0}_{-2.0}$ & $64.7^{+1.3}_{-0.8}$ & $3.22^{+1.64}_{-0.49}$ & $-2.46^{+0.09}_{-0.1}$ \\
Reticulum3      & 56.36 & -60.45 & $19.81^{+0.31}_{-0.31}$ & $2.4^{+0.9}_{-0.8}$ & --- & --- & $274.2^{+7.5}_{-7.4}$ & --- & --- \\
Sagittarius2    & 298.1688 & -22.0681 & $19.32^{+0.03}_{-0.02}$ & $1.7^{+0.05}_{-0.05}$ & $0.06^{+0.06}_{-0.06}$ & $103.0^{+28.0}_{-17.0}$ & $-177.3^{+1.2}_{-1.2}$ & $2.7^{+1.3}_{-1.0}$ & $-2.28^{+0.03}_{-0.03}$ \\
Sculptor         & 15.0392 & -33.7092 & $19.67^{+0.14}_{-0.14}$ & $12.33^{+0.05}_{-0.05}$ & $0.37^{+0.01}_{-0.01}$ & $94.0^{+1.0}_{-1.0}$ & $111.4^{+0.1}_{-0.1}$ & $9.2^{+1.4}_{-1.4}$ & $-1.68^{+0.01}_{-0.01}$ \\
Segue1         & 151.7667 & 16.0819 & $16.8^{+0.2}_{-0.2}$ & $3.95^{+0.48}_{-0.48}$ & $0.34^{+0.11}_{-0.11}$ & $75.0^{+16.0}_{-16.0}$ & $208.5^{+0.9}_{-0.9}$ & $3.9^{+0.8}_{-0.8}$ & $-2.72^{+0.4}_{-0.4}$ \\
Segue2           & 34.8167 & 20.1753 & $17.7^{+0.1}_{-0.1}$ & $3.64^{+0.29}_{-0.29}$ & $0.21^{+0.07}_{-0.07}$ & $166.0^{+15.0}_{-15.0}$ & $-39.2^{+2.5}_{-2.5}$ & $3.4^{+2.5}_{-1.2}$ & $-2.22^{+0.13}_{-0.13}$ \\
Sextans1       & 153.2625 & -1.6147 & $19.67^{+0.1}_{-0.1}$ & $27.8^{+1.2}_{-1.2}$ & $0.35^{+0.05}_{-0.05}$ & $56.0^{+5.0}_{-5.0}$ & $224.2^{+0.1}_{-0.1}$ & $7.9^{+1.3}_{-1.3}$ & $-1.93^{+0.01}_{-0.01}$ \\
Triangulum2     & 33.3225 & 36.1783 & $17.4^{+0.1}_{-0.1}$ & $3.9^{+1.1}_{-0.9}$ & $0.21^{+0.17}_{-0.21}$ & $56.0^{+16.0}_{-24.0}$ & $-381.7^{+1.1}_{-1.1}$ & --- & $-2.24^{+0.05}_{-0.05}$ \\
Tucana2          & 342.9796 & -58.5689 & $18.8^{+0.2}_{-0.2}$ & $9.83^{+1.66}_{-1.11}$ & $0.39^{+0.1}_{-0.2}$ & $107.0^{+18.0}_{-18.0}$ & $-129.1^{+3.5}_{-3.5}$ & $8.6^{+4.4}_{-2.7}$ & $-2.23^{+0.18}_{-0.12}$ \\
Tucana3         & 359.15 & -59.6 & $17.01^{+0.16}_{-0.16}$ & $6.0^{+0.8}_{-0.6}$ & --- & --- & $-102.3^{+2.4}_{-2.4}$ & $0.1^{+0.7}_{-0.1}$ & $-2.42^{+0.07}_{-0.08}$ \\
Tucana4         & 0.73 & -60.85 & $18.41^{+0.19}_{-0.19}$ & $9.3^{+1.4}_{-0.9}$ & $0.39^{+0.07}_{-0.1}$ & $27.0^{+9.0}_{-8.0}$ & $15.9^{+1.8}_{-1.7}$ & $4.3^{+1.7}_{-1.0}$ & $-2.49^{+0.15}_{-0.16}$ \\
Tucana5         & 354.35 & -63.27 & $18.71^{+0.34}_{-0.34}$ & $2.1^{+0.6}_{-0.4}$ & $0.51^{+0.09}_{-0.18}$ & $29.0^{+11.0}_{-11.0}$ & $-36.3^{+2.5}_{-2.2}$ & --- & $-2.17^{+0.23}_{-0.23}$ \\
UrsaMajor1     & 158.72 & 51.92 & $19.93^{+0.1}_{-0.1}$ & $8.34^{+0.34}_{-0.34}$ & $0.57^{+0.03}_{-0.03}$ & $67.0^{+2.0}_{-2.0}$ & $-55.3^{+1.4}_{-1.4}$ & $7.6^{+1.0}_{-1.0}$ & $-2.18^{+0.04}_{-0.04}$ \\
UrsaMajor2       & 132.875 & 63.13 & $17.5^{+0.3}_{-0.3}$ & $13.95^{+0.46}_{-0.46}$ & $0.56^{+0.03}_{-0.03}$ & $-77.0^{+2.0}_{-2.0}$ & $-116.5^{+1.9}_{-1.9}$ & $6.7^{+1.4}_{-1.4}$ & $-2.47^{+0.06}_{-0.06}$ \\
UrsaMinor        & 227.2854 & 67.2225 & $19.4^{+0.1}_{-0.1}$ & $17.32^{+0.11}_{-0.11}$ & $0.55^{+0.01}_{-0.01}$ & $50.0^{+1.0}_{-1.0}$ & $-246.9^{+0.1}_{-0.1}$ & $9.5^{+1.2}_{-1.2}$ & $-2.13^{+0.01}_{-0.01}$ \\
{\it Virgo1}           & 180.0379 & 0.681 & $19.8^{+0.2}_{-0.1}$ & $1.76^{+0.49}_{-0.4}$ & $0.59^{+0.12}_{-0.14}$ & $62.0^{+8.0}_{-13.0}$ & --- & --- & --- \\
Willman1         & 162.3375 & 51.05 & $17.9^{+0.4}_{-0.4}$ & $2.53^{+0.22}_{-0.22}$ & $0.47^{+0.06}_{-0.06}$ & $74.0^{+4.0}_{-4.0}$ & $-12.3^{+2.5}_{-2.5}$ & $4.3^{+2.3}_{-1.3}$ & $-2.1$ \\
\end{tabular*}
\caption{All Milky Way dwarf galaxies and candidates that are considered in this paper, along with relevant positional, structural, radial velocity and metallicity parameters. Objects in italics are those satellites for which we are unable to derive proper motions.}\label{gals}
\end{center} }
\end{table*}

We use the updated and curated list of nearby galaxies from
\cite{mcconnachie2012} to define our
target sample\footnote{\url{http://www.astro.uvic.ca/~alan/Nearby\_Dwarf\_Database.html}}. This list contains both confirmed dwarf galaxies as
well as possible candidates. The latter category includes systems
that may actually be globular clusters (such as Eridanus 3; see
\citealt{conn2018b}), or objects that may not actually be 
stellar systems (e.g., see the discussion of Tucana
5 and Cetus 2 in \citealt{simon2019} and \citealt{conn2018a}, respectively). We stress that it is not the purpose of this paper to discuss the reality or otherwise of all these systems, but rather to determine if a systemic proper motion can be derived on the {\it assumption} that they are real stellar systems.

We consider all galaxies within 450kpc of the Sun (that is, out to and
including Leo T). We do not consider the Magellanic Clouds; excellent
and comprehensive studies of the proper motions of these galaxies can
be found in \cite{helmi2018b} as well as \cite{kallivayalil2013} and related work. Similarly, we do not consider the Sagitarrius
dwarf galaxy, whose stars extend across the entire sky and which has
recently been explored using Gaia DR2 by \cite{ibata2020}. Further, we do
not consider the Canis Major (\citealt{martin2004a}), Hydra 1
(\citealt{grillmair2011}) and Bootes 3 (\citealt{grillmair2009}) stellar overdensities.  The
nature of these structures, especially for the first two, remain
uncertain (but see \citealt{hargis2016}). For the latter, Bootes 3 is likely the projenitor of the
Styx stellar stream (\citealt{carlin2018}). For all three systems,
reliable structural parameters are not available, which is a
prerequisite for our method.

\begin{figure*}
  \centering
    \includegraphics[width=\textwidth]{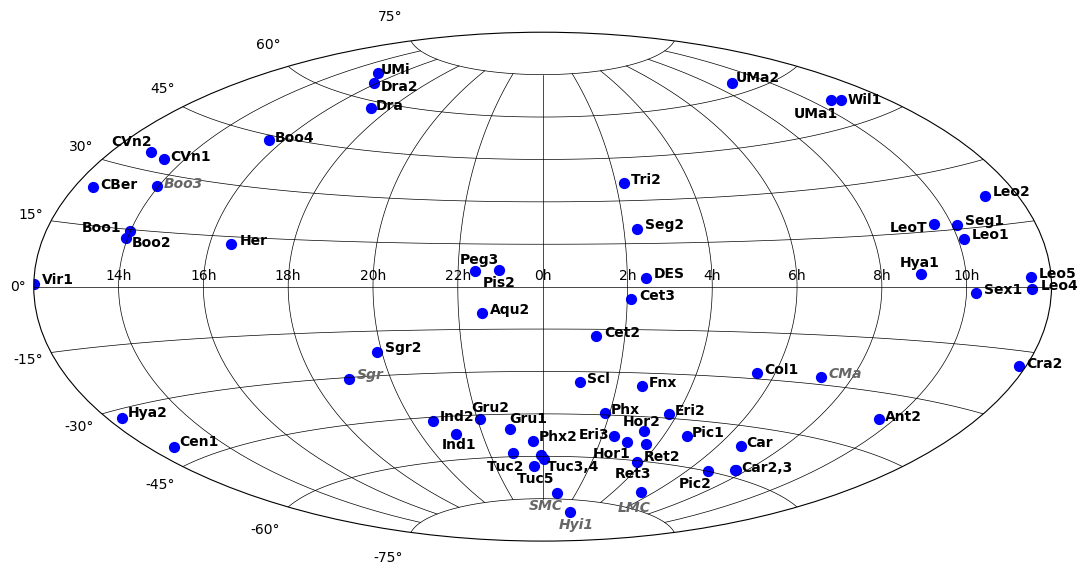}
    \caption{Aitoff projection of all Milky Way dwarf galaxies and candidates within a distance of 450\,kpc from the Sun.  Objects labelled with grey italic font are not considered in this paper.}\label{distribution}
\end{figure*}

Table~\ref{gals} gives a list of all Milky Way satellites considered
in this paper, 
along with positional information, structural parameters, mean
metallicities, and radial velocity information, where these
data exist. Figure~\ref{distribution} shows an Aitoff projection of the spatial
distribution of all the satellites in equatorial coordinates.

\subsection{Selection of stars from Gaia}

In this era of Gaia (\citealt{gaia2016}), any star bright enough to be detected by the spacecraft has
information available regarding its spatial position $(\xi, \eta)_i$, 
color and magnitude $(G, BP - RP)_i$ and proper motion
$(\mu_\alpha \cos\delta, \mu_\delta)_i$. Note that $(\xi, \eta)_i$ are
the coordinates of star $i$ projected on a tangent plane to the
celestial sphere, and these are a function of the right ascension and
declination $(\alpha, \delta)_i$ as measured by Gaia.

For every satellite listed in Table~\ref{gals}, we consider all stellar
sources in Gaia DR2 (\citealt{gaia2018b}) that are within 2 degrees of the center
of the galaxy. In 53 out of the 59 cases, this area corresponds to
a much larger area than subtended by the dwarf itself (i.e., hundreds or more
times its half light radius). However, for six of the larger 
(in terms of subtended area) satellites, we consider
sources within $4 - 8$ degrees from their centers 
(Antlia 2, Carina, Crater 2, Fornax, Sextans 1 and Ursa Minor).

As the colour-magnitude distribution of sources is critical to our method, 
we require the $G, BP, RP$ photometry to be reliable, and therefore
follow \cite{lindegren2018} (their Equation C.2) to consider only
stars that meet the following criterion:

\begin{equation}
 1.0 + 0.015 (BP - RP)^2 <  E < 1.3+0.06 (BP - RP)^2 \label{fluxexcess}
\end{equation}

\noindent where $E$ is the flux excess factor, as defined in \cite{lindegren2018}. 
We use dereddened Gaia DR2 stellar magnitudes in our analysis, using the extinction maps of \cite{schlegel1998} and following the definition of the 
relevant extinction coefficients described in Equation~1 and Table~1 of \cite{babusiaux2018}. 

We only consider stars with full five parameter astrometric
solutions, and high quality astrometry as defined via the renormalised
unit weighted error ({\tt ruwe}; see \citealt{lindegren2018} and 
discussion in the Gaia DR2 Documentation Release 1.2). Inspection of the distribution
of {\tt ruwe} for sources that otherwise meet our selection criteria motivated us to
adopt {\tt ruwe} $ < 1.3$.

Finally, only stars with parallaxes that are consistent with the distance to the
satellite are included. Specifically, the $3\,\sigma$ parallax range measured by
Gaia DR2 must overlap the $3\,\sigma$ parallax range implied by the distance modulus of
the dwarf, as given in Table~\ref{gals}.  
We correct for the global zero-point of the parallax in Gaia DR2 of -0.029mas (\citealt{lindegren2018}). 

\subsection{Radial velocity data}

Member stars in Milky Way dwarf galaxies are generally too faint to have been
observed with the Gaia/RVS. Therefore,  unlike photometry and proper motions, radial velocities are only
available for the stars in dwarf galaxies that were 
specifically targeted for followed-up spectroscopy from ground-based
observatories. There are a wealth of
such studies in the literature; a comprehensive compilation of 
the known radial velocity publications for the satellites in this paper 
is given in Table~\ref{radvel} (see also Table~1 of
\citealt{fritz2018}). For all except the brightest objects, these
references are intended to be a complete list of published radial velocity
measurements for these satellites, and any omissions are
unintentional. For the brightest (``classical'') objects - Fornax, Ursa Minor, Carina,
Sextans 1, Draco, Sculptor, Leo I and Leo 2 - the references are not
comprehensive. These bright dwarfs typically have hundreds-to-thousands of stars with radial
velocity information, compared to typically tens-to-hundreds of stars in the
fainter systems.

For each paper in Table~\ref{radvel}, we have cross-matched
the stars with measured radial velocities with Gaia DR2. 
For those papers listed in Table~\ref{radvel} that
give multiple measurements per star, we have taken the weighted mean
velocity for each star. For those satellites that have been observed
by multiple groups, we have combined datasets, and taken the weighted
mean velocities of any stars in common.

\begin{table*}
\begin{center}
\begin{tabular*}{0.65\textwidth}{l|c|l|c}
Galaxy & References & Galaxy & References\\
\hline
Antlia2 &\cite{torrealba2019}&                         Leo1& \cite{mateo2008}\\				                  
Aquarius2  &\cite{torrealba2016b} &		       Leo2&\cite{spencer2017}\\				  
Bootes1  & \cite{munoz2006b}&			     Leo4  & \cite{simon2007}\\ 				  
               & \cite{martin2007b}&		     Leo5  &\cite{walker2009a}\\ 				  
               & \cite{norris2008}&		               &\cite{collins2017}\\ 			  
               & \cite{norris2010a}&		      LeoT  & \cite{simon2007}\\				  
               & \cite{koposov2011}&		     Pegasus3  & \cite{kim2016} \\					  
Bootes2  &\cite{koch2009}&			       Phoenix&\cite{kacharov2017}\\				  
Bootes4          &  --- &			     Phoenix2  &\cite{fritz2019}\\ 				  
CanesVenatici2   &   \cite{simon2007}&		     Pictor1  & --- \\					  
CanesVenatici2  &  \cite{martin2007b}&		     Pictor2 & --- \\					  
       &\cite{simon2007}&			     Pisces2  & \cite{kirby2015}\\				  
  Carina&\cite{walker2009b}&			     Reticulum 2 & \cite{koposov2015}\\			  
Carina2  &\cite{li2018b}&			                         & \cite{simon2015}\\		  
Carina3  &\cite{li2018b}&			                         & \cite{walker2015a}\\		  
Centaurus1  & --- & 				     Reticulum3  &\cite{fritz2019}\\				  
 Cetus2  & --- &				     Sagittarius2  & \cite{longeard2020a, longeard2020b}\\			  
Cetus3  & --- &					       Sculptor& \cite{walker2009b}\\			  
Columba1  &\cite{fritz2019}&			     Segue1  & \cite{simon2011}\\				  
ComaBerenices    &  \cite{simon2007}&		            & \cite{norris2010a}\\				  
  Crater2 &\cite{caldwell2017b}&		       &\cite{geha2009}\\					  
  DES J0225+0304 & --- &			     Segue2           &  \cite{kirby2013}\\			  
  Draco& \cite{walker2015b}&			       Sextans&\cite{walker2009b}\\				  
Draco2           & \cite{longeard2018}&		     Triangulum2      &  \cite{kirby2015}\\			  
                          & \cite{martin2016c}&	                               & \cite{martin2016b}\\	  
Eridanus2  & \cite{li2017}&			                               &\cite{kirby2017}\\		  
Eridanus3  & --- &				     Tucana2  & \cite{walker2016}\\				  
  Fornax &\cite{walker2009b}&			                    & \cite{chiti2018}\\  			  
Grus1  & \cite{walker2016}& 			     Tucana3  &\cite{simon2017}\\				  
Grus2  & \cite{simon2019}& 			                     &\cite{li2018a}\\			  
Hercules  & \cite{simon2007}&			     Tucana4  & \cite{simon2019}\\				  
                & \cite{aden2009a}&		     Tucana5  &  \cite{simon2019} \\				  
                & \cite{deason2012a}&		     UrsaMajor1       &  \cite{martin2007b}\\		  
Horologium1  & \cite{koposov2015}&		                               &\cite{simon2007}\\		  
                       & \cite{nagasawa2018}&	     UrsaMajor2       &  \cite{martin2007b}\\		  
Horologium2  & \cite{fritz2019}&		                               &\cite{simon2007}\\		  
Hydra2  & \cite{kirby2015}&			       UrsaMinor& \cite{spencer2018}\\			  
Hydrus1  &\cite{koposov2018}&			     Virgo1  & --- \\ 					  
Indus1  & --- & 				     Willman1         &    \cite{martin2007b}\\		  
Indus2  & --- &					                               &\cite{willman2011}\\          
  
\end{tabular*}
\caption{Radial velocity data for all the Milky Way satellites under consideration. For Triangulum 2, only velocities from
  \cite{kirby2017} were used. References for the ``classical'' satellites - Fornax, Ursa Minor, Carina, Sextans 1, Draco, Sculptor, Leo I and Leo 2 - are not complete.}\label{radvel}
\end{center}
\end{table*}

\section{Methodology}\label{sec:method}

\subsection{Overview}

In general, the membership of a star in a dwarf galaxy can be judged
by the position of the star (is the star near the dwarf?), by the
color or metallicity of the star (does the star appear to have
color/metallicity properties consistent with the stellar populations
of the dwarf?), and by the dynamics of the star (is the motion of the
star consistent with the motion of the rest of the dwarf?). The latter
can be broken up into both a proper motion component and a radial
velocity component. All approaches
to the derivation of the systemic proper motion of satellites consider
some weighted combination of these criteria.

One of the ultimate science goals of our analysis is to develop
comprehensive membership lists for the faint Milky Way satellites 
(these shall be discussed in forthcoming contributions). As such, we
favor approaches in the derivation of the systemic proper motions that
simultaneously allow us to calculate stellar membership
probabilities (as opposed to cuts, which imply either a star is
definitely a member or definitely not a member). Furthermore, we want to
incorporate as much information as is available in the data, while
being aware that some types of data (i.e., radial velocities) are only
available for a subset of stars. We are also aware that the
uncertainties on some important parameters (e.g., the projected size
of the satellite) are quite uncertain in some cases.

Our adopted approach is inspired by \cite{pace2019}. As described in
that paper, for any star in
Gaia, it is either a member of the Milky Way satellite being studied, or it is a member of the
Milky Way foreground/background. We can therefore define the total
likelyhood $\mathcal{L}$ for a star as

\begin{equation}
\mathcal{L} = f_{sat}\mathcal{L}_{sat} + (1 - f_{sat})
\mathcal{L}_{MW} \label{maxl}
\end{equation}

\noindent where $\mathcal{L}_{sat}$ and $\mathcal{L}_{MW}$ are the likelihoods
for the satellite and MW foreground/background, respectively, and
$f_{sat}$ is the fraction of stars in the satellite. $\mathcal{L}_{sat}$ can be broken down as

\begin{equation}
\mathcal{L}_{sat} = \mathcal{L}_{s}\mathcal{L}_{CM}\mathcal{L}_{PM} \label{breakdown}
\end{equation}

\noindent where $\mathcal{L}_s, \mathcal{L}_{CM}$ and
$\mathcal{L}_{PM}$ are the likelihoods from the spatial information,
color-magnitude information, and proper motion information,
respectively. $\mathcal{L}_{MW}$ can similarly be broken down into the
product of its three constituent likelihoods.

Within a Bayesian framework, the probability of the data, $D$, given a set of model parameters, $\theta$, is given by:

\begin{equation}
P\left( D | \theta \right) \propto \mathcal{L} \times P(\theta)~. \label{bayes}
\end{equation}

\noindent $P(\theta)$ is
our prior on the model parameters. We aim to determine the set of model parameters that
maximizes $P\left( D | \theta \right)$. 

If it was the case that all stars in our sample also had radial
velocity information, then the radial velocity information could be
incorporated in Equation~\ref{bayes} via the incorporation of a fourth
term, $\mathcal{L}_{RV}$, in Equation~\ref{breakdown}. However,
only a tiny fraction of our data has radial velocity information. As such, only the
spatial, color-magnitude and proper motion information are
incorporated into Equation~\ref{bayes} via the likelihood,
$\mathcal{L}$. The radial velocity data, where it exists, will be
incorporated into Equation~\ref{bayes} via the prior, $P(\theta)$.

We note that once all of the relevant likelihoods have been calculated, the
probability that a star is a member of the satellite is given by

\begin{equation}
 P_{sat} =  \frac{f_{sat}\mathcal{L}_{sat}}{f_{sat}\mathcal{L}_{sat} +
   (1 - f_{sat})  \mathcal{L}_{MW}} \label{members}
\end{equation}

Within this framework, the problem of obtaining systemic proper
motions for satellites (and the related problem of identifying member
stars within satellites) becomes one of defining the appropriate likelihood functions
for the satellite and the background. Here, we are conscious of two
driving considerations. The first is that many of the satellites 
are intrinsically faint, and as such there are
significant uncertainties on their basic parameters (especially
distance). Thus, any model of their structural and color-magnitude
properties should seek to incorporate these uncertainties. The second
consideration is that the structure of the Milky Way foreground and
background is, to put it mildly, immensely complex. While variations in the
global structure as a function of position can potentially be
parameterized \citep[e.g., see][]{pace2019}, 
smaller scale variations due to known or unknown structures or
substructures are especially problematic for proper motion
analyses. Thus, we make absolutely no attempt to construct a
parameterized model of the foreground/background. Instead, 
an empirical model based entirely on the data in hand is
constructed with no unknown parameters.

Our model parameters, $\theta$, therefore consist only of the unknown systematic proper motion components, $(\mu_\alpha\cos\delta, \mu_\delta)_{sat}$, in addition to $f_{sat}$ introduced in Equation~\ref{maxl}. This parameter space is explored using {\tt emcee} (\citealt{foremanmackey2013, foremanmackey2019}). We now discuss our likelihood models for each term in Equation~\ref{breakdown}, and the form of our adopted prior, $P(\theta)$.

\subsection{Spatial distribution}

\subsubsection{Satellite}

\begin{figure*}
  \centering
    \includegraphics[width=0.8\textwidth]{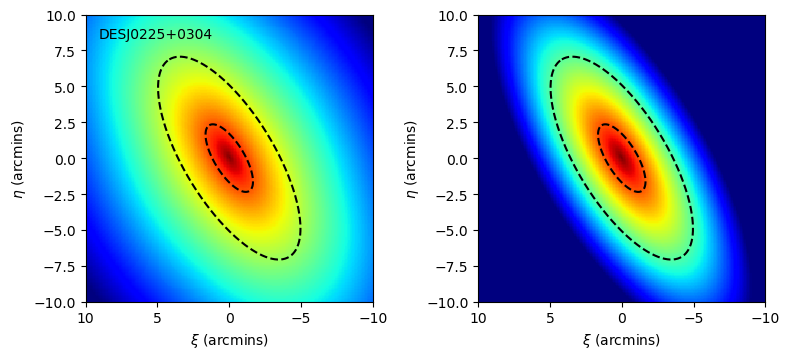}
    \caption{Left panel: the spatial likelihood map for DES J0225+0304, shown with logarithmic scaling. Right panel: same as left panel (with identical scaling) but without the uncertainties in the structural parameters being taken into account. For reference, the black ellipses in each panel correspond to 1 and 3 half-light radii from the center of the satellite.}\label{spatial}
\end{figure*}

The spatial likelihood function for a star to be a member of a dwarf galaxy is determined from the 
structural parameterization of the host galaxy, as estimated from its resolved stellar distribution. 
These are generally adequately described by an exponential function, adopted here for simplicity. 
The necessary parameters to (completely) describe the relevant two-dimensional 
exponential distributions are given in Table~\ref{gals}.

To account for (large) uncertainties, we construct a two-dimensional lookup map for the satellite spatial likelihood. This is made by co-adding a thousand realisations of the dwarf galaxy stellar density distribution, where each realisation uses parameters drawn from Gaussian distributions, centered on the reported values for $r_h$, $e$ and $\theta$ in Table~\ref{gals}. The standard deviation is set by the reported uncertainties on each parameter.  Systems that do not have reported values of ellipticity or position angle are assumed to be circular. We note that the centers of these satellites are assumed to be fixed, although a few of the faintest satellites have relatively large uncertainties in their positions. However, these satellites also have large uncertainties in their other structural parameters, such that inclusion of the uncertainties in their positions does not change their spatial likelihood functions significantly, and does not affect the systemic proper motions that are calculated.

An example of the resulting likelihood function is shown in
the left panel of Figure~\ref{spatial} for DES J0225+0304. For comparison, the right
panel shows what the spatial likelihood function would look like if
the uncertainties had not been considered. DES J0225+0304 has relatively poorly defined structural parameters; for systems with well
measured stuctural parameters, the distinction between the two panels
becomes negligible.

\subsubsection{Foreground/background}

The Milky Way contamination is assumed to be spatially uniform over the area subtended by the dwarf. For satellites at low Galactic latitude, there is a gradient due to the disk. Further, there is also the possibility of small substructures in the (projected) vicinity of the dwarfs. However, the assumption of uniformity appears to be reasonable given the results we obtain, especially the minimal amounts of foreground contamination that we measure (see discussion in Section~\ref{sec:cont}).

\subsection{Color-magnitude distribution}

\begin{figure*}
  \centering
    \includegraphics[width=0.3\textwidth]{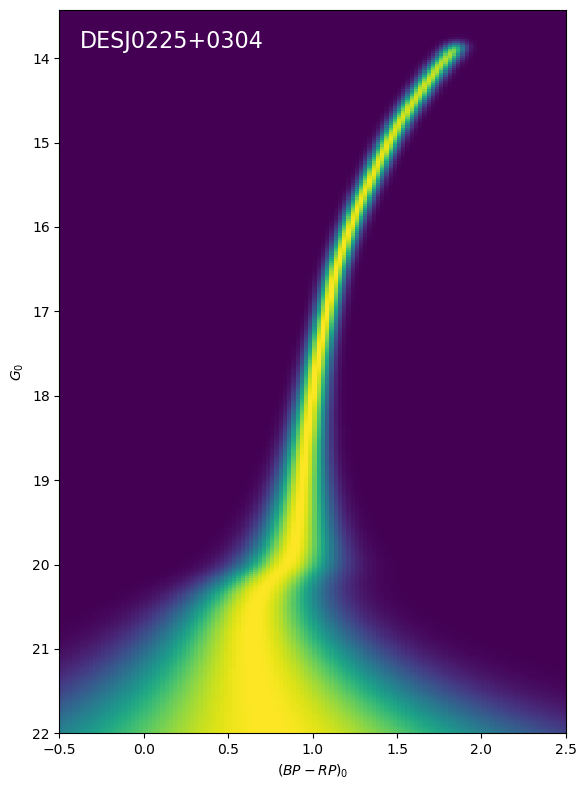}
    \includegraphics[width=0.3\textwidth]{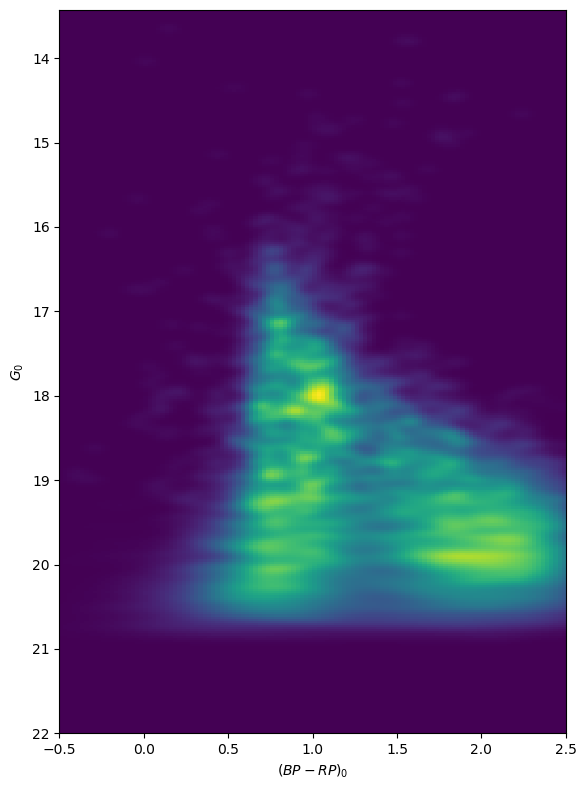}
    \caption{Left panel: the color-magnitude likelihood map for DES J0225+0304, shown with linear scaling. Right panel: the color-magnitude likelihood map for the Milky Way contamination in the field of DES J0225+0304, also with linear scaling.}\label{cmdlikelihood}
\end{figure*}

\subsubsection{Satellite}

The majority of the satellites under consideration are
dominated by relatively old, metal poor stellar populations. Mean
metallicities (estimated using a variety of techniques) exist for most of them, and
so it is relatively easy to construct a simple model of the expected
color-magnitude likelihood function for the satellite.

In the first instance, we adopt a 12\,Gyr Padova isochrone (\citealt{girardi2002})
in the Gaia photometric bands using the filter definitions from \cite{weiler2018}.
For each dwarf satellite, the mean metallicity in Table~\ref{gals} is adopted,
and the isochrone shifted to the appropriate distance. 
Post-Helium flash stars are not considered.
At every magnitude, the stellar population is modeled as a Gaussian function 
with an intrinsic full width
at half maximum of 0.1 magnitudes in color, centered on the color of
the isochrone. This is combined with the mean uncertainty in color at
each magnitude for the stars under consideration. 
For a few of the dwarfs (Carina, Fornax,
Leo 1, Leo 2 and Phoenix), a younger age is adopted
for the isochrone to better match the color of the red giant branch stars.

In a similar way to the spatial distribution, a 2D lookup
map is constructed for the color-magnitude likelihood function using the above methodology
(with stars brighter than the tip of the red giant branch having zero
likelihood of being a member). However, we co-add a thousand
realisations, where the isochrone is moved to a distance
selected from a Gaussian distribution centered
on the recorded distance modulus of the dwarf, with a standard
deviation equal to the uncertainty on the distance modulus. In this
way, the (sometimes significant) distance uncertainties on the ultra-faint 
dwarf galaxies are taken into account, which otherwise makes
a single realisation of color-magnitude space unsatisfactory.

The left panel of Figure~\ref{cmdlikelihood}
shows an example of the resulting satellite likelihood model for color-magnitude
space for the case of DES J0225+0304. 
We only consider stars defined within our likelihood grid; 
$-0.5 < (BP - RP) < 2.5$ and $22 < G < G_{TRGB}+ 5\,\sigma_{(m-M)_o}$, 
where $\sigma_{(m-M)_o}$ is the uncertainty in the distance modulus, given in Table~\ref{gals}

Padova isochrones for metallicities below [Fe/H] $= -2.19$\,dex were not available. 
As such, we adopted this, the most metal poor isochrone, for any galaxy that was 
more metal poor than this limit.  This is acceptable as the change in color at 
these low metallicities is very small. 
For any satellites that lack a metallicity estimate, $<$[Fe/H]$> = -2$ was adopted.
We note that it would also be possible to draw the metallicity from a 
Gaussian function (also for age); however, this does not have a significant
impact on our the results. 
Only the red giant branch and upper main sequence are considered when
constructing these models, and we ignore stars on the horizontal branch and
asymptotic giant branch. 
Thus, by construction, a star lying on an isochrone at any magnitude is 
considered a likely member of the dwarf satellite. 
It is clearly possible to develop a more sophisticated model,
e.g., taking into account additional stellar populations and the
relative number of sources as a function of magnitude or stellar
evolutionary phase; however, we did not find this to be necessary 
for the task in hand.

\subsubsection{Foreground/background}

For the color-magnitude likelihood function of the Milky Way contamination, 
we assume the contamination at the position of the satellite has the same 
statistical properties as the Milky Way population near the satellite. 
In practice, a hole is excised in the Gaia DR2 catalog for each satellite 
corresponding to five half light radii centered on the satellite. 
A 2D lookup map of the color-magnitude distribution for all remaining stars
is constructed, where each star is mapped as a bivariate Gaussian, 
with standard deviations corresponding to the 
recorded uncertainties in color and magnitude. 
Suitably normalised, this 2D map of the background becomes our likelihood function. 
The right panel of Figure~\ref{cmdlikelihood}
shows an example of the Milky Way likelihood function for
the color-magnitude space of DES J0225+0304. 

\subsection{Proper motion distribution}

\subsubsection{Satellite}

For the likelihood of the proper motion of the satellite, 
all members are assumed to share the same systemic proper motion,
$\left(\mu_\alpha\cos\delta, \mu_\delta\right)_{sat}$. The spread in recorded proper motion values for stars in each satellite is assumed to be entirely due to measurement errors, since these dominate over the  expected intrinsic spread in proper motions for each satellite. Thus, the satellite proper motion likelihood function
is described by a bivariate Gaussian with a covariance matrix defined
by the reported proper motion uncertainties and their correlation for
each star under consideration.  

\subsubsection{Foreground/background}

\begin{figure}
  \centering
    \includegraphics[width=0.9\columnwidth]{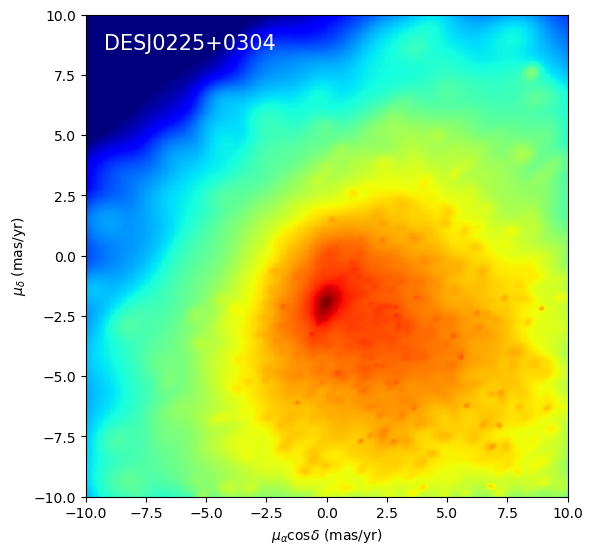}
    \caption{The proper motion likelihood map for the Milky Way contamination in the field of DES J0225+0304, shown with logarithmic scaling.}\label{pmlikelihood}
\end{figure}

A 2D lookup map for the Milky Way proper motion likelihood function is constructed
using the same approach as for the Milky Way CMD likelihood function (above). 
Specifically, we excise the same area as before and construct a proper motion map,
where each remaining star is a bivariate Gaussian, using the reported stellar proper
motion uncertainties and their correlation to define the relevant covariance
matrices. Figure~\ref{pmlikelihood}
shows an example of the Milky Way proper motion likelihood function, for the case of DES J0225+0304. 
Note that we only consider stars in the likelihood grid with an absolute 
proper motion less than or equal to 10\,mas/yr in each direction. For the closest satellites in our sample, this corresponds to tangential velocities of around 1000\,km\,s$^{-1}$, ensuring that the possible proper motion space for the satellite is fully explored.

\subsection{Priors on the likelihood and incorporation of radial
  velocity data}\label{prior}

There are three unknown parameters in our model: $f_{sat}$ (the
fraction of stars belonging to the satellite), and the two components
of the systemic proper motion of the satellite,
$\left(\mu_\alpha\cos\delta, \mu_\delta\right)_{sat}$. For the
former, we adopt a uniform prior between 0 and 1.

Two priors are selected on the systemic proper motions:

\begin{enumerate}
\item {\it Prior A:} The first prior assumes that the dispersion in the set of tangential
  velocities of the satellites is somewhat similar to the dispersion in the radial
  velocities of known halo tracer populations (including the
  satellites). The radial velocity dispersion profile of the outer Galaxy
  has been quantified by numerous authors (e.g., \citealt{deason2012b, bhattacharjee2014}) and is
  reasonably characterized as a roughly constant velocity dispersion at
  $\sigma_r \simeq 100$\,km\,s$^{-1}$. Therefore, we adopt a
  Gaussian prior on the two dimensional proper motion, such that it is
  centered on the equivalent of a Galactocentric velocity of zero
  (after the reflex motion of the Sun has been taken into account) and
  has a dispersion equivalent to 100\,km\,s$^{-1}$ at the distance of
  the satellite. This
  prior is not particularly restrictive, as its primary purpose is to
  inhibit unphysically large tangential velocities; for example,
  velocities greatly in excess of the escape velocity of the Galaxy. However, it does have the effect of creating an ``envelope'' of $\sim 100$\,km\,s$^{-1}$ in the maximum uncertainties on the systemic proper motions that we derive, where this value is driven by the prior, not the data. In addition, we note that for the three most distant galaxies in our sample - Eridanus 2, Phoenix and Leo T - it is not clear if these objects should trace the tangential velocity dispersion of the halo of the Milky Way. As such, for these three galaxies, we provide estimates with and without this prior.

\item {\it Prior B:} For satellites with radial velocity follow-up of
  individual stars, we have additional information to assess whether the
  stars are members of the satellite. For those stars that appear
  likely members, we modify the above prior to favor values of the
  systemic proper motion that will increase the probability of those
  stars being considered members.

We search for stars with radial velocities 
that are within $2\,\sigma$ of the mean radial velocity of the
satellite. For the satellites in Table~\ref{gals}, where the mean radial 
velocities are known but the radial velocity dispersions are unknown, 
we assume $\sigma_v = 5$\,km\,s$^{-1}$. 
Simultaneously, a $2\,\sigma$ cut around the isochrone 
in the color-magnitude likelihood is adopted, as well as a cut inside two half-light radii in the spatial likelihood 
map.

For stars that satisfy these three criteria (radial velocity
likelihood, CMD likelihood, spatial likelihood), the
weighted mean of their proper motions is calculated. 
One iteration of sigma-clipping is performed to remove any 
obviously deviant proper motions. This is in line with the 
approach taken by many authors to calculate the proper motions 
of satellites from radial velocity data 
\citep[e.g.,][]{fritz2018, simon2019}.
The resulting mean and uncertainities are interpreted as a
bivariate Gaussian, that we dub ``the velocity prior''. 
This velocity prior is multiplied into Prior A to create Prior B. 
This will heavily favor results that are consistent with stars that 
appear to be members in all of their other characteristics, 
including radial velocities, while still maintaining consistency 
with our original prior.  Importantly, stars without radial velocities still
contribute to the selection of the preferred model through the
likelihood function, ensuring that the full Gaia dataset 
is still utilized.

\end{enumerate}

\section{Results}\label{sec:results}

\begin{table*} {\scriptsize
\begin{center}
\begin{tabular*}{0.9\textwidth}{l|rrr|rrr}
&\multicolumn{3}{|c|}{Prior A}&  \multicolumn{3}{|c}{Prior B}\\   
Galaxy &$f_{sat}$ & $\mu_\alpha\cos\delta$ (mas/yr)&$\mu_\delta$ (mas/yr)& $f_{sat}$ & $\mu_\alpha\cos\delta$ (mas/yr)&$\mu_\delta$ (mas/yr)\\
  \hline
  Antlia2 &  $ 0.00007 \pm 0.00001 $  &  $ -0.03 \pm 0.05 $  &  $ 0.05 \pm 0.06 $  &  $ 0.00007 \pm 0.00001 $  &  $ -0.05 \pm 0.04 $  &  $ 0.04^{+0.04}_{-0.05}$ \\
Aquarius2 &  $ 0.0007^{+0.0004}_{-0.0003} $  &  $ 0.03 \pm 0.16 $  &  $ -0.24 \pm 0.16 $  &  $ 0.0007^{+0.0004}_{-0.0003} $  &  $ -0.0 \pm 0.16 $  &  $ -0.2^{+0.16}_{-0.15} $ \\
Bootes1 &  $ 0.007 \pm 0.001 $  &  $ -0.46 \pm 0.05 $  &  $ -1.06 \pm 0.04 $  &  $ 0.007 \pm 0.001 $  &  $ -0.47 \pm 0.04 $  &  $ -1.07 \pm 0.03 $ \\
Bootes2 &  $ 0.0011^{+0.0004}_{-0.0003} $  &  $ -2.09^{+0.25}_{-0.23} $  &  $ -0.63^{+0.18}_{-0.17} $  &  $ 0.0011^{+0.0004}_{-0.0003} $  &  $ -2.25 \pm 0.21 $  &  $ -0.63 \pm 0.15 $ \\
CanesVenatici1 &  $ 0.017 \pm 0.002 $  &  $ -0.23 \pm 0.06 $  &  $ -0.06 \pm 0.04 $  &  $ 0.017 \pm 0.002 $  &  $ -0.26 \pm 0.05 $  &  $ -0.06 \pm 0.03 $ \\
CanesVenatici2 &  $ 0.002 \pm 0.001 $  &  $ -0.28^{+0.12}_{-0.11} $  &  $ -0.34 \pm 0.11 $  &  $ 0.002 \pm 0.001 $  &  $ -0.34 \pm 0.11 $  &  $ -0.35 \pm 0.1 $ \\
  Carina &  $ 0.0066 \pm 0.0002 $  &  $ 0.48 \pm 0.01 $  &  $ 0.13 \pm 0.01 $  &  $ 0.0066 \pm 0.0002 $  &  $ 0.48 \pm 0.01 $  &  $ 0.13 \pm 0.01 $ \\
Carina2 &  $ 0.0003 \pm 0.0001 $  &  $ 1.83 \pm 0.03 $  &  $ 0.11 \pm 0.03 $  &  $ 0.0003 \pm 0.0001 $  &  $ 1.84 \pm 0.03 $  &  $ 0.11 \pm 0.03 $ \\
Carina3 &  $ 0.0001^{+0.00005}_{-0.00004} $  &  $ 2.95^{+0.12}_{-1.4} $  &  $ 1.42^{+0.14}_{-1.27} $  &  $ 0.00011^{+0.00005}_{-0.00004} $  &  $ 2.99 \pm 0.08 $  &  $ 1.49^{+0.1}_{-0.09} $ \\
Centaurus1 &  $ 0.00011^{+0.00005}_{-0.00004} $  &  $ -0.13 \pm 0.13 $  &  $ -0.17 \pm 0.14 $  &  ---  &  ---  &  --- \\
Cetus2 &  $ 0.0007^{+0.0004}_{-0.0003} $  &  $ 2.24^{+0.63}_{-1.05} $  &  $ 0.37^{+0.16}_{-0.66} $  &   ---  &  ---  &  --- \\
Columba1 &  $ 0.0004^{+0.0002}_{-0.0001} $  &  $ 0.2 \pm 0.09 $  &  $ -0.1 \pm 0.1 $  &  $ 0.0004^{+0.0002}_{-0.0001} $  &  $ 0.21 \pm 0.09 $  &  $ -0.14^{+0.1}_{-0.09} $ \\
ComaBerenices &  $ 0.004 \pm 0.001 $  &  $ 0.49 \pm 0.05 $  &  $ -1.65 \pm 0.04 $  &  $ 0.004 \pm 0.001 $  &  $ 0.5 \pm 0.05 $  &  $ -1.67 \pm 0.04 $ \\
Crater2 &  $ 0.003 \pm 0.0003 $  &  $ -0.14 \pm 0.05 $  &  $ -0.08 \pm 0.03 $  &  $ 0.0029 \pm 0.0003 $  &  $ -0.17 \pm 0.04 $  &  $ -0.09 \pm 0.02 $ \\
DESJ0225+0304 &  $ 0.0003^{+0.0003}_{-0.0002} $  &  $ 1.31^{+0.83}_{-0.76} $  &  $ -1.13^{+0.85}_{-0.97} $  &   ---  &  ---  &  --- \\
Draco &  $ 0.046 \pm 0.002 $  &  $ -0.01 \pm 0.01 $  &  $ -0.14 \pm 0.01 $  &  $ 0.047^{+0.002}_{-0.001} $  &  $ -0.01 \pm 0.01 $  &  $ -0.14 \pm 0.01 $ \\
Draco2 &  $ 0.0014 \pm 0.0004 $  &  $ 1.02^{+0.23}_{-0.22} $  &  $ 0.99^{+0.23}_{-0.24} $  &  $ 0.0014 \pm 0.0004 $  &  $ 1.06 \pm 0.17 $  &  $ 0.96 \pm 0.18 $ \\
Eridanus2 &  $ 0.003 \pm 0.001 $  &  $ 0.1 \pm 0.05 $  &  $ -0.06 \pm 0.05 $  &  $ 0.003 \pm 0.001 $  &  $ 0.11 \pm 0.05 $  &  $ -0.06 \pm 0.05 $ \\
Eridanus3 &  $ 0.0002^{+0.0002}_{-0.0001} $  &  $ 0.73 \pm 0.18 $  &  $ -0.35 \pm 0.17 $  &   ---  &  ---  &  --- \\
  Fornax &  $ 0.233 \pm 0.002 $  &  $ 0.380 \pm 0.003 $  &  $ -0.416 \pm 0.004 $  &  $ 0.234 \pm 0.002 $  &  $ 0.380 \pm 0.002 $  &  $ -0.416^{+0.004}_{-0.003}$ \\
Grus1 &  $ 0.0005 \pm 0.0002 $  &  $ -0.03^{+0.13}_{-0.12} $  &  $ -0.39 \pm 0.14 $  &  $ 0.0005 \pm 0.0002 $  &  $ -0.05 \pm 0.12 $  &  $ -0.41 \pm 0.14 $ \\
Grus2 &  $ 0.0009 \pm 0.0003 $  &  $ 0.45^{+0.09}_{-0.08} $  &  $ -1.45^{+0.13}_{-0.19} $  &  $ 0.0008 \pm 0.0003 $  &  $ 0.48 \pm 0.06 $  &  $ -1.41^{+0.08}_{-0.09} $ \\
Hercules &  $ 0.0006^{+0.0002}_{-0.0001} $  &  $ -0.15 \pm 0.09 $  &  $ -0.39 \pm 0.07 $  &  $ 0.0006^{+0.0002}_{-0.0001} $  &  $ -0.13 \pm 0.07 $  &  $ -0.39 \pm 0.06 $ \\
Horologium1 &  $ 0.0016^{+0.0005}_{-0.0004} $  &  $ 0.91 \pm 0.07 $  &  $ -0.55 \pm 0.06 $  &  $ 0.0016^{+0.0005}_{-0.0004} $  &  $ 0.87 \pm 0.05 $  &  $ -0.58 \pm 0.05 $ \\
Horologium2 &  $ 0.0002^{+0.0003}_{-0.0002} $  &  $ 0.59^{+0.24}_{-0.25} $  &  $ -0.16 \pm 0.25 $  &  $ 0.0003^{+0.0003}_{-0.0002} $  &  $ 0.89^{+0.22}_{-0.23} $  &  $ -0.21 \pm 0.25 $ \\
Hydra2 &  $ 0.0003 \pm 0.0001 $  &  $ -0.27 \pm 0.14 $  &  $ -0.04^{+0.13}_{-0.12} $  &  $ 0.0003 \pm 0.0001 $  &  $ -0.26 \pm 0.13 $  &  $ -0.05 \pm 0.12 $ \\
Hydrus1 &  $ 0.0017 \pm 0.0003 $  &  $ 3.79^{+0.03}_{-0.04} $  &  $ -1.53^{+0.04}_{-0.03} $  &  $ 0.0017 \pm 0.0003 $  &  $ 3.77 \pm 0.03 $  &  $ -1.55 \pm 0.03 $ \\
  Indus1 &  $ 0.0001^{+0.00008}_{-0.00005} $  &  $ 0.21^{+0.16}_{-0.19} $  &  $ -0.72^{+0.31}_{-0.19} $  &   ---  &  ---  &  --- \\
Leo1 &  $ 0.008 \pm 0.002 $  &  $ -0.05 \pm 0.08 $  &  $ -0.18 \pm 0.08 $  &  $ 0.008 \pm 0.002 $  &  $ -0.06 \pm 0.07 $  &  $ -0.18 \pm 0.08 $ \\
Leo2 &  $ 0.024^{+0.003}_{-0.002} $  &  $ -0.11 \pm 0.06 $  &  $ -0.18 \pm 0.06 $  &  $ 0.024^{+0.003}_{-0.002} $  &  $ -0.12 \pm 0.06 $  &  $ -0.17 \pm 0.06 $ \\
Leo4 &  $ 0.0004^{+0.0003}_{-0.0002} $  &  $ -0.17 \pm 0.13 $  &  $ -0.26 \pm 0.13 $  &  $ 0.0004^{+0.0003}_{-0.0002} $  &  $ -0.2 \pm 0.13 $  &  $ -0.26 \pm 0.12 $ \\
Leo5 &  $ 0.0005^{+0.0004}_{-0.0003} $  &  $ -0.1 \pm 0.11 $  &  $ -0.21 \pm 0.1 $  &  $ 0.0005^{+0.0004}_{-0.0003} $  &  $ -0.11^{+0.1}_{-0.11} $  &  $ -0.21 \pm 0.1 $ \\
LeoT &  $ 0.002 \pm 0.001 $  &  $ -0.01 \pm 0.05 $  &  $ -0.11 \pm 0.05 $  &  $ 0.002 \pm 0.001 $  &  $ -0.01 \pm 0.05 $  &  $ -0.11 \pm 0.05 $ \\
Phoenix &  $ 0.009 \pm 0.002 $  &  $ 0.08 \pm 0.05 $  &  $ -0.08 \pm 0.05 $  &  $ 0.009 \pm 0.002 $  &  $ 0.08 \pm 0.05 $  &  $ -0.08 \pm 0.05 $ \\
Phoenix2 &  $ 0.0006^{+0.0003}_{-0.0002} $  &  $ 0.41 \pm 0.09 $  &  $ -1.0 \pm 0.11 $  &  $ 0.0006^{+0.0003}_{-0.0002} $  &  $ 0.44^{+0.08}_{-0.07} $  &  $ -1.03 \pm 0.09 $ \\
Pictor1 &  $ 0.0005 \pm 0.0002 $  &  $ 0.18 \pm 0.13 $  &  $ 0.0 \pm 0.15 $  &   ---  &  ---  &  --- \\
Pictor2 &  $ 0.00008^{+0.00007}_{-0.00005} $  &  $ 1.18^{+0.14}_{-0.47} $  &  $ 1.15^{+0.13}_{-0.75} $  &   ---  &  ---  &  --- \\
Pisces2 &  $ 0.0002^{+0.0002}_{-0.0001} $  &  $ 0.08^{+0.12}_{-0.11} $  &  $ -0.22 \pm 0.11 $  &  $ 0.0002^{+0.0002}_{-0.0001} $  &  $ 0.07 \pm 0.11 $  &  $ -0.26 \pm 0.11 $ \\
Reticulum2 &  $ 0.004 \pm 0.001 $  &  $ 2.38 \pm 0.04 $  &  $ -1.3 \pm 0.04 $  &  $ 0.004 \pm 0.001 $  &  $ 2.39 \pm 0.03 $  &  $ -1.3 \pm 0.03 $ \\
Reticulum3 &  $ 0.0002^{+0.0002}_{-0.0001} $  &  $ 0.2^{+0.24}_{-0.23} $  &  $ -0.09 \pm 0.22 $  &  $ 0.0002^{+0.0002}_{-0.0001} $  &  $ 0.05 \pm 0.21 $  &  $ -0.09 \pm 0.2 $ \\
Sagittarius2 &  $ 0.00011^{+0.00004}_{-0.00003} $  &  $ -0.62 \pm 0.08 $  &  $ -0.94 \pm 0.05 $  &$ 0.00011^{+0.00004}_{-0.00003} $  &  $ -0.65 \pm 0.07 $  &  $ -0.96 \pm 0.04 $   \\
Sculptor &  $ 0.315 \pm 0.005 $  &  $ 0.082 \pm 0.005 $  &  $ -0.133 \pm 0.005 $  &  $ 0.315 \pm 0.005 $  &  $ 0.081 \pm 0.005 $  &  $ -0.136 \pm 0.004 $ \\
Segue1 &  $ 0.0011^{+0.0006}_{-0.0005} $  &  $ -1.67^{+0.46}_{-0.37} $  &  $ -3.43^{+0.44}_{-0.33} $  &  $ 0.0011^{+0.0006}_{-0.0005} $  &  $ -1.59^{+0.23}_{-0.22} $  &  $ -3.5^{+0.2}_{-0.21} $ \\
Segue2 &  $ 0.0007^{+0.0004}_{-0.0003} $  &  $ 1.6^{+0.19}_{-0.2} $  &  $ 0.05 \pm 0.13 $  &  $ 0.0006 \pm 0.0003 $  &  $ 1.68^{+0.12}_{-0.13} $  &  $ 0.12 \pm 0.08 $ \\
  Sextans1 &  $ 0.0116 \pm 0.0005 $  &  $ -0.43 \pm 0.02 $  &  $ 0.09 \pm 0.02 $  &  $ 0.0116 \pm 0.0005 $  &  $ -0.44 \pm 0.02 $  &  $ 0.09 \pm 0.02 $ \\
Triangulum2 &  $ 0.00013^{+0.00010}_{-0.00007} $  &  $ 0.76^{+0.23}_{-0.18} $  &  $ 0.33^{+0.15}_{-0.27} $  &  $ 0.00013^{+0.00010}_{-0.00007} $  &  $ 0.62^{+0.13}_{-0.15} $  &  $ 0.42 \pm 0.11 $ \\
Tucana2 &  $ 0.0011 \pm 0.0003 $  &  $ 0.92 \pm 0.06 $  &  $ -1.14 \pm 0.08 $  &  $ 0.0011 \pm 0.0003 $  &  $ 0.94 \pm 0.05 $  &  $ -1.22 \pm 0.06 $ \\
Tucana3 &  $ 0.003 \pm 0.001 $  &  $ -0.03 \pm 0.04 $  &  $ -1.66 \pm 0.04 $  &  $ 0.003 \pm 0.001 $  &  $ -0.02 \pm 0.03 $  &  $ -1.67 \pm 0.03 $ \\
Tucana4 &  $ 0.0004^{+0.0003}_{-0.0002} $  &  $ 0.68^{+0.62}_{-0.29} $  &  $ -1.36^{+0.44}_{-0.22} $  &  $ 0.0004^{+0.0003}_{-0.0002} $  &  $ 0.63^{+0.13}_{-0.12} $  &  $ -1.54^{+0.1}_{-0.11} $ \\
Tucana5 &  $ 0.0002^{+0.0002}_{-0.0001} $  &  $ -0.04^{+0.18}_{-0.12} $  &  $ -1.02^{+0.3}_{-0.11} $  &  $ 0.0002^{+0.0002}_{-0.0001} $  &  $ -0.1^{+0.09}_{-0.2} $  &  $ -1.01^{+0.23}_{-0.1} $ \\
UrsaMajor1 &  $ 0.003 \pm 0.001 $  &  $ -0.5 \pm 0.07 $  &  $ -0.65 \pm 0.09 $  &  $ 0.003 \pm 0.001 $  &  $ -0.56 \pm 0.06 $  &  $ -0.68 \pm 0.08 $ \\
UrsaMajor2 &  $ 0.002 \pm 0.001 $  &  $ 1.73 \pm 0.05 $  &  $ -1.87 \pm 0.06 $  &  $ 0.002 \pm 0.001 $  &  $ 1.72 \pm 0.05 $  &  $ -1.84 \pm 0.06 $ \\
UrsaMinor &  $ 0.024 \pm 0.001 $  &  $ -0.16 \pm 0.01 $  &  $ 0.06 \pm 0.01 $  &  $ 0.024 \pm 0.001 $  &  $ -0.16 \pm 0.01 $  &  $ 0.06 \pm 0.01 $ \\
Willman1 &  $ 0.0004^{+0.0003}_{-0.0002} $  &  $ 0.34 \pm 0.15 $  &  $ -1.06^{+0.24}_{-0.25} $  &  $ 0.0004^{+0.0003}_{-0.0002} $  &  $ 0.36 \pm 0.1 $  & $ -1.04 \pm 0.18 $ \\
\end{tabular*}
\caption{Median, 14th and 86th percentiles of the probability density functions for our three unknown parameters, for 54 out of 59 of the satellites in our sample.  We do not derive robust solutions for Bootes 4, Cetus 3, Indus 2, Pegasus 3 and Virgo 1, and we argue that this is due to a lack of member stars with reliable data in Gaia DR2. Columns 2, 3 and 4 are the derived results using Prior A, and columns 5, 6 and 7 are the derived results using Prior B.} 
\label{pmresults}
\end{center} }
\end{table*}

\begin{table*} {\scriptsize
\begin{center}
\begin{tabular*}{0.9\textwidth}{l|rrrrrr}
Galaxy &$\mu_\alpha\cos\delta$ (mas/yr)&$\mu_\delta$ (mas/yr)& $v_\alpha\cos\delta$ (km\,s$^{-1}$) & $v_\delta$ (km\,s$^{-1}$)& $v_t$ (km\,s$^{-1}$)& $v_r$ (km\,s$^{-1}$)\\
\hline 
Antlia2\tablenotemark{\scriptsize{a}} &   $ -0.05 \pm 0.04 $   &   $ 0.04^{+0.04}_{-0.05}$ \ & $ -2 \pm 25 $ &  $ 72^{+25}_{-31} $ & 72 & 55 \\ 
Aquarius2 &   $ -0.0 \pm 0.16 $   &   $ -0.2^{+0.16}_{-0.15} $ \ & $ -69 \pm 82 $ &  $ 98^{+82}_{-77} $ & 120 & 46 \\ 
Bootes1\tablenotemark{\scriptsize{a}} &   $ -0.47 \pm 0.04 $   &   $ -1.07 \pm 0.03 $ \ & $ 11 \pm 13 $ &  $ -155 \pm 9 $ & 155 & 107 \\ 
Bootes2 &   $ -2.25 \pm 0.21 $   &   $ -0.63 \pm 0.15 $ \ & $ -286 \pm 41 $ &  $ 58 \pm 30 $ & 292 & -116 \\ 
CanesVenatici1\tablenotemark{\scriptsize{a}} &   $ -0.26 \pm 0.05 $   &   $ -0.06 \pm 0.03 $ \ & $ -117 \pm 52 $ &  $ 120 \pm 31 $ & 167 & 80 \\ 
CanesVenatici2 &   $ -0.34 \pm 0.11 $   &   $ -0.35 \pm 0.1 $ \ & $ -115 \pm 83 $ &  $ -73 \pm 76 $ & 137 & -94 \\ 
Carina\tablenotemark{\scriptsize{a}} &   $ 0.48 \pm 0.01 $   &   $ 0.13 \pm 0.01 $ \ & $ 150 \pm 5 $ &  $ 71 \pm 5 $ & 166 & -2 \\ 
Carina2\tablenotemark{\scriptsize{a}} &   $ 1.84 \pm 0.03 $   &   $ 0.11 \pm 0.03 $ \ & $ 261 \pm 5 $ &  $ -17 \pm 5 $ & 262 & 245 \\ 
Carina3 &   $ 2.99 \pm 0.08 $   &   $ 1.49^{+0.1}_{-0.09} $ \ & $ 341 \pm 11 $ &  $ 160^{+13}_{-12} $ & 377 & 52 \\ 
Centaurus1 &   $ -0.13 \pm 0.13 $   &   $ -0.17 \pm 0.14 $ \ & $ 64 \pm 72 $ &  $ -19 \pm 77 $ & 66 & --- \\ 
Cetus2 &   $ 2.24^{+0.63}_{-1.05} $   &   $ 0.37^{+0.16}_{-0.66} $  \ & $ 130^{+79}_{-131} $ &  $ 236^{+20}_{-82} $ & 270 & --- \\ 
Columba1 &   $ 0.21 \pm 0.09 $   &   $ -0.14^{+0.1}_{-0.09} $ \ & $ 55 \pm 78 $ &  $ -12^{+86}_{-78} $ & 56 & -21 \\ 
ComaBerenices\tablenotemark{\scriptsize{a}} &   $ 0.5 \pm 0.05 $   &   $ -1.67 \pm 0.04 $ \ & $ 234 \pm 10 $ &  $ -143 \pm 8 $ & 275 & 81 \\ 
Crater2\tablenotemark{\scriptsize{a}} &   $ -0.17 \pm 0.04 $   &   $ -0.09 \pm 0.02 $ \ & $ 18 \pm 22 $ &  $ 82 \pm 11 $ & 84 & -80 \\ 
DESJ0225+0304 &   $ 1.31^{+0.83}_{-0.76} $   &   $ -1.13^{+0.85}_{-0.97} $ \ & $ -14^{+94}_{-86} $ &  $ 49^{+96}_{-109} $ & 52 & --- \\ 
Draco\tablenotemark{\scriptsize{a}} &   $ -0.01 \pm 0.01 $   &   $ -0.14 \pm 0.01 $ \ & $ 127 \pm 4 $ &  $ -38 \pm 4 $ & 133 & -88 \\ 
Draco2 &   $ 1.06 \pm 0.17 $   &   $ 0.96 \pm 0.18 $ \ & $ 266 \pm 17 $ &  $ 137 \pm 18 $ & 300 & -164 \\ 
Eridanus2 &   $ 0.11 \pm 0.05 $   &   $ -0.06 \pm 0.05 $ \ & $ 39 \pm 90 $ &  $ -4 \pm 90 $ & 39 & -73 \\ 
Eridanus3 &   $ 0.73 \pm 0.18 $   &   $ -0.35 \pm 0.17 $  \ & $ 140 \pm 74 $ &  $ -18 \pm 70 $ & 141 & --- \\ 
Fornax\tablenotemark{\scriptsize{a}} &   $ 0.380 \pm 0.002 $   &   $ -0.416^{+0.004}_{-0.003}$ \ & $ 102 \pm 1 $ &  $ -138^{+3}_{-2} $ & 172 & -37 \\ 
Grus1 &   $ -0.05 \pm 0.12 $   &   $ -0.41 \pm 0.14 $ \ & $ -112 \pm 68 $ &  $ -12 \pm 80 $ & 112 & -187 \\ 
Grus2\tablenotemark{\scriptsize{a}} &   $ 0.48 \pm 0.06 $   &   $ -1.41^{+0.08}_{-0.09} $ \ & $ 71 \pm 15 $ &  $ -119^{+20}_{-23} $ & 138 & -132 \\ 
Hercules\tablenotemark{\scriptsize{a}} &   $ -0.13 \pm 0.07 $   &   $ -0.39 \pm 0.06 $ \ & $ 68 \pm 44 $ &  $ -85 \pm 37 $ & 108 & 150 \\ 
Horologium1\tablenotemark{\scriptsize{a}} &   $ 0.87 \pm 0.05 $   &   $ -0.58 \pm 0.05 $ \ & $ 164 \pm 19 $ &  $ -115 \pm 19 $ & 201 & -32 \\ 
Horologium2 &   $ 0.89^{+0.22}_{-0.23} $   &   $ -0.21 \pm 0.25 $ \ & $ 167^{+81}_{-85} $ &  $ 24 \pm 92 $ & 168 & 22 \\ 
Hydra2 &   $ -0.26 \pm 0.13 $   &   $ -0.05 \pm 0.12 $ \ & $ -37 \pm 83 $ &  $ 66 \pm 76 $ & 76 & 123 \\ 
Hydrus1\tablenotemark{\scriptsize{a}} &   $ 3.77 \pm 0.03 $   &   $ -1.55 \pm 0.03 $ \ & $ 330 \pm 4 $ &  $ -152 \pm 4 $ & 363 & -91 \\ 
Indus1 &   $ 0.21^{+0.16}_{-0.19} $   &   $ -0.72^{+0.31}_{-0.19} $  \ & $ 89^{+76}_{-90} $ &  $ -103^{+147}_{-90} $ & 136 & --- \\ 
Leo1 &   $ -0.06 \pm 0.07 $   &   $ -0.18 \pm 0.08 $ \ & $ -20 \pm 84 $ &  $ -9 \pm 96 $ & 22 & 169 \\ 
Leo2\tablenotemark{\scriptsize{a}} &   $ -0.12 \pm 0.06 $   &   $ -0.17 \pm 0.06 $ \ & $ -39 \pm 66 $ &  $ 27 \pm 66 $ & 48 & 21 \\ 
Leo4 &   $ -0.2 \pm 0.13 $   &   $ -0.26 \pm 0.12 $ \ & $ -42 \pm 95 $ &  $ -13 \pm 88 $ & 44 & 5 \\ 
Leo5 &   $ -0.11^{+0.1}_{-0.11} $   &   $ -0.21 \pm 0.1 $ \ & $ 1^{+93}_{-102} $ &  $ -12 \pm 93 $ & 12 & 54 \\ 
LeoT &   $ -0.01 \pm 0.05 $   &   $ -0.11 \pm 0.05 $ \ & $ 9 \pm 99 $ &  $ 0 \pm 99 $ & 9 & -63 \\ 
Phoenix &   $ 0.08 \pm 0.05 $   &   $ -0.08 \pm 0.05 $ \ & $ -2 \pm 97 $ &  $ 3 \pm 97 $ & 3 & -114 \\ 
Phoenix2 &   $ 0.44^{+0.08}_{-0.07} $   &   $ -1.03 \pm 0.09 $ \ & $ 65^{+32}_{-28} $ &  $ -203 \pm 35 $ & 213 & -41 \\ 
Pictor1 &   $ 0.18 \pm 0.13 $   &   $ 0.0 \pm 0.15 $   \ & $ -47 \pm 71 $ &  $ 56 \pm 82 $ & 73 & --- \\ 
Pictor2 &   $ 1.18^{+0.14}_{-0.47} $   &   $ 1.15^{+0.13}_{-0.75} $  \ & $ 168^{+30}_{-102} $ &  $ 219^{+28}_{-163} $ & 276 & --- \\ 
Pisces2 &   $ 0.07 \pm 0.11 $   &   $ -0.26 \pm 0.11 $ \ & $ -24 \pm 95 $ &  $ -63 \pm 95 $ & 67 & -69 \\ 
Reticulum2\tablenotemark{\scriptsize{a}} &   $ 2.39 \pm 0.03 $   &   $ -1.3 \pm 0.03 $ \ & $ 182 \pm 4 $ &  $ -105 \pm 4 $ & 210 & -97 \\ 
Reticulum3 &   $ 0.05 \pm 0.21 $   &   $ -0.09 \pm 0.2 $ \ & $ -137 \pm 91 $ &  $ 17 \pm 87 $ & 138 & 101 \\ 
Sagittarius2\tablenotemark{\scriptsize{a}} &   $ -0.65 \pm 0.07 $   &   $ -0.96 \pm 0.04 $   \ & $ -182 \pm 24 $ &  $ -108 \pm 14 $ & 212 & -98 \\ 
Sculptor\tablenotemark{\scriptsize{a}} &   $ 0.081 \pm 0.005 $   &   $ -0.136 \pm 0.004 $ \ & $ -111 \pm 2 $ &  $ 136 \pm 2 $ & 175 & 77 \\ 
Segue1 &   $ -1.59^{+0.23}_{-0.22} $   &   $ -3.5^{+0.2}_{-0.21} $ \ & $ -121^{+25}_{-24} $ &  $ -166^{+22}_{-23} $ & 206 & 109 \\ 
Segue2 &   $ 1.68^{+0.12}_{-0.13} $   &   $ 0.12 \pm 0.08 $ \ & $ 115^{+20}_{-21} $ &  $ 178 \pm 13 $ & 212 & 45 \\ 
Sextans1\tablenotemark{\scriptsize{a}} &   $ -0.44 \pm 0.02 $   &   $ 0.09 \pm 0.02 $ \ & $ -124 \pm 8 $ &  $ 210 \pm 8 $ & 244 & 66 \\ 
Triangulum2 &   $ 0.62^{+0.13}_{-0.15} $   &   $ 0.42 \pm 0.11 $ \ & $ -72^{+19}_{-21} $ &  $ 187 \pm 16 $ & 200 & -253 \\ 
Tucana2\tablenotemark{\scriptsize{a}} &   $ 0.94 \pm 0.05 $   &   $ -1.22 \pm 0.06 $ \ & $ 176 \pm 14 $ &  $ -119 \pm 16 $ & 212 & -207 \\ 
Tucana3\tablenotemark{\scriptsize{a}} &   $ -0.02 \pm 0.03 $   &   $ -1.67 \pm 0.03 $ \ & $ -119 \pm 4 $ &  $ -11 \pm 4 $ & 120 & -198 \\ 
Tucana4 &   $ 0.63^{+0.13}_{-0.12} $   &   $ -1.54^{+0.1}_{-0.11} $ \ & $ 24^{+30}_{-27} $ &  $ -168^{+23}_{-25} $ & 169 & -86 \\ 
Tucana5 &   $ -0.1^{+0.09}_{-0.2} $   &   $ -1.01^{+0.23}_{-0.1} $ \ & $ -133^{+24}_{-52} $ &  $ -74^{+60}_{-26} $ & 152 & -140 \\ 
UrsaMajor1\tablenotemark{\scriptsize{a}} &   $ -0.56 \pm 0.06 $   &   $ -0.68 \pm 0.08 $ \ & $ -187 \pm 28 $ &  $ -86 \pm 37 $ & 206 & -6 \\ 
UrsaMajor2\tablenotemark{\scriptsize{a}} &   $ 1.72 \pm 0.05 $   &   $ -1.84 \pm 0.06 $ \ & $ 256 \pm 7 $ &  $ -50 \pm 9 $ & 261 & -31 \\ 
UrsaMinor\tablenotemark{\scriptsize{a}} &   $ -0.16 \pm 0.01 $   &   $ 0.06 \pm 0.01 $ \ & $ 105 \pm 4 $ &  $ 81 \pm 4 $ & 133 & -78 \\ 
  Willman1 &   $ 0.36 \pm 0.1 $   &  $ -1.04 \pm 0.18 $  & $ 144 \pm 18 $ &  $ 36 \pm 32 $ & 148 & 36 \\
  \hline
Eridanus2 &   $ 0.35 ^{+0.21}_{-0.20}$   &   $ -0.08 \pm 0.25 $ \ & $ 472^{+378}_{-360} $ &  $ -40 \pm 451 $ & 473 & -73 \\
Phoenix &   $ 0.08 \pm 0.15 $   &   $ -0.08 \pm 0.18 $ \ & $ -2 \pm 291 $ &  $ 3 \pm 349 $ & 3 & -114 \\  
LeoT &   $ 0.10^{+0.67}_{-0.69} $   &   $ 0.01 ^{+0.57}_{-0.56} $ \ & $ 227^{+1324}_{-1364} $ &  $ 237^{+1126}_{-1107} $ & 328 & -63 \\ 
\end{tabular*}
\tablenotetext{\scriptsize{a}}{Systematic uncertainties are a significant or major contributor to the proper motion error budget for this satellite. See \cite{lindegren2018}}
\caption{Preferred systemic proper motion estimates for the 54 out of 59 galaxies in our sample (corresponding to Prior B from Table~\ref{pmresults} if available, Prior A otherwise).  As in Table 3, no results were able to be derived for Bootes 4, Cetus 3, Indus 2, Pegasus 3 and Virgo 1. The corresponding tangential velocity components in a Galactocentric frame of reference 
($v_\alpha\cos\delta, v_\delta$) are listed, as well as the overall tangential velocity (v$_t$). 
The implied Galactocentric radial velocities are listed in the last column for comparison, and are converted from the heliocentric radial velocities listed in Table~\ref{gals} (see text for details). The final three rows of the table present the proper motions for the three most distant galaxies in our sample derived without the requirement that they are bound to the Galaxy.}
\label{preferred}
\end{center} }
\end{table*}

For each satellite listed in Table~\ref{gals}, we construct the spatial, color-magnitude and proper motion likelihood maps as described in Section~3 and explore parameter space using {\tt emcee} for those values of $f_{sat}$ 
and $\left(\mu_\alpha\cos\delta, \mu_\delta\right)_{sat}$ that maximise the probability of the data.
Table~\ref{pmresults} lists the median values and 14th and 86th percentiles for the three parameters in our model, $f_{sat}$ 
and $\left(\mu_\alpha\cos\delta, \mu_\delta\right)_{sat}$ under the assumptions of our two different priors 
(Prior A - Columns 2, 3 and 4; Prior B - Columns 5, 6, and 7) for 54 of the satellites listed in Table~\ref{gals}. 
We are unable to determine systemic proper motions from Gaia DR2 for Bootes 4, Cetus 3, Indus 2, Pegasus 3, and Virgo 1, and we argue in Section~\ref{atthelimits} that this is due to a lack of any member stars with reliable data in Gaia DR2. The estimates for Cetus 2, Pictor 2 and Leo T 
are derived using looser cuts on the Gaia data in order to identify more member stars, and these systems are also discussed in more detail in Section~\ref{atthelimits}.

Table~\ref{preferred} lists the adopted proper motions for each
satellite, and the corresponding tangential velocity components
assuming $(R_{\odot}, V_c) = (8.122$\,kpc, 229\,km\,s$^{-1})$
and 
$(U_\odot, V_\odot, W_\odot) = (11.1, 12.24, 7.25)$\,km\,s$^{-1}$
(\citealt{gravity2018, schonrich2010}). The adopted proper motion for each
satellite corresponds to that derived using Prior B if it is
available, and Prior A otherwise. For the three most distant galaxies in our sample - Eridanus 2, Phoenix and Leo T - the last three rows of Table 4 also give the results obtained in the absence of the prior on the expected tangential velocity dispersion of the halo.

All error bars describe random errors only, and systematic uncertainties are not included. \cite{lindegren2018} show that on scales of a degree or less, the systematic uncertainty in each component of the proper motion is approximately 0.066\,mas/yr, although \cite{helmi2018b} suggest that, for the larger dwarfs ($\gtrsim 0.2^\circ$), the systematic uncertainty per component is 0.035 mas/yr. Adopting a constant systematic uncertainty for each dwarf of 0.066\,mas/yr, and comparing to the random errors listed in Tables~\ref{pmresults} and \ref{preferred}, suggests that systematics are a major or dominant component in the  uncertainty on the proper motions of Antlia 2, Bootes I, Canes Venatici 1, Carina, Carina 2, Coma Berenices, Crater 2, Draco, Fornax, Grus 2, Hercules,  Horologium 1, Hydrus 1, Leo 2, Reticulum 2, Sagittarius 2,  Sculptor, Sextans, Tucana 2, Tucana 3, Ursa Major 1, Ursa Major 2 and Ursa Minor.

\subsection{Internal consistency}

\subsubsection{Comparison of results with Prior A and B}

\begin{figure*}
  \centering
    \includegraphics[width=\textwidth]{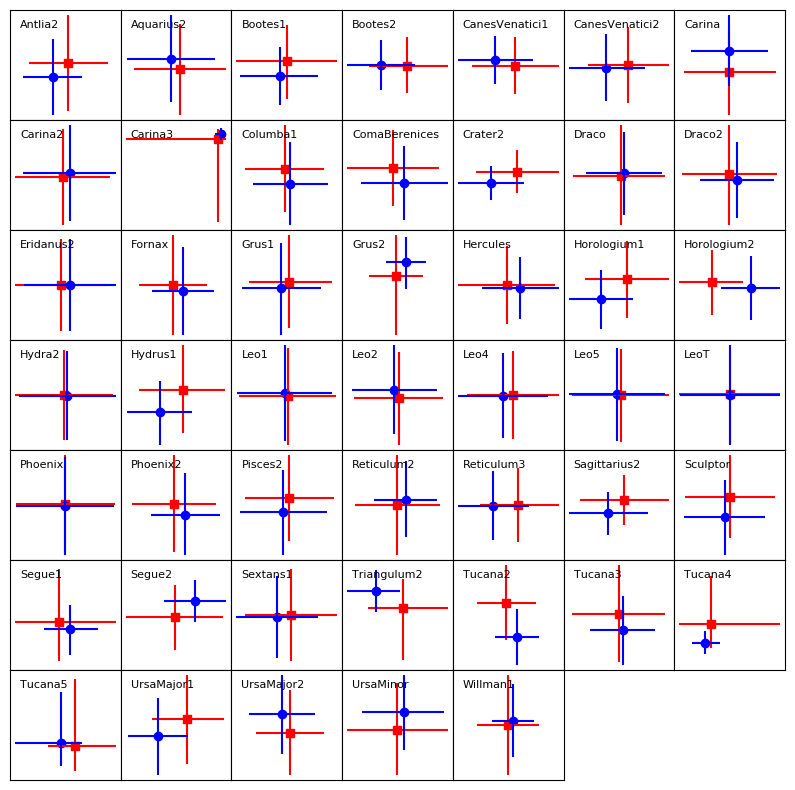}
    \caption{Comparison of systemic proper motions  for satellites derived without using radial velocity information (Prior A; red squares), and incorporating radial velocity information (Prior B; blue circles). $\mu_\alpha\cos\delta$ is shown on the x-axis, and $\mu_\delta$ is shown on the y-axis. Error bars show $1\,\sigma$ uncertainties, excluding systematic errors. The scales are the same on each axis in each panel, but are different between panels. }\label{internal}
\end{figure*}

The internal consistency of our approach is examined for satellites with radial velocity information. In Figure~\ref{internal},
a comparison of the systemic proper motions for satellites derived with (Prior B; blue circles)
and without (Prior A; red squares) radial velocity information is shown.
$\mu_\alpha\cos\delta$ is shown on the x-axis, and $\mu_\delta$ is shown on the y-axis. 
Error bars show $1\,\sigma$ uncertainties, excluding systematic errors. The scales are the 
same on each axis in each panel, but are different between panels, and were chosen to encompass all of the relevant points and their uncertainties. This figure is intended 
to highlight the relative agreement, or otherwise, between estimates.

Without exception, Figure~\ref{internal} shows extremely good agreement between the two sets of estimates. In all cases, they
agree to within their combined $1\,\sigma$ uncertainties. 
This is extremely encouraging and suggests that
the adopted methodology is robust in the absence of radial velocity data, in line with the findings from \cite{pace2019},  
This will be of increasing importance in the coming years, with the expected  flood of discoveries 
from the Legacy Survey of Space and Time (LSST, e.g., \citealt{hargis2014}). 
Inspection of Table~\ref{radvel} shows that even with the current sample, 
radial velocity follow-up is not immediate or complete.

\begin{figure*}
  \centering
    \includegraphics[width=0.45\textwidth]{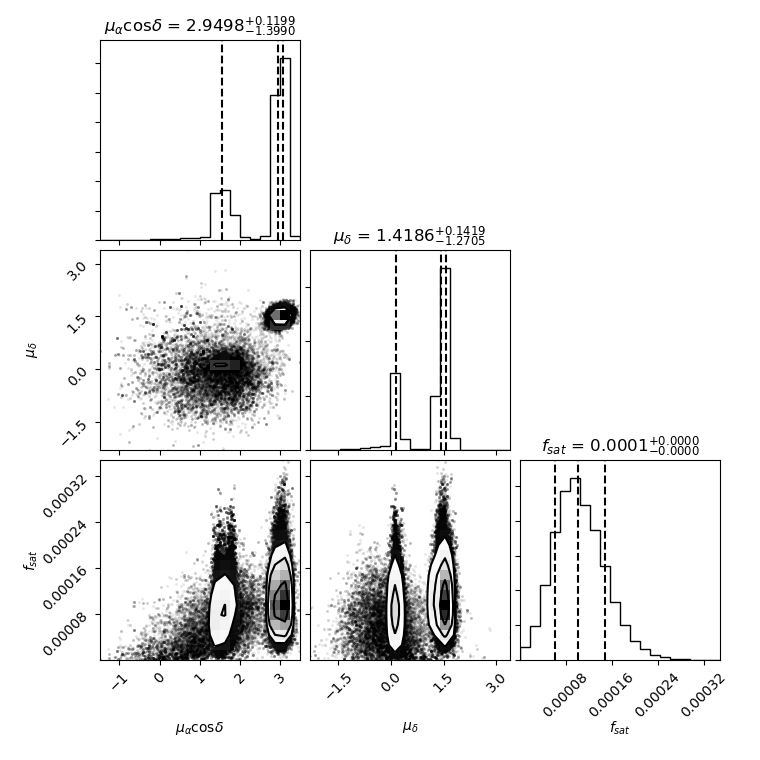}
    \includegraphics[width=0.45\textwidth]{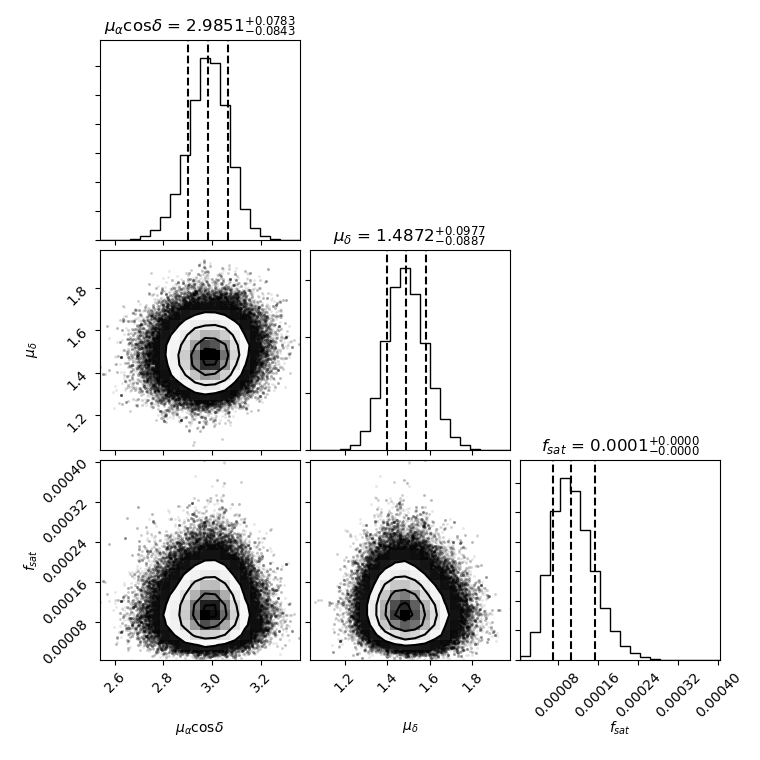}
    \caption{Corner plots for our three unknown parameters for the case of Carina 3 (a) without using radial velocity information (Prior A, left panel) (b) using radial velocity information (Prior B, right panel).}\label{carina3}
\end{figure*}

In general, it appears that radial velocity information is not required 
to obtain a reasonable estimate of the proper motions of dwarf satellites, 
but it can play an important role in some cases.  For example, the large 
error bars on the systemic proper motion for Carina 3 using Prior A is 
the result of a bimodal probability distribution function (PDF). 
Carina 3 is at low latitude ($b \simeq 18^\circ$), where there is considerable 
foreground contamination in the field, including the satellite Carina 2, which is responsible for the second peak in the PDF. Incorporation of a velocity prior 
in this case helps to break the degeneracy in the solutions, and leads to a 
well-defined single-peaked solution. 
This is demonstrated in Figure~\ref{carina3}, which shows the corner plots for Carina 3 in the case of Prior A (left panels) and Prior B (right panels). The same is also true, to a less dramatic extent, for Segue 1, Triangulum 2 and Tucana 4. For these systems, the PDFs without velocity information show some bimodality (though the two peaks are much closer to each other than for Carina 3), but the incorporation of radial velocity information via Prior B collapses these to single-peaked solutions. 

Despite the obvious benefits, we find that radial velocity information should be used with 
caution for systems with only a few radial velocities. 
It is entirely possible for a star to have a consistent radial velocity (and color-magnitude position, and spatial location), 
but not actually be a member of the satellite. Indeed, it is because of this consideration that we created Prior B by 
{\it multiplying} the radial velocity prior with Prior A, rather than a straight 
{\it substitution} of the  radial velocity prior for Prior A. 
For a galaxy with radial velocity follow-up of a large number of stars, 
this is not generally an issue, 
but it can be a concern for systems with only a few radial velocity members. 
For example, the velocity prior for Horologium 2 is based on a single 
star with $(\mu_\alpha\cos\delta, \mu_\delta) = (5.81 \pm 0.97, -2.05 \pm 1.58)$\,mas/yr,
which passes all of our quality control criteria, including parallax, CMD location within $2\,\sigma$ of the isochrone, 
the mean radial velocity, and on-sky location within 2 half-light radii. 
As a member of Horologium 2, this star would imply a tangential velocity greatly in excess of the escape velocity of the Milky Way. 
However, by weighting the velocity prior with the original Prior A, our algorithm is able to converge to a more acceptable value. 
Horologium 2 is the most extreme example of this effect in our analysis, 
but the principle remains true for each of the systems under study.

\subsubsection{Sanity checks using confirmed radial velocity members}

\label{weighted}

An important sanity-check of our results is obtained by examining the implied membership of systems for which individual radial velocities are available.

We confirm that the mean (weighted) proper motion of all stars with radial velocities that are confirmed as members ($P_{sat} \ge 0.5$) is consistent with the systemic proper motions listed in Table~\ref{preferred}. When we originally did this check for Tucana 4 and Ursa Major 2, we found that the mean proper motions of members with radial velocities were consistent with results using Prior A, not Prior B. The origin of this inconsistency was easily traced to the inclusion of a single deviant star in each system, which were used to create the velocity priors. These two stars had small uncertainties on their proper motions, 
but their proper motions were notably different to the other stars. 
Exclusion of these two stars from the radial velocity priors led to the solutions tabulated in Tables~\ref{pmresults} and \ref{preferred}. Again, this emphasises the need to carefully examine the radial velocity data for poorly populated systems, where a single measurement (particularly if it has a small uncertainty in its proper motion) can have a significant effect. It also emphasises the robustness of results derived using Prior A.

Using the values of the systemic proper motions tabulated in Table~\ref{preferred}, the weighted mean proper motions of all member stars with radial velocities is within $1\,\sigma$ of our derived values for each of the 47 satellites that use Prior B. Exceptions are Reticulum 3 and Pisces 2:

\begin{itemize}
\item For Reticulum 3, the weighted mean proper motion is based upon two faint ($G = 20.1, 19.6$) radial velocity members, which have $(\mu_\alpha\cos\delta, \mu_\delta)_1 = (-0.78 \pm 0.89, -1.05 \pm 1.12)$ and $(\mu_\alpha\cos\delta, \mu_\delta)_2 = (-0.78 \pm 0.72, 0.30 \pm 0.83)$\,mas/yr, respectively. These individual measurements are highly uncertain and the 
individual error bars comfortably overlap with the derived systemic proper motion reported in Table~\ref{preferred}. 
Their combined mean proper motion in the RA and declination directions are within $2\,\sigma$ and $1\,\sigma$ of the derived systemic proper motion for Reticulum 3, respectively;
\item For Pisces 2, the weighted ``mean'' proper motion is based upon a single faint ($G = 19.1$) radial velocity member, which has a proper motion $(\mu_\alpha\cos\delta, \mu_\delta)_1 = (-0.62 \pm 0.74, -1.38 \pm 0.59)$\,mas/yr. 
This is within $1\,\sigma$ and $2\,\sigma$ of the derived systemic proper motion for Pisces 2 in the RA and declination directions, respectively.
\end{itemize}

We conclude from this analysis that our derived systemic proper motions are internally self-consistent.
 
\subsection{Contamination and completeness}

We now turn our attention to examining the  contamination and completeness fractions of this new technique. 
This provides an indirect test on the robustness of our proper motion estimates.
For example, if this technique assigned many stars as members that cannot be members (or many stars as non-members which are clearly members), then this technique would not be trustworthy.

\subsubsection{Contamination: radial velocity non-members}
\label{sec:cont}

\begin{table*} {\scriptsize
  \begin{center}
    \begin{tabular*}{0.6\textwidth}{l|cccccc}
  Galaxy & $n_{DR2}$&$n_{QC} $ &$n_{QC,cont} $ &$n_{m}$ &$n_{m,cont}$ & $F_{cont}$\\
      \hline\\
      Overall &14675 & 8912 & 3280 & 5173 & 173 & 0.05\\
      \multicolumn{7}{l}{  }\\
      
      \multicolumn{7}{l}{Excluding galaxies with $n_{m}>100$:}\\
  &5595 & 2387 & 1592 & 462 & 25 & 0.02\\
      \hline\\
 Fornax & 2482 & 1951 & 37 & 1922 & 28 & 0.76\\
  Sculptor & 1457 & 1243 & 70 & 1185 & 30 & 0.43\\
  Carina & 1854 & 1246 & 676 & 544 & 56 & 0.08\\
  Draco & 1419 & 950 & 467 & 414 & 6 & 0.01\\
  UrsaMinor & 958 & 580 & 172 & 377 & 4 & 0.02\\
  Sextans1 & 910 & 555 & 266 & 269 & 24 & 0.09\\
  Leo2 & 219 & 59 & 3 & 56 & 1 & 0.33\\
  Crater2 & 404 & 189 & 122 & 55 & 2 & 0.02\\
  CanesVenatici1 & 144 & 55 & 10 & 45 & 2 & 0.2\\
  Bootes1 & 275 & 122 & 75 & 37 & 7 & 0.09\\
  Hydrus1 & 139 & 95 & 60 & 30 & 3 & 0.05\\
  Reticulum2 & 59 & 38 & 12 & 24 & 0 & 0.0\\
  Hercules & 81 & 39 & 15 & 18 & 0 & 0.0\\
  Tucana3 & 675 & 294 & 265 & 18 & 5 & 0.02\\
  Leo1 & 381 & 22 & 2 & 15 & 0 & 0.0\\
  Antlia2 & 221 & 202 & 57 & 11 & 0 & 0.0\\
  UrsaMajor2 & 128 & 39 & 24 & 11 & 2 & 0.08\\
  ComaBerenices & 47 & 21 & 6 & 10 & 0 & 0.0\\
  Tucana2 & 95 & 53 & 34 & 10 & 0 & 0.0\\
  UrsaMajor1 & 107 & 32 & 13 & 10 & 0 & 0.0\\
  CanesVenatici2 & 41 & 15 & 6 & 9 & 0 & 0.0\\
  Eridanus2 & 42 & 17 & 7 & 9 & 0 & 0.0\\
  Phoenix & 116 & 13 & 3 & 9 & 0 & 0.0\\
  Draco2 & 44 & 18 & 11 & 7 & 0 & 0.0\\
  Grus2 & 254 & 105 & 83 & 7 & 0 & 0.0\\
  Carina2 & 283 & 217 & 200 & 6 & 1 & 0.0\\
  Horologium1 & 19 & 12 & 6 & 6 & 0 & 0.0\\
  Segue1 & 310 & 113 & 104 & 6 & 2 & 0.02\\
  Sagittarius2 & 120 & 31 & 24 & 5 & 0 & 0.0 \\
  Grus1 & 70 & 21 & 14 & 5 & 0 & 0.0\\
  Hydra2 & 18 & 10 & 4 & 5 & 0 & 0.0\\
  LeoT & 39 & 10 & 4 & 5 & 0 & 0.0\\
  Bootes2 & 8 & 4 & 0 & 4 & 0 & \\
  Phoenix2 & 75 & 18 & 4 & 4 & 0 & 0.0\\
  Segue2 & 214 & 55 & 33 & 4 & 0 & 0.0\\
  Columba1 & 49 & 29 & 21 & 3 & 0 & 0.0\\
  Tucana4 & 209 & 82 & 58 & 3 & 0 & 0.0\\
  Aquarius2 & 10 & 3 & 1 & 2 & 0 & 0.0\\
  Carina3 & 283 & 219 & 213 & 2 & 0 & 0.0\\
  Leo5 & 124 & 33 & 29 & 2 & 0 & 0.0\\
  Reticulum3 & 45 & 26 & 22 & 2 & 0 & 0.0\\
  Tucana5 & 29 & 17 & 12 & 2 & 0 & 0.0\\
  Willman1 & 66 & 11 & 3 & 2 & 0 & 0.0\\
  Leo4 & 25 & 5 & 4 & 1 & 0 & 0.0\\
  Pisces2 & 7 & 4 & 3 & 1 & 0 & 0.0\\
  Triangulum2 & 28 & 19 & 14 & 1 & 0 & 0.0\\
  Horologium2 & 92 & 20 & 11 & 0 & 0 & 0.0\\
\end{tabular*}
\caption{Summary of membership contamination rates estimated from radial velocity data. $n_{DR2}$ is the number of Gaia stars with ground-based radial velocity data for each satellite; $n_{QC}$ is the number of these stars that pass our quality control cuts; $n_{QC,cont}$ is the number of these stars that have radial velocities more than $3\,\sigma$ from the mean velocity of the satellite; $n_{m}$ is the number of stars that are considered high probability members ($P_{sat} \ge 0.5$) by our algorithm; $n_{m,cont}$ is the number of these stars with radial velocities that are more than $3\,\sigma$ different from the mean radial velocity of the satellite. $F_{cont}$ is the percentage contamination per galaxy, estimated as the ratio of $n_{m,cont}$ to $n_{QC,cont}$.}\label{vstatus}
\end{center} }
\end{table*}

Most stars that are identified as a member of a dwarf satellite, and which also have a radial velocity measurement, should have a radial velocity that is  consistent with membership. The number of stars with deviant radial velocities helps us to estimate of our contamination fraction, $F_{cont}$.

Some radial velocity information is available for 47 satellites for which we derive systemic proper motions.
(see Table~\ref{radvel}), for a total of $n_{DR2} = 14675$ stars that are also present in Gaia DR2. 
Of these, $n_{QC} = 8912$ stars pass our quality cuts and lie within our likelihood grids (defined in Section 3). However, these stars are not evenly distributed across the satellites: 6525 of them are found in only 6 galaxies (Fornax, Sculptor, Carina, Draco, Ursa Minor and Sextans). Table~\ref{vstatus} describes the distribution of radial velocity measurements between galaxies.

Defining member stars as those stars with $P_{sat} \ge 0.5$, there are a total of $n_m = 5173$ member stars with radial velocities across all satellites. Of these, only $n_{m,cont}$ = 173 of them have velocities that are more than $3\,\sigma$ from the mean systemic radial velocity (as given in Table~\ref{gals}). 
Considering only the dwarf satellite galaxies with fewer than 100 members with radial velocity measurements (i.e., excluding the 6 galaxies mentioned previously), there are a total 25 contaminants out of a sample of 462 stars. 
To understand what this means for the average contamination rate, it is important to recognise that each galaxy had different levels of contamination prior to the application of the algorithm. In particular, there were a total of $n_{QC,cont} = 3280$ stars which passed our initial cuts and which had velocities more than $3\,\sigma$ from the mean systemic radial velocity of the satellites. Since only 173 of these remain after application of the algorithm, then we find $F_{cont} = n_{m,cont}/n_{QC,cont} = 0.05$. Ignoring the brightest satellites, then $F_{cont} = 0.016$\footnote{There are other ways to define contamination, for example using the ratio of obvious contaminants to confirmed members, $n_{m,cont}/n_{m} = 173/5173 = 0.03$.} 

Of course, there can be significant variations in these results between the dwarf satellite galaxies. 
Fornax and Sculptor have high values of $F_{cont}$, but the radial velocity sample we are using is already extremely clean, 
and so the algorithm cannot significantly improve on the high purity. In contrast, Draco initially has 467 stars out of 950 with discrepant velocities, and our algorithm identifies only 6 stars out of 414 members with discrepant velocities. 
Table~\ref{vstatus} shows the break-down of these numbers on a galaxy-by-galaxy basis. Based on these numbers, we have high confidence in the ability of our algorithm to select member stars with only modest ($\leq$1 in 20) contamination. 
This also implies that the systemic proper motion estimates are robust to contamination effects.

\subsubsection{Completeness: Confirmed members via high resolution spectroscopy}
\label{sec:chem}

\begin{longrotatetable}
\begin{deluxetable*}{l|rrccccrcr}
\tablecaption{Membership probabilities for stars in ultra-faint dwarf galaxies studied in the literature.This table includes all stars with $P_{sat}< 0.5$. See Table~\ref{allmems} for stars with $P_{sat}\ge 0.5$.\label{chem} }
\tablewidth{0pt}
\tablehead{
\colhead{Galaxy} & \colhead{Reference} & \colhead{Star ID} & \colhead{RA} & \colhead{Dec} & \colhead{G} & \colhead{$r/r_{h}$} & \colhead{Gaia DR2 source ID} & \colhead{$P_{sat}$} & \colhead{Comments}
}
\tabletypesize{\scriptsize}
  
\startdata
 Bootes1  &  \cite{frebel2016}  &  Boo-980  &  13:59:12.68  &  +13:42:55.7  &  17.88  & 4.4 &   1230649834160653440  &  0.01  &   At several $r_h$   \\
  Carina2  &  \cite{ji2020}  &  CarII-V3  &  07:35:09.12  &  -57:57:14.8  &  18.47  & 1.7 &   5293940924860019584  &  $<10^{-10}$  &   RR Lyrae variable   \\
  Carina2  &  \cite{li2018b}  &  J073646.47-575910.2  &  07:36:46.47  &  -57:59:10.1  &  19.41  & 0.5 &   5293948273547045248  &  0.36  &  Binary star    \\
  Carina2  &  \cite{li2018b}   &  J073729.30-580447.8  &  07:37:29.28  &  -58:04:47.7  &  19.14  & 1.5 &   5293900513510499584  &  0.08  &      \\
  Carina2  &  \cite{li2018b}   &  J073737.04-574925.5  &  07:37:37.02  &  -57:49:25.4  &  19.79  & 2.2 &   5293959680980519424  &  $<10^{-14}$  &  BHB    \\
  Carina2  &  \cite{li2018b}  &  J073745.86-580406.7  &  07:37:45.85  &  -58:04:06.7  &  18.85  & 1.8 &   5293900857107911808  &  $<10^{-31}$  &   BHB   \\
  Hercules  &  \cite{koch2013}  &  12175  &  16:31:15.82  &  +12:34:56.5  &  18.34  & 6.0 &   4460295812884988672  &  0.01  &  At several $r_h$    \\
  Hercules  &  \cite{koch2013}  &  12729  &  16:31:07.49  &  +12:31:33.7  &  19.6  & 8.1 &   4460293953161686656  &  0.00033  &   At several $r_h$   \\
  Horologium2  &  \cite{fritz2019}  &  horo2\_2\_48  &  03:17:21.08  &  -50:03:40.6  &  18.74  & 8.2 &   4749201525396641408  &  0.00094  &  At several $r_h$    \\
  Segue1  &  \cite{frebel2014}  &  J100702+155055  &  10:07:02.46  &  +15:50:55.2  &  18.0  & 5.2 &   621924492960734464  &  0.00142  &  At several $r_h$    \\
  Segue1  &  \cite{frebel2014}  &  J100742+160106  &  10:07:42.71  &  +16:01:06.9  &  18.16  & 3.1 &   621933460851966976  &  0.3  &      \\
  Segue1  &  \cite{frebel2014}  &  J100639+160008  &  10:06:39.33  &  +16:00:08.5  &  19.04  & 2.1 &   621941535390988928  &  0.37  &      \\
  Segue1  &  \cite{norris2010b}  &  SegueI-7  &  10:08:14.44  &  +16:05:01.1  &  17.45  & 4.5 &   621934805177217920  &  0.00024  &   At several $r_h$   \\
  Triangulum2  &  \cite{venn2017, ji2019}  &  Star46  &  02:13:21.54  &  +36:09:57.4  &  18.84  & 0.4 &   331086526201161088  &  0.04  &  Binary star\\
  Tucana3  &  \cite{marshall2019} &  J000549-593406  &  00:05:48.72  &  -59:34:06.2  &  16.2  & 11.7 &   4917991682841282176  &  $<10^{-7}$  &  At many $r_h$ (in tails)    \\
  Tucana3  &  \cite{marshall2019}  &  J234351-593926  &  23:43:50.85  &  -59:39:25.8  &  17.22  & 16.1 &   6488511925629603200  &  $<10^{-8}$  &   At many $r_h$ (in tails)   \\
  \enddata

      \end{deluxetable*}
\end{longrotatetable}

To the extent that it is possible to ``know'' that a star is a member of a satellite, stars that have had their 
stellar parameters determined and chemical abundances derived from high resolution spectroscopy are the gold standard. 
Prior to Gaia DR2, such stars were usually targeted for follow-up spectroscopy based on spatial coincidence, color-magnitude 
consistency, and radial velocity matching (post Gaia DR2, proper motion consistency is also usually required). 
For stars with published high resolution spectroscopy, their subsequent analyses would confirm their memberships, and 
provide metallicity and abundance characteristics. 
We examine the completeness of our algorithm by determining our ability to retrieve these ``known'' members.

A search of the literature was carried out for all stars in the ultra-faint satellites (here defined as $M_V \lesssim -8$) for which high resolution spectroscopic analyses exist, and which are present in Gaia DR2. 
This list is intended to be complete up to May 2020, and contains a total of 91 stars. These were cross-matchd 
with our stellar catalogs to calculate the implied values of $P_{sat}$ for each star based on our algorithm. 
Of these 91 stars, 16 stars are assigned $P_{sat}< 0.5$ and are listed in Table~\ref{chem}; 
the other 75 stars are listed in a table in the Appendix, Table~\ref{allmems}, in the same format.

Analysis of the stars in Table~\ref{chem} reveals the following relevant details for membership considerations:

\begin{itemize}
\item 8 of these stars are at radii corresponding to more than 4 half-light radii (accounting for the elliptical shape of the satellites). Assuming that the spatial distribution of stars in these satellites are very well described by the parameters listed in Table~\ref{gals}, then it is increasingly likely that the most distant stars are not members. However, the structure of faint dwarf galaxies, especially at large radius, can be very complex, and it is known that some of these systems have extended tidal features that are not well captured by the simple parameterisations in Table~\ref{gals} (especially Tucana 3; see \citealt{li2018a}). As currently constructed, this algorithm will preferentially assign low probability measurements to stars at large radius, and this may include stars that are actually members. If these stars are subsequently shown to be members of the satellite, then this will imply the satellite's structure is more complex than currently parameterized in the spatial likelihood functions. 

\item Three stars are horizontal branch stars, either blue horizontal branch or RR Lyrae. As our algorithm considers membership of the satellite in color-magnitude space in relation to the proximity of the star to the main isochrone locus (main sequence, sub-giant or red giant branch stars), then it is not surprising that these horizontal branch stars are not identified as members.

\item Two stars appear to be in binary systems, one in Carina 2 \citep{li2018b} and another in Triangulum 2 \citep{venn2017}.
It is unclear if (or how) these affected the ability of the algorithm to determine their memberships.
\end{itemize}

If we consider stars with spatial and color-magnitude characteristics that should be well captured by the algorithm (specifically, stars within four half-light radii and which are not on the horizontal branch), there are 80 relevant stars in Tables~\ref{chem} and \ref{allmems}. Of these, 75 stars (94\,\% of the sample) have $P_{sat} \ge 0.5$. Interestingly, two of the five stars missed by the algorithm in this case are known binaries. With the necessary caveat relating to the structure of galaxies at large radii, we conclude that the algorithm is reasonably complete in correctly identifying main sequence/sub-giant/red giant branch stars as members of the dwarf satellites under consideration.

\subsection{Comparison to literature estimates}

\begin{figure*}
  \centering
    \includegraphics[width=\textwidth]{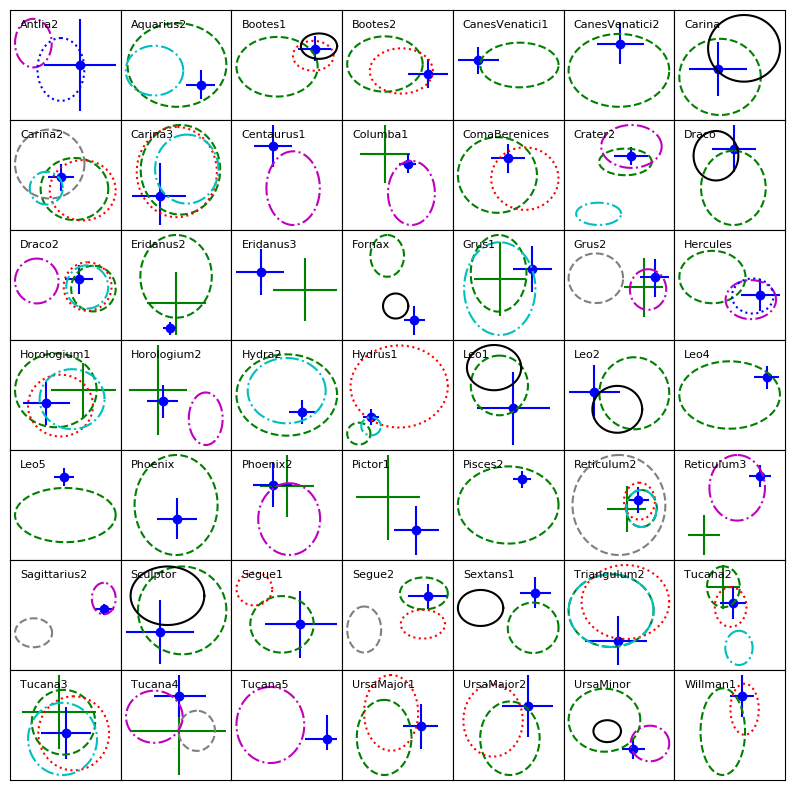}
    \caption{Comparison of the preferred systemic proper motions derived in this paper (blue circles) to previous estimates derived from Gaia DR2 data in the literature, for those satellites where previous estimates exist. $\mu_\alpha\cos\delta$ is shown on the x-axis, and $\mu_\delta$ is shown on the y-axis. The scales are the same in each panel, but different between panels. Error bars or semi-axis lengths correspond to $1\,\sigma$ uncertainties, and exclude systematic uncertainties. Literature estimates are from \cite{helmi2018b} (black solid ellipses), \cite{simon2018} (red dotted ellipses), \cite{fritz2018} (green dashed ellipses), \cite{kallivayalil2018} (cyan dot-dashed ellipses), \cite{massari2018} (grey dashed ellipses), \cite{pace2019} (green error bars), \cite{fritz2019, simon2019, longeard2018, longeard2020a, torrealba2019, fu2019, mau2020, pace2020}  (magenta dot-dashed ellipses), \cite{chakrabarti2019, gregory2020} (blue dotted ellipses).}\label{literature}
  \end{figure*}

Figure~\ref{literature} shows a comparison between the adopted systemic proper motion from Table~\ref{preferred} and all literature estimates for those satellites based on Gaia DR2 for which previous estimates are published. $\mu_\alpha\cos\delta$ is shown on the x-axis, and $\mu_\delta$ is shown on the y-axis. The adopted systemic proper motions derived in this paper are shown as blue points with blue error bars. Error bars or the semi-axes of the ellipses correspond to $1\,\sigma$ uncertainties, excluding systematic errors. The scales are the same on each axis in each panel, but different between panels, and were chosen to encompass all of the relevant points and their uncertainties. This figure is intended only to highlight the relative agreement, or otherwise, between estimates. Literature estimates are shown either as error bars or ellipses:

\begin{itemize}
\item black solid ellipses correspond to estimates from \cite{helmi2018b}; 
\item red dotted ellipses correspond to estimates from \cite{simon2018};
\item green dashed ellipses correspond to estimates from \cite{fritz2018};
\item cyan dot-dashed ellipses correspond to estimates from \cite{kallivayalil2018};
\item grey dashed ellipses correspond to estimates from \cite{massari2018};
 \item green error bars correspond to estimates from \cite{pace2019};
\item  magenta dot-dashed ellipses correspond to estimates from
  \cite{fritz2019} (Horologium 2, Reticulum 3, Columba 1, Phoenix 2),
  \cite{simon2019} (Grus 2, Tucana 4, Tucana 5), \cite{longeard2018}
  (Draco 2), \cite{longeard2020a} (Sagittarius 2), \cite{torrealba2019} (Antlia 2), \cite{mau2020}
  (Centaurus 1), \cite{fu2019} (Crater 2, Hercules) and \cite{pace2020} (Ursa Minor); 
\item  blue dotted ellipses correspond to estimates from
  \cite{gregory2020} (Hercules) and \cite{chakrabarti2019} (Antlia 2).
  \end{itemize}

Inspection of Figure~\ref{literature} shows good consistency
between literature estimates for the satellites using Gaia DR2 and the new estimates
derived here. Indeed, while there can be considerable spread in the measurements for each satellite between different studies, none of the values derived here are further than $\sim 1\,\sigma$ from at least one of the literature estimates. It is also interesting to note that two of the brightest satellites - Fornax and Sextans - appear to have some of the most discrepant measurements in this plot. However, this is largely a result of the absence of a scale in the panels of Figure~\ref{literature}. Fornax and Sextans have mean random uncertainities in Table~\ref{preferred} of $\sim 2\,\mu$as/yr and $20\,\mu$as/yr, respectively. Systematic uncertainties in these proper motions are of order several tens of $\mu$as/yr (\citealt{lindegren2018}) i.e., an order of magnitude larger than the random errors for Fornax, and the same order as the random errors for Sextans.

Almost all of the measurements that are highlighted by ellipses in Figure~\ref{literature} are derived from the weighted mean proper motion of a subset of stars already identified as members. In contrast, our approach is inspired by \cite{pace2019} (green error bars), in which determination of the most likely systemic proper motion is made by the analysis of the full dataset in the vicinity of the satellite, members and non-members alike. As such, our uncertainties are in general smaller. Note also that the maximum uncertainties that we measure on the systemic proper motions are of order 100\,km\,s$^{-1}$ in each direction, and this is a result of our prior, as discussed in Section~\ref{prior}. While this is reasonable for most of the objects under discussion, it is important to realise that Eridanus 2, Phoenix and Leo T are suffciently distant that the assumption that they trace the velocity dispersion of the Milky Way may be incorrect. As such, we also list in Table~\ref{preferred} the systemic proper motions for these three galaxies in the absence of this prior. It is notable that the uncertainties increase significantly, especially for Leo T.

While our uncertainties are generally smaller, our estimate for the systemic proper motion of Antlia 2 is less precise than the literature values (\citealt{torrealba2019, chakrabarti2019}). We think this is due to the extreme foreground contamination for Antlia 2; i.e., the value of $f_{sat}$ that we calculate corresponds to only a few in every million stars that we would consider to be actual members. Thus, for individual stars, the probability that it is a member of Antlia 2 is much lower than for most other galaxies, and this propagates through to our estimate of  the systematic proper motion. For those satellites which have estimates by \cite{pace2019}, the difference in the size of the uncertainties between this study and \cite{pace2019} 
is likely a result of our Milky Way contamination model, in which we do not have additional parameters that need to be marginalized over.

Only estimates using Gaia DR2 are shown in Figure~\ref{literature}. Many of the brighter satellites also have previous estimates of their proper motions derived using Hubble Space Telescope observations. These are compared to estimates from Gaia DR2 in Figure~15 of \cite{helmi2018b}. These authors note agreement between their estimates and previous estimates at the 2$-\sigma$ level, and also note the much smaller error bars that are made possible by the Gaia DR2 data. Encouraging, they also find that the Gaia-based estimates agree best with those galaxies for which the space-based data have the longest baselines. The results from \cite{helmi2018b} are shown in Figure~\ref{literature} as black ellipses, and it is clear that we agree closely with their estimates, especially when systematic uncertainties are taken into account.

Finally, we compare the $1\,\sigma$ (random) uncertainties in our estimates of the systemic proper motions to the uncertainties on the previous literature estimates using Gaia DR2. For each satellite in Figure~\ref{literature}, we selected the literature estimate that had the smallest quoted random uncertainties (averaged in both directions), $\sigma_{lit}$. This value is compared to our average uncertainty for each galaxy, $\sigma_{McVenn}$. In the median, $\sigma_{McVenn} \simeq 0.74 \, \sigma_{lit}$ ($\sigma_{McVenn} \simeq 0.77 \, \sigma_{lit}$ excluding those galaxies for which the prior is imposing a maximum uncertainty of around 100\,km\,s$^{-1}$). This suggests that our algorithm provides robust proper motion values with significantly improved random uncertainties.

\subsection{At the limits of Gaia: new systemic proper motions}

\label{atthelimits}

\begin{figure*}
  \centering
    \includegraphics[width=\textwidth]{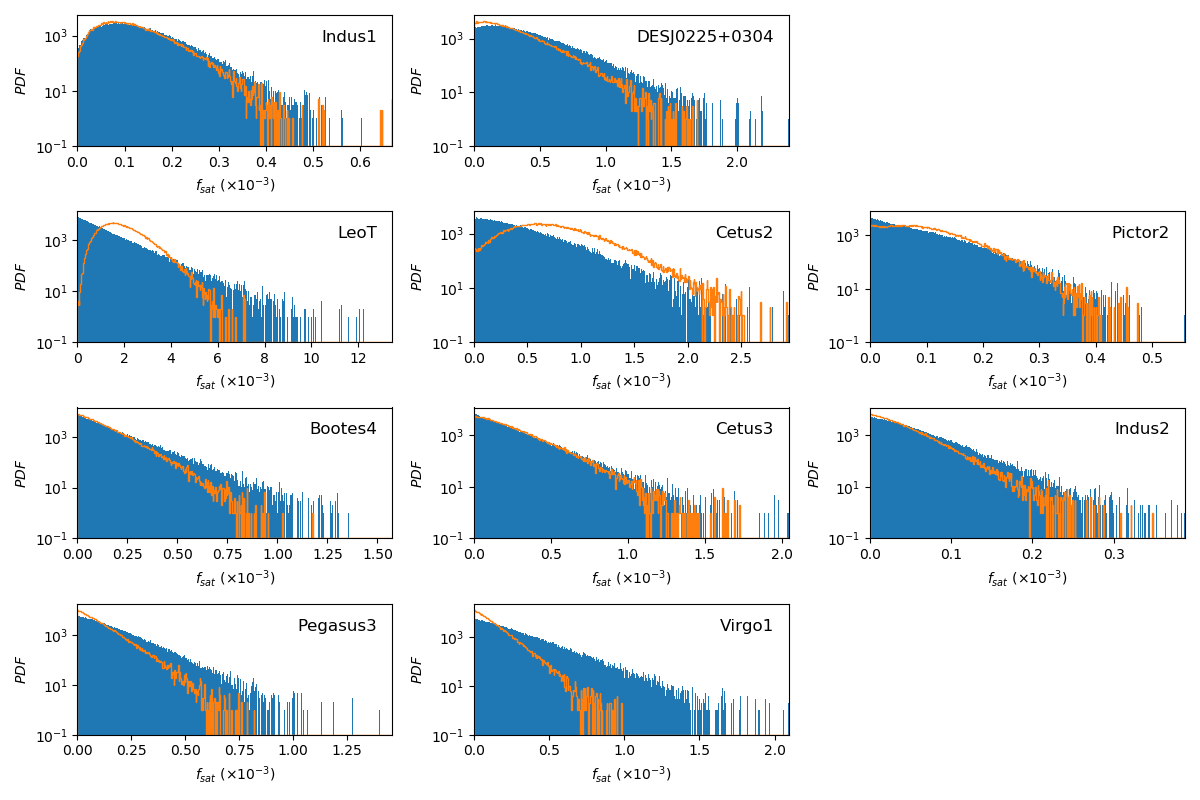}
    \caption{For each of the 10 satellites with new or no systemic proper motions, the blue (filled) histograms show the one-dimensional PDFs for $f_{sat}$. The orange (unfilled) histograms show the same PDFs derived using looser cuts on the Gaia data (specifically, {\tt ruwe}$ < 1.4$ and no cuts on the photometric quality). The top row shows those systems with new proper motions that we argue are robust, the second row shows those systems with new proper motions that are less robust, and the remaining rows show those systems for which no proper motion is derived.}\label{fsat}
\end{figure*}

In addition to the 49 satellites shown in Figure~\ref{literature} which have previous estimates from the literature, there are an additional 5 satellites for which we derive estimates of their systemic proper motions for the first time, namely Indus 1, DES J0225+0304, Leo T, Cetus 2 and Pictor 2. The latter three are less robust than for the other systems.  
In addition, for Bootes 4, Cetus 3, Indus 2, Pegasus 3 and Virgo 1, we cannot resolve any signal for the systemic proper motions of these satellites in Gaia DR2. Here, we discuss each of these systems in more detail.

In Figure~\ref{fsat}, the one-dimensional PDFs for $f_{sat}$ for the 10 satellites with new or no systemic proper motions are shown as the blue (filled) histograms. The top row shows the systems with new proper motions that we argue are robust, the second row shows those systems with new proper motions that are less robust (for reasons we will soon discuss), and the remaining rows show the systems for which no proper motion is derived.
As $f_{sat}$ is the fraction of stars that belong to the satellite, then when the most likely value for $f_{sat}$ is zero, this implies that no stars belong to that satellite, and we cannot derive a robust systemic proper motion.

For the satellites in the top row of Figure~\ref{fsat}, it is clear that the most likely value for $f_{sat}$ is greater than zero (the peak of the PDF is resolved). This is especially clear for Indus 1. For DES J0255+0304, the peak is more marginally resolved, but it is at a value greater than zero. For all of the remaining satellites, the most likely value of $f_{sat}$ is zero.

As $f_{sat}$ was not resolved for several satellites, we decided to rerun the algorithm for all of the 10 satellites in Figure~\ref{fsat}, but with looser selections on the data under consideration. 
In particular, the cut on the astrometric quality of the data was loosened (very slightly), such that {\tt ruwe} $< 1.4$, and the cut on the photometric quality of the data was removed. This latter change is the most significant: by no longer applying the cut described in Equation~\ref{fluxexcess}, many more stars were included for the analysis, including potential member stars of satellites. However, we did this knowing that there are concerns with the consistency of the $G, BP$ and $RP$ photometry, which may affect the robustness of the CMD likelihood analysis.

The orange (unfilled) histogram in each panel of Figure~\ref{fsat} shows the corresponding PDFs for $f_{sat}$ using this revised dataset. For the satellites in the top row, the position of the  peak of the PDF is relatively unchanged. In each of these two cases, the corresponding systemic proper motion is statistically the same as previously, giving us confidence in the robustness of the result, especially for DES J0255+0304. This system is relatively close at only $\sim 20$\,kpc, 
suggesting that there are just not very many stars in this galaxy in the Gaia magnitude range.

The situation for DES J0255+0304 should be contrasted with the satellites in the second row of Figure~\ref{fsat}, where there is a clear difference in the PDF for $f_{sat}$ between the two versions of the dataset. For Leo T and Cetus 2, the peak in $f_{sat}$ is clearly resolved with the looser cuts on the data. For Pictor 2, the situation is more ambiguous, but even here it appears that the favored model has $f_{sat} > 0$.

Leo T has radial velocity data available, and the use of the looser quality cuts enables a solution to be determined using both Prior A and Prior B. Indeed, the weighted mean proper motion of those members ($P_{sat} > 0.5$) with radial velocities is entirely consistent with the derived value (as discussed for all the satellites in Section~\ref{weighted}). However, in all cases, the individual uncertainties in the proper motions of each member star are very large (greater than 1mas/yr), and the relatively small uncertainty in the systemic proper motion derived under the assumption that Leo T traces the tangential velocity dispersion of the halo is due to the prior (as comparison to the final entries in Table~\ref{preferred} makes clear). Nevertheless, no stars are assigned membership of Leo T that have inconsistent radial velocities. Most encouraging of all, all five stars with radial velocities that we determine to be members are also, completely independently, determined to be members by \cite{simon2007}, where they use a combination of SDSS photometry and a selection of spectral characteristics to assign membership. These findings give us confidence that, despite the poorer quality photometry, our estimates of the proper motion of Leo T are reasonable given the data and priors.

\begin{table*}
  \begin{center}
    \begin{tabular*}{0.75\textwidth}{l|rcccc}
  Galaxy &Gaia DR2 Source ID & RA & Dec & G & $P_{sat}$ \\
\hline
 Cetus2 &  2355341476508455680 & 01:17:53.76 & -17:25:58.1 & 20.2 & 0.97\\
 Cetus2 &  2355341583883100160 & 01:17:55.18 & -17:25:16.3 & 20.9 & 0.91\\
 Cetus2 &  2355341824400788352 & 01:17:54.72 & -17:24:16.2 & 20.9 & 0.88\\
 Cetus2 &  2355341137206398592 & 01:17:40.96 & -17:27:55.5 & 20.0 & 0.8\\
 Cetus2 &  2355366181160358528 & 01:17:48.77 & -17:22:23.5 & 20.6 & 0.65\\
 DESJ0225+0304 &  2515347012786866688 & 02:25:42.32 & +03:03:33.6 & 19.6 & 0.68\\
 Indus1 &  6476977533258082304 & 21:08:45.01 & -51:10:18.2 & 19.6 & 0.91\\
 Indus1 &  6476965472989876864 & 21:09:02.65 & -51:12:34.8 & 17.6 & 0.85\\
 Indus1 &  6476977468833947392 & 21:08:51.20 & -51:09:29.6 & 20.9 & 0.8\\
  LeoT &  620750528074286592 & 09:34:51.14 & +17:02:17.5 & 20.4 & 0.99\\
  LeoT &  620753654810753024 & 09:34:54.50 & +17:04:17.9 & 20.5 & 0.96\\
  LeoT &  620750558139087744 & 09:34:53.95 & +17:02:17.8 & 20.7 & 0.95\\
  LeoT &  620753684875529728 & 09:34:49.78 & +17:04:30.9 & 20.3 & 0.9\\
  LeoT &  620750489419612544 & 09:34:57.28 & +17:02:21.8 & 20.6 & 0.8\\
  LeoT &  620750287556115712 & 09:34:58.64 & +17:01:40.1 & 20.5 & 0.56\\
 Pictor2 &  5480249356255194112 & 06:44:54.69 & -59:55:03.0 & 17.1 & 1.0\\
 Pictor2 &  5480252925370608000 & 06:44:17.12 & -59:53:56.4 & 17.7 & 0.86\\
    \end{tabular*}
\caption{Stars identified as members ($P_{sat} \ge 0.5$) by our algorithm in systems with newly derived proper motions.}\label{tab:mems}
\end{center}
\end{table*}

We conclude from this analysis that the measurements for Indus 1 and DES J0255+0304 are robust. We present the measurements for Leo T, Cetus 2 and Pictor 2 using the looser cuts on the data alongside the other measurements in Tables~\ref{pmresults} and \ref{preferred}. We stress that the proper motions for Leo T, Cetus 2 and Pictor 2 are based in part on photometry which may not be reliable. Any analyses that use these three measurements should proceed with caution, and we expect that future data releases from Gaia - potentially including the imminent EDR3 -  may contain sufficently improved photometry to improve these three measurements. Table~\ref{tab:mems} lists all those stars in these 5 objects for which $P_{sat} \ge 0.5$.

\section{Summary}\label{sec:summ}

 We have presented a new derivation of systemic proper motions for most of the Milky Way dwarf galaxy satellite population, using a maximum likelihood
  approach inspired by the work of \cite{pace2019}. Our approach differs insofar as we examine simultaneously the likelihood of the spatial,
  color-magnitude, and proper motion distribution of sources, and adopt empirical models for the unknown Milky Way contamination instead of constructing models that need to be marginalized over. 
In addition, radial velocity information (where available) is incorporated into the analysis through a prior on the model.

  Analysis of the implied membership distribution of the satellites suggests that we accurately identify member stars with a contamination rate of $\leq$1 in 20. The associated uncertainties on the systemic proper motions are on average a factor of $\sim 1.4$ smaller than existing literature values. Systemic proper motions are derived for the first time for some of the faintest and most distant satellites, namely Indus 1, DES J0225+0304, Cetus 2, Pictor 2 and Leo~T.

The coming months and years will see the ongoing study of the orbits of the Milky Way satellites, firstly through dynamical studies enabled  by Gaia EDR3, and subsequent data releases, and then with the dwarf galaxy discovery power of LSST. It is a testament to the success of Gaia that the challenge for the observer in this coming era is the exact opposite of what it has been for the past 30 or 40 years. The study of the orbits of the faint and distant satellites of the Milky Way in the 2020s will  not be limited by an observer's inability to make the intrinsically complex measurement of the change in the mean position of a set of stars, but rather by our ability to obtain telescope time to make the intrinsically simple measurements of the Doppler shifts of those same stars.

\appendix

We include here the results of our literature search for stars in ultra-faint satellites which have high resolution spectroscopic follow-up, including the relevant identification in Gaia DR2 and the membership probability assigned by our algorithm. Table~\ref{chem} listed the 16 stars for which $P_{sat}<0.5$, and this appendix contains Table~\ref{allmems}, which lists the 75 stars for which $P_{sat}\ge 0.5$. We refer the reader to the relevant discussion in Section~\ref{sec:chem}.

\begin{longrotatetable}
\begin{deluxetable*}{l|rrccccrcr}
\tablecaption{Membership probabilities for stars in ultra-faint dwarf galaxies studied in the literature.This table includes all stars with $P_{sat}\ge 0.5$. See Table~\ref{chem} for stars with $P_{sat}< 0.5$.\label{allmems} }
\tablewidth{0pt}
\tablehead{
\colhead{Galaxy} & \colhead{Reference} & \colhead{Star ID} & \colhead{RA} & \colhead{Dec} & \colhead{G} & \colhead{$r/r_{h}$} & \colhead{Gaia DR2 source ID} & \colhead{$P_{sat}$} 
}
\tabletypesize{\scriptsize}
  
\startdata
 Bootes1  &  \cite{feltzing2009}  &  Boo-007  &  13:59:35.52  &  +14:20:23.7  &  17.62  & 1.2 &   1230826335841709184  &  0.96 \\
  Bootes1  &  \cite{feltzing2009}  &  Boo-033  &  14:00:11.72  &  +14:25:01.4  &  17.51  & 0.5 &   1230830222787069056  &  1.0 \\
  Bootes1  &  \cite{feltzing2009}  &  Boo-094  &  14:00:31.50  &  +14:34:03.6  &  16.62  & 0.8 &   1230835892143883008  &  1.0 \\
  Bootes1  &  \cite{feltzing2009}  &  Boo-117  &  14:00:10.49  &  +14:31:45.5  &  17.46  & 0.2 &   1230835578610843648  &  1.0 \\
  Bootes1  &  \cite{feltzing2009}  &  Boo-121  &  14:00:36.52  &  +14:39:27.3  &  17.11  & 1.2 &   1230861043472371712  &  1.0 \\
  Bootes1  &  \cite{feltzing2009}  &  Boo-127  &  14:00:14.56  &  +14:35:52.7  &  17.4  & 0.6 &   1230837163454204672  &  1.0 \\
  Bootes1  &  \cite{feltzing2009}  &  Boo-911  &  14:00:01.07  &  +14:36:51.5  &  17.18  & 0.6 &   1230848914484728064  &  1.0 \\
  Bootes1  &  \cite{norris2010a}  &  Boo-1137  &  13:58:33.81  &  +14:21:08.5  &  17.37  & 2.7 &   1230744529599617280  &  0.94 \\
  Bootes1  &  \cite{frebel2016}  &  Boo-9  &  13:59:48.80  &  +14:19:42.9  &  17.13  & 1.0 &   1230823247760125184  &  1.0 \\
  Bootes1  &  \cite{frebel2016}  &  Boo-41  &  14:00:25.83  &  +14:26:07.6  &  17.7  & 0.7 &   1230831047420766464  &  0.99 \\
  Bootes1  &  \cite{frebel2016} &  Boo-119  &  14:00:09.85  &  +14:28:23.0  &  17.73  & 0.2 &   1230833585745987968  &  0.98 \\
  Bootes1  &  \cite{frebel2016}  &  Boo-130  &  13:59:48.97  &  +14:30:06.2  &  17.51  & 0.5 &   1230834826991993088  &  1.0 \\
  Bootes2  &  \cite{francois2016}  &  J135801.42+125105.0  &  13:58:01.42  &  +12:51:05.0  &  19.1  & 0.1 &   3727827241204558720  &  0.98 \\
  Bootes2  &  \cite{francois2016}  &  J135751.18+125136.9  &  13:57:51.18  &  +12:51:36.9  &  18.76  & 0.7 &   3727827382938730496  &  0.97 \\
  CanesVenatici  &  \cite{francois2016}  &  J125713.63+341846.9  &  12:57:13.64  &  +34:18:47.0  &  18.98  & 1.0 &   1515697257293619584  &  1.0 \\
  Carina2  &  \cite{ji2020}  &  CarII-6544  &  07:36:51.11  &  -58:01:46.4  &  15.07  & 0.6 &   5293947247051916544  &  1.0 \\
  Carina2  &  \cite{ji2020}  &  CarII-7872  &  07:36:51.89  &  -58:16:39.2  &  15.5  & 2.0 &   5293894539213647872  &  1.0 \\
  Carina2  &  \cite{ji2020}  &  CarII-5664  &  07:38:08.51  &  -58:09:35.1  &  16.33  & 2.5 &   5293896360279425664  &  0.99 \\
  Carina2  &  \cite{ji2020}  &  CarII-0064  &  07:36:21.26  &  -57:58:00.2  &  16.78  & 0.2 &   5293951473299720064  &  1.0 \\
  Carina2  &  \cite{ji2020}  &  CarII-9296  &  07:37:39.79  &  -58:05:06.9  &  17.72  & 1.7 &   5293900827045399296  &  0.88 \\
  Carina2  &  \cite{ji2020}  &  CarII-4704  &  07:35:37.67  &  -58:01:51.7  &  17.4  & 1.2 &   5293928074318184704  &  0.99 \\
  Carina2  &  \cite{ji2020}  &  CarII-2064  &  07:36:01.33  &  -57:58:43.8  &  18.22  & 0.6 &   5293951881319592064  &  0.61 \\
  Carina2  &  \cite{ji2020}  &  CarII-4928  &  07:36:24.99  &  -57:57:14.2  &  18.42  & 0.3 &   5293951503362524928  &  0.91 \\
  Carina3  &  \cite{ji2020}  &  CarIII-1120  &  07:38:22.30  &  -57:53:02.1  &  17.46  & 0.5 &   5293955665187701120  &  1.0 \\
  Carina3  &  \cite{ji2020}  &  CarIII-8144  &  07:38:34.93  &  -57:57:05.3  &  17.65  & 1.0 &   5293907630273478144  &  1.0 \\
  ComaBerenices  &  \cite{frebel2010}  &  S1  &  12:26:43.47  &  +23:57:02.5  &  17.91  & 0.8 &   3959869107138860800  &  1.0 \\
  ComaBerenices  &  \cite{frebel2010}  &  S2  &  12:26:55.46  &  +23:56:09.8  &  17.81  & 0.4 &   3959874634761013120  &  1.0 \\
  ComaBerenices  &  \cite{frebel2010}  &  S3  &  12:26:56.66  &  +23:56:11.8  &  17.26  & 0.5 &   3959873883142494208  &  1.0 \\
  Grus1  &  \cite{ji2019}  &  Gru1-032  &  22:56:58.07  &  -50:13:58.1  &  17.63  & 3.8 &   6513255850697323648  &  0.87 \\
  Grus1  &  \cite{ji2019}  &  Gru1-038  &  22:56:29.93  &  -50:04:33.5  &  18.4  & 3.9 &   6514762009827996032  &  0.78 \\
  Hercules  &  \cite{koch2008b}  &  Her\_2\_42241  &  16:30:57.23  &  +12:47:20.3  &  18.3  & 0.3 &   4460319761622689152  &  1.0 \\
  Hercules  &  \cite{koch2008b,koch2014}  &  Her\_3\_41082  &  16:31:22.96  &  +12:44:47.9  &  18.66  & 1.1 &   4460316012113878528  &  0.94 \\
  Hercules  &  \cite{koch2013}  &  40789  &  16:31:29.77  &  +12:44:25.0  &  19.19  & 1.3 &   4460315840315183360  &  0.86 \\
  Hercules  &  \cite{koch2013}  &  40993  &  16:31:25.04  &  +12:45:29.2  &  19.4  & 1.0 &   4460316115193097856  &  0.93 \\
  Hercules  &  \cite{koch2013}  &  41460  &  16:31:14.06  &  +12:45:26.6  &  19.27  & 0.8 &   4460319001411691776  &  0.94 \\
  Hercules  &  \cite{koch2013}  &  41743  &  16:31:08.12  &  +12:48:06.1  &  19.11  & 0.6 &   4460319379374101760  &  0.88 \\
  Hercules  &  \cite{koch2013}  &  42096  &  16:31:00.63  &  +12:49:31.8  &  19.26  & 1.0 &   4460320410161056640  &  0.96 \\
  Hercules  &  \cite{koch2013}  &  42149  &  16:30:59.32  &  +12:47:25.6  &  18.86  & 0.2 &   4460319727263070208  &  0.99 \\
  Hercules  &  \cite{koch2013}  &  42324  &  16:30:55.46  &  +12:46:10.8  &  19.43  & 0.9 &   4460319585527257216  &  0.91 \\
  Hercules  &  \cite{koch2013}  &  42795  &  16:30:44.50  &  +12:49:47.8  &  19.19  & 0.9 &   4460367036326028160  &  0.87 \\
  Horologium1  &  \cite{nagasawa2018}  &  J025540-540807  &  02:55:40.32  &  -54:08:07.2  &  18.03  & 1.5 &   4740853586442400256  &  1.0 \\
  Horologium1  &  \cite{nagasawa2018}  &  J025543-544349  &  02:55:43.81  &  -54:05:19.6  &  17.55  & 1.7 &   4740854273637197952  &  1.0 \\
  Horologium1  &  \cite{nagasawa2018}  &  J025535-540643  &  02:55:35.22  &  -54:06:44.0  &  16.82  & 0.4 &   4740853865615992320  &  1.0 \\
  Leo4  &  \cite{simon2010}  &  S1  &  11:32:56.00  &  -00:30:27.8  &  18.87  & 0.6 &   3797163406525039232  &  0.96 \\
  Leo4  &  \cite{francois2016}  &  LeoIV  &  11:32:58.70  &  -00:34:50.0  &  19.75  & 1.1 &   3797154095035913216  &  0.75 \\
  Phoenix2  &  \cite{fritz2019}   &  phx2\_8\_64  &  23:39:55.80  &  -54:22:08.4  &  17.89  & 1.5 &   6497793143797836032  &  1.0 \\
  Pisces2  &  \cite{spite2018}  &  PiscesII10694  &  22:58:25.08  &  +05:57:20.1  &  19.14  & 1.2 &   2663690544626120448  &  0.68 \\
  Reticulum2  &  \cite{ji2016}  &  DESJ033523-540407  &  03:35:23.86  &  -54:04:07.6  &  15.73  & 0.5 &   4732507472849709568  &  1.0 \\
  Reticulum2  &  \cite{ji2016}  &  DESJ033607-540235  &  03:36:07.77  &  -54:02:35.5  &  16.91  & 0.8 &   4732598457436901760  &  1.0 \\
  Reticulum2  &  \cite{ji2016}  &  DESJ033447-540525  &  03:34:47.95  &  -54:05:25.0  &  17.03  & 1.5 &   4732506785654950016  &  1.0 \\
  Reticulum2  &  \cite{ji2016}  &  DESJ033531-540148  &  03:35:31.16  &  -54:01:48.2  &  17.11  & 0.7 &   4732507983952648704  &  1.0 \\
  Reticulum2  &  \cite{ji2016}  &  DESJ033548-540349  &  03:35:48.06  &  -54:03:49.8  &  17.79  & 0.5 &   4732504724070638208  &  1.0 \\
  Reticulum2  &  \cite{ji2016}  &  DESJ033537-540401  &  03:35:37.08  &  -54:04:01.2  &  18.13  & 0.4 &   4732507605992924672  &  0.99 \\
  Reticulum2  &  \cite{ji2016}  &  DESJ033556-540316  &  03:35:56.30  &  -54:03:16.2  &  18.46  & 0.5 &   4732598487502150016  &  0.98 \\
  Reticulum2  &  \cite{ji2016}  &  DESJ033457-540531  &  03:34:57.59  &  -54:05:31.4  &  18.5  & 1.3 &   4732506820014686464  &  0.94 \\
  Reticulum2  &  \cite{ji2016}  &  DESJ033454-540558  &  03:34:54.26  &  -54:05:58.0  &  18.51  & 1.4 &   4732506716935472384  &  0.98 \\
  Sagittarius2  &  \cite{longeard2020a}  &  298.16146  &  19:52:38.75  &  -22:04:57.6  &  18.72  & 0.6 &   6864047618136214528  &  0.9 \\
  Segue1  &  \cite{frebel2014}  &  J100714+160154  &  10:07:14.58  &  +16:01:54.5  &  18.33  & 1.4 &   621939679965123456  &  0.9 \\
  Segue1  &  \cite{frebel2014}  &  J100710+160623  &  10:07:10.07  &  +16:06:23.9  &  18.71  & 0.6 &   621949747368482304  &  0.71 \\
  Segue1  &  \cite{frebel2014}  &  J100652+160235  &  10:06:52.33  &  +16:02:35.8  &  18.4  & 1.0 &   621943184658438784  &  0.97 \\
  Segue2  &  \cite{roederer2014}  &  J021933.13+200830.2  &  02:19:33.13  &  +20:08:30.2  &  16.18  & 1.4 &   87202026681250432  &  1.0 \\
  Triangulum2  &  \cite{venn2017, ji2019}  &  Star40  &  02:13:16.55  &  +36:10:45.8  &  17.02  & 0.0 &   331089446778920576  &  0.99 \\
  Tucana2  &  \cite{chiti2018}  &  TucII-006  &  22:51:43.08  &  -58:32:33.8  &  18.27  & 0.3 &   6503772322389658880  &  0.99 \\
  Tucana2  &  \cite{chiti2018}  &  TucII-011  &  22:51:50.30  &  -58:37:40.3  &  17.68  & 0.6 &   6503770402539968128  &  1.0 \\
  Tucana2  &  \cite{chiti2018}  &  TucII-033  &  22:51:08.33  &  -58:33:08.2  &  18.13  & 0.6 &   6503774693211601280  &  0.97 \\
  Tucana2  &  \cite{chiti2018}  &  TucII-052  &  22:50:51.64  &  -58:34:32.7  &  18.32  & 0.9 &   6503774246534991360  &  0.93 \\
  Tucana2  &  \cite{chiti2018}  &  TucII-078  &  22:50:41.07  &  -58:31:08.4  &  18.12  & 1.0 &   6503775689644031104  &  0.91 \\
  Tucana2  &  \cite{chiti2018, chiti2020}  &  TucII-206  &  22:54:36.66  &  -58:36:58.1   &  18.31  & 2.2 &   6491771187332385408  &  0.74 \\
  Tucana2  &  \cite{chiti2018}  &  TucII-203  &  22:50:08.92  &  -58:29:59.2  &  18.27  & 1.5 &   6503787337595350400  &  0.57 \\
  Tucana3  &  \cite{hansen2017}  &  J235532-593115  &  23:55:32.69  &  -59:31:15.1  &  15.46  & 1.6 &   6494298968859977216  &  1.0 \\
  Tucana3  &  \cite{marshall2019}  &  J235738-593612  &  23:57:38.51  &  -59:36:11.8  &  16.67  & 1.3 &   6494252892450764800  &  1.0 \\
  Tucana3  &  \cite{marshall2019}  &  J235550-593300  &  23:55:49.93  &  -59:32:59.7  &  16.94  & 1.1 &   6494251964737889536  &  1.0 \\
  UrsaMajor2  &  \cite{frebel2010}  &  S1  &  08:49:53.47  &  +63:08:21.6  &  17.86  & 0.8 &   1043842293206190848  &  0.96 \\
  UrsaMajor2  &  \cite{frebel2010}  &  S2  &  08:52:33.51  &  +63:05:01.1  &  17.37  & 0.6 &   1043920324172211584  &  0.96 \\ 
  UrsaMajor2  &  \cite{frebel2010}  &  S3  &  08:52:59.04  &  +63:05:54.6  &  16.43  & 0.7 &   1043873491848815104  &  1.0 \\
  \enddata
  
      \end{deluxetable*}
\end{longrotatetable}

\acknowledgements

Thanks to the entire team of the Gemini High Resolution Optical Spectrograph (GHOST) for creating a fantastic instrument, that is waiting patiently for us all to commission and use. In the meantime, you have provided us with the motivation to tackle the target selection issue, for which this paper has been a very enjoyable spin-off. 

AWM and KAV would like to acknowledge funding from the National Science and Engineering Research Council 
Discovery Grants program. We thank Andrew Pace, Ting Li and Josh Simon for very useful feedback, and we thank the referee for very useful and constructive comments.

This work has made use of data from the European Space Agency (ESA) mission
{\it Gaia} (\url{https://www.cosmos.esa.int/gaia}), processed by the {\it Gaia}
Data Processing and Analysis Consortium (DPAC,
\url{https://www.cosmos.esa.int/web/gaia/dpac/consortium}). Funding for the DPAC
has been provided by national institutions, in particular the institutions
participating in the {\it Gaia} Multilateral Agreement.


\begin{thebibliography}{107}
\expandafter\ifx\csname natexlab\endcsname\relax\def\natexlab#1{#1}\fi

\bibitem[{{Ad{\'e}n} {et~al.}(2009){Ad{\'e}n}, {Feltzing}, {Koch}, {Wilkinson},
  {Grebel}, {Lundstr{\"o}m}, {Gilmore}, {Zucker}, {Belokurov}, {Evans}, \&
  {Faria}}]{aden2009a}
{Ad{\'e}n}, D., {Feltzing}, S., {Koch}, A., {Wilkinson}, M.~I., {Grebel},
  E.~K., {Lundstr{\"o}m}, I., {Gilmore}, G.~F., {Zucker}, D.~B., {Belokurov},
  V., {Evans}, N.~W., \& {Faria}, D. 2009, \aap, 506, 1147

\bibitem[{{Bhattacharjee} {et~al.}(2014){Bhattacharjee}, {Chaudhury}, \&
  {Kundu}}]{bhattacharjee2014}
{Bhattacharjee}, P., {Chaudhury}, S., \& {Kundu}, S. 2014, \apj, 785, 63

\bibitem[{{Caldwell} {et~al.}(2017){Caldwell}, {Walker}, {Mateo}, {Olszewski},
  {Koposov}, {Belokurov}, {Torrealba}, {Geringer-Sameth}, \&
  {Johnson}}]{caldwell2017b}
{Caldwell}, N., {Walker}, M.~G., {Mateo}, M., {Olszewski}, E.~W., {Koposov},
  S., {Belokurov}, V., {Torrealba}, G., {Geringer-Sameth}, A., \& {Johnson},
  C.~I. 2017, \apj, 839, 20

\bibitem[{{Carlin} \& {Sand}(2018)}]{carlin2018}
{Carlin}, J.~L. \& {Sand}, D.~J. 2018, \apj, 865, 7

\bibitem[{{Cautun} {et~al.}(2020){Cautun}, {Ben{\'\i}tez-Llambay}, {Deason},
  {Frenk}, {Fattahi}, {G{\'o}mez}, {Grand}, {Oman}, {Navarro}, \&
  {Simpson}}]{cautun2020}
{Cautun}, M., {Ben{\'\i}tez-Llambay}, A., {Deason}, A.~J., {Frenk}, C.~S.,
  {Fattahi}, A., {G{\'o}mez}, F.~A., {Grand}, R. J.~J., {Oman}, K.~A.,
  {Navarro}, J.~F., \& {Simpson}, C.~M. 2020, \mnras, 494, 4291

\bibitem[{{Chakrabarti} {et~al.}(2019){Chakrabarti}, {Chang}, {Price-Whelan},
  {Read}, {Blitz}, \& {Hernquist}}]{chakrabarti2019}
{Chakrabarti}, S., {Chang}, P., {Price-Whelan}, A.~M., {Read}, J., {Blitz}, L.,
  \& {Hernquist}, L. 2019, \apj, 886, 67

\bibitem[{{Chiti} {et~al.}(2020){Chiti}, {Frebel}, {Jerjen}, {Kim}, \&
  {Norris}}]{chiti2020}
{Chiti}, A., {Frebel}, A., {Jerjen}, H., {Kim}, D., \& {Norris}, J.~E. 2020,
  \apj, 891, 8

\bibitem[{{Chiti} {et~al.}(2018){Chiti}, {Frebel}, {Ji}, {Jerjen}, {Kim}, \&
  {Norris}}]{chiti2018}
{Chiti}, A., {Frebel}, A., {Ji}, A.~P., {Jerjen}, H., {Kim}, D., \& {Norris},
  J.~E. 2018, \apj, 857, 74

\bibitem[{{Collins} {et~al.}(2017){Collins}, {Tollerud}, {Sand}, {Bonaca},
  {Willman}, \& {Strader}}]{collins2017}
{Collins}, M. L.~M., {Tollerud}, E.~J., {Sand}, D.~J., {Bonaca}, A., {Willman},
  B., \& {Strader}, J. 2017, \mnras, 467, 573

\bibitem[{{Conn} {et~al.}(2018{\natexlab{a}}){Conn}, {Jerjen}, {Kim}, \&
  {Schirmer}}]{conn2018b}
{Conn}, B.~C., {Jerjen}, H., {Kim}, D., \& {Schirmer}, M. 2018{\natexlab{a}},
  \apj, 852, 68

\bibitem[{{Conn} {et~al.}(2018{\natexlab{b}}){Conn}, {Jerjen}, {Kim}, \&
  {Schirmer}}]{conn2018a}
---. 2018{\natexlab{b}}, \apj, 857, 70

\bibitem[{{Deason} {et~al.}(2012{\natexlab{a}}){Deason}, {Belokurov}, {Evans},
  {Koposov}, {Cooke}, {Pe{\~n}arrubia}, {Laporte}, {Fellhauer}, {Walker}, \&
  {Olszewski}}]{deason2012b}
{Deason}, A.~J., {Belokurov}, V., {Evans}, N.~W., {Koposov}, S.~E., {Cooke},
  R.~J., {Pe{\~n}arrubia}, J., {Laporte}, C.~F.~P., {Fellhauer}, M., {Walker},
  M.~G., \& {Olszewski}, E.~W. 2012{\natexlab{a}}, \mnras, 425, 2840

\bibitem[{{Deason} {et~al.}(2012{\natexlab{b}}){Deason}, {Belokurov}, {Evans},
  {Watkins}, \& {Fellhauer}}]{deason2012a}
{Deason}, A.~J., {Belokurov}, V., {Evans}, N.~W., {Watkins}, L.~L., \&
  {Fellhauer}, M. 2012{\natexlab{b}}, \mnras, 425, L101

\bibitem[{{Eadie} \& {Juri{\'c}}(2019)}]{eadie2019}
{Eadie}, G. \& {Juri{\'c}}, M. 2019, \apj, 875, 159

\bibitem[{{Erkal} \& {Belokurov}(2020)}]{erkal2020}
{Erkal}, D. \& {Belokurov}, V.~A. 2020, \mnras, 495, 2554

\bibitem[{{Feltzing} {et~al.}(2009){Feltzing}, {Eriksson}, {Kleyna}, \&
  {Wilkinson}}]{feltzing2009}
{Feltzing}, S., {Eriksson}, K., {Kleyna}, J., \& {Wilkinson}, M.~I. 2009, \aap,
  508, L1

\bibitem[{{Foreman-Mackey} {et~al.}(2019){Foreman-Mackey}, {Farr}, {Sinha},
  {Archibald}, {Hogg}, {Sanders}, {Zuntz}, {Williams}, {Nelson}, {de
  Val-Borro}, {Erhardt}, {Pashchenko}, \& {Pla}}]{foremanmackey2019}
{Foreman-Mackey}, D., {Farr}, W., {Sinha}, M., {Archibald}, A., {Hogg}, D.,
  {Sanders}, J., {Zuntz}, J., {Williams}, P., {Nelson}, A., {de Val-Borro}, M.,
  {Erhardt}, T., {Pashchenko}, I., \& {Pla}, O. 2019, The Journal of Open
  Source Software, 4, 1864

\bibitem[{{Foreman-Mackey} {et~al.}(2013){Foreman-Mackey}, {Hogg}, {Lang}, \&
  {Goodman}}]{foremanmackey2013}
{Foreman-Mackey}, D., {Hogg}, D.~W., {Lang}, D., \& {Goodman}, J. 2013, \pasp,
  125, 306

\bibitem[{{Fran{\c{c}}ois} {et~al.}(2016){Fran{\c{c}}ois}, {Monaco},
  {Bonifacio}, {Moni Bidin}, {Geisler}, \& {Sbordone}}]{francois2016}
{Fran{\c{c}}ois}, P., {Monaco}, L., {Bonifacio}, P., {Moni Bidin}, C.,
  {Geisler}, D., \& {Sbordone}, L. 2016, \aap, 588, A7

\bibitem[{{Frebel} {et~al.}(2016){Frebel}, {Norris}, {Gilmore}, \&
  {Wyse}}]{frebel2016}
{Frebel}, A., {Norris}, J.~E., {Gilmore}, G., \& {Wyse}, R. F.~G. 2016, \apj,
  826, 110

\bibitem[{{Frebel} {et~al.}(2010){Frebel}, {Simon}, {Geha}, \&
  {Willman}}]{frebel2010}
{Frebel}, A., {Simon}, J.~D., {Geha}, M., \& {Willman}, B. 2010, \apj, 708, 560

\bibitem[{{Frebel} {et~al.}(2014){Frebel}, {Simon}, \& {Kirby}}]{frebel2014}
{Frebel}, A., {Simon}, J.~D., \& {Kirby}, E.~N. 2014, \apj, 786, 74

\bibitem[{{Fritz} {et~al.}(2018){Fritz}, {Battaglia}, {Pawlowski},
  {Kallivayalil}, {van der Marel}, {Sohn}, {Brook}, \& {Besla}}]{fritz2018}
{Fritz}, T.~K., {Battaglia}, G., {Pawlowski}, M.~S., {Kallivayalil}, N., {van
  der Marel}, R., {Sohn}, S.~T., {Brook}, C., \& {Besla}, G. 2018, \aap, 619,
  A103

\bibitem[{{Fritz} {et~al.}(2019){Fritz}, {Carrera}, {Battaglia}, \&
  {Taibi}}]{fritz2019}
{Fritz}, T.~K., {Carrera}, R., {Battaglia}, G., \& {Taibi}, S. 2019, \aap, 623,
  A129

\bibitem[{{Fritz} {et~al.}(2020){Fritz}, {Di Cintio}, {Battaglia}, {Brook}, \&
  {Taibi}}]{fritz2020}
{Fritz}, T.~K., {Di Cintio}, A., {Battaglia}, G., {Brook}, C., \& {Taibi}, S.
  2020, \mnras, 494, 5178

\bibitem[{{Fu} {et~al.}(2019){Fu}, {Simon}, \& {Alarc{\'o}n Jara}}]{fu2019}
{Fu}, S.~W., {Simon}, J.~D., \& {Alarc{\'o}n Jara}, A.~G. 2019, \apj, 883, 11

\bibitem[{{Gaia Collaboration} {et~al.}(2018{\natexlab{a}}){Gaia
  Collaboration}, {Babusiaux}, {van Leeuwen}, {Barstow}, {Jordi}, {Vallenari},
  {Bossini}, {Bressan}, {Cantat-Gaudin}, {van Leeuwen}, {Brown}, {Prusti}, {de
  Bruijne}, {Bailer-Jones}, {Biermann}, {Evans}, {Eyer}, {Jansen}, {Klioner},
  {Lammers}, {Lindegren}, {Luri}, {Mignard}, {Panem}, {Pourbaix}, {Randich},
  {Sartoretti}, {Siddiqui}, {Soubiran}, {Walton}, {Arenou}, {Bastian},
  {Cropper}, {Drimmel}, {Katz}, {Lattanzi}, {Bakker}, {Cacciari},
  {Casta{\~n}eda}, {Chaoul}, {Cheek}, {De Angeli}, {Fabricius}, {Guerra},
  {Holl}, {Masana}, {Messineo}, {Mowlavi}, {Nienartowicz}, {Panuzzo},
  {Portell}, {Riello}, {Seabroke}, {Tanga}, {Th{\'e}venin}, {Gracia-Abril},
  {Comoretto}, {Garcia-Reinaldos}, {Teyssier}, {Altmann}, {Andrae}, {Audard},
  {Bellas-Velidis}, {Benson}, {Berthier}, {Blomme}, {Burgess}, {Busso},
  {Carry}, {Cellino}, {Clementini}, {Clotet}, {Creevey}, {Davidson}, {De
  Ridder}, {Delchambre}, {Dell'Oro}, {Ducourant},
  {Fern{\'a}ndez-Hern{\'a}ndez}, {Fouesneau}, {Fr{\'e}mat}, {Galluccio},
  {Garc{\'\i}a-Torres}, {Gonz{\'a}lez-N{\'u}{\~n}ez}, {Gonz{\'a}lez-Vidal},
  {Gosset}, {Guy}, {Halbwachs}, {Hambly}, {Harrison}, {Hern{\'a}ndez},
  {Hestroffer}, {Hodgkin}, {Hutton}, {Jasniewicz}, {Jean-Antoine-Piccolo},
  {Jordan}, {Korn}, {Krone-Martins}, {Lanzafame}, {Lebzelter}, {L{\"o}ffler},
  {Manteiga}, {Marrese}, {Mart{\'\i}n-Fleitas}, {Moitinho}, {Mora}, {Muinonen},
  {Osinde}, {Pancino}, {Pauwels}, {Petit}, {Recio-Blanco}, {Richards},
  {Rimoldini}, {Robin}, {Sarro}, {Siopis}, {Smith}, {Sozzetti}, {S{\"u}veges},
  {Torra}, {van Reeven}, {Abbas}, {Abreu Aramburu}, {Accart}, {Aerts},
  {Altavilla}, {{\'A}lvarez}, {Alvarez}, {Alves}, {Anderson}, {Andrei},
  {Anglada Varela}, {Antiche}, {Antoja}, {Arcay}, {Astraatmadja}, {Bach},
  {Baker}, {Balaguer-N{\'u}{\~n}ez}, {Balm}, {Barache}, {Barata}, {Barbato},
  {Barblan}, {Barklem}, {Barrado}, {Barros}, {Bartholom{\'e} Mu{\~n}oz},
  {Bassilana}, {Becciani}, {Bellazzini}, {Berihuete}, {Bertone}, {Bianchi},
  {Bienaym{\'e}}, {Blanco-Cuaresma}, {Boch}, {Boeche}, {Bombrun}, {Borrachero},
  {Bouquillon}, {Bourda}, {Bragaglia}, {Bramante}, {Breddels}, {Brouillet},
  {Br{\"u}semeister}, {Brugaletta}, {Bucciarelli}, {Burlacu}, {Busonero},
  {Butkevich}, {Buzzi}, {Caffau}, {Cancelliere}, {Cannizzaro}, {Carballo},
  {Carlucci}, {Carrasco}, {Casamiquela}, {Castellani}, {Castro-Ginard},
  {Charlot}, {Chemin}, {Chiavassa}, {Cocozza}, {Costigan}, {Cowell}, {Crifo},
  {Crosta}, {Crowley}, {Cuypers}, {Dafonte}, {Damerdji}, {Dapergolas}, {David},
  {David}, {de Laverny}, {De Luise}, {De March}, {de Martino}, {de Souza}, {de
  Torres}, {Debosscher}, {del Pozo}, {Delbo}, {Delgado}, {Delgado}, {Diakite},
  {Diener}, {Distefano}, {Dolding}, {Drazinos}, {Dur{\'a}n}, {Edvardsson},
  {Enke}, {Eriksson}, {Esquej}, {Eynard Bontemps}, {Fabre}, {Fabrizio},
  {Faigler}, {Falc{\~a}o}, {Farr{\`a}s Casas}, {Federici}, {Fedorets},
  {Fernique}, {Figueras}, {Filippi}, {Findeisen}, {Fonti}, {Fraile}, {Fraser},
  {Fr{\'e}zouls}, {Gai}, {Galleti}, {Garabato}, {Garc{\'\i}a-Sedano},
  {Garofalo}, {Garralda}, {Gavel}, {Gavras}, {Gerssen}, {Geyer}, {Giacobbe},
  {Gilmore}, {Girona}, {Giuffrida}, {Glass}, {Gomes}, {Granvik}, {Gueguen},
  {Guerrier}, {Guiraud}, {Guti{\'e}}, {Haigron}, {Hatzidimitriou}, {Hauser},
  {Haywood}, {Heiter}, {Helmi}, {Heu}, {Hilger}, {Hobbs}, {Hofmann}, {Holland},
  {Huckle}, {Hypki}, {Icardi}, {Jan{\ss}en}, {Jevardat de Fombelle}, {Jonker},
  {Juh{\'a}sz}, {Julbe}, {Karampelas}, {Kewley}, {Klar}, {Kochoska}, {Kohley},
  {Kolenberg}, {Kontizas}, {Kontizas}, {Koposov}, {Kordopatis},
  {Kostrzewa-Rutkowska}, {Koubsky}, {Lambert}, {Lanza}, {Lasne}, {Lavigne}, {Le
  Fustec}, {Le Poncin-Lafitte}, {Lebreton}, {Leccia}, {Leclerc},
  {Lecoeur-Taibi}, {Lenhardt}, {Leroux}, {Liao}, {Licata}, {Lindstr{\o}m},
  {Lister}, {Livanou}, {Lobel}, {L{\'o}pez}, {Managau}, {Mann}, {Mantelet},
  {Marchal}, {Marchant}, {Marconi}, {Marinoni}, {Marschalk{\'o}}, {Marshall},
  {Martino}, {Marton}, {Mary}, {Massari}, {Matijevi{\v{c}}}, {Mazeh},
  {McMillan}, {Messina}, {Michalik}, {Millar}, {Molina}, {Molinaro},
  {Moln{\'a}r}, {Montegriffo}, {Mor}, {Morbidelli}, {Morel}, {Morris},
  {Mulone}, {Muraveva}, {Musella}, {Nelemans}, {Nicastro}, {Noval},
  {O'Mullane}, {Ord{\'e}novic}, {Ord{\'o}{\~n}ez-Blanco}, {Osborne}, {Pagani},
  {Pagano}, {Pailler}, {Palacin}, {Palaversa}, {Panahi}, {Pawlak},
  {Piersimoni}, {Pineau}, {Plachy}, {Plum}, {Poggio}, {Poujoulet},
  {Pr{\v{s}}a}, {Pulone}, {Racero}, {Ragaini}, {Rambaux}, {Ramos-Lerate},
  {Regibo}, {Reyl{\'e}}, {Riclet}, {Ripepi}, {Riva}, {Rivard}, {Rixon},
  {Roegiers}, {Roelens}, {Romero-G{\'o}mez}, {Rowell}, {Royer}, {Ruiz-Dern},
  {Sadowski}, {Sagrist{\`a} Sell{\'e}s}, {Sahlmann}, {Salgado}, {Salguero},
  {Sanna}, {Santana-Ros}, {Sarasso}, {Savietto}, {Schultheis}, {Sciacca},
  {Segol}, {Segovia}, {S{\'e}gransan}, {Shih}, {Siltala}, {Silva}, {Smart},
  {Smith}, {Solano}, {Solitro}, {Sordo}, {Soria Nieto}, {Souchay}, {Spagna},
  {Spoto}, {Stampa}, {Steele}, {Steidelm{\"u}ller}, {Stephenson}, {Stoev},
  {Suess}, {Surdej}, {Szabados}, {Szegedi-Elek}, {Tapiador}, {Taris}, {Tauran},
  {Taylor}, {Teixeira}, {Terrett}, {Teyssand ier}, {Thuillot}, {Titarenko},
  {Torra Clotet}, {Turon}, {Ulla}, {Utrilla}, {Uzzi}, {Vaillant}, {Valentini},
  {Valette}, {van Elteren}, {Van Hemelryck}, {Vaschetto}, {Vecchiato},
  {Veljanoski}, {Viala}, {Vicente}, {Vogt}, {von Essen}, {Voss}, {Votruba},
  {Voutsinas}, {Walmsley}, {Weiler}, {Wertz}, {Wevers}, {Wyrzykowski},
  {Yoldas}, {{\v{Z}}erjal}, {Ziaeepour}, {Zorec}, {Zschocke}, {Zucker},
  {Zurbach}, \& {Zwitter}}]{babusiaux2018}
{Gaia Collaboration}, {Babusiaux}, C., {van Leeuwen}, F., {Barstow}, M.~A.,
  {Jordi}, C., {Vallenari}, A., {Bossini}, D., {Bressan}, A., {Cantat-Gaudin},
  T., {van Leeuwen}, M., {Brown}, A.~G.~A., {Prusti}, T., {de Bruijne},
  J.~H.~J., {Bailer-Jones}, C.~A.~L., {Biermann}, M., {Evans}, D.~W., {Eyer},
  L., {Jansen}, F., {Klioner}, S.~A., {Lammers}, U., {Lindegren}, L., {Luri},
  X., {Mignard}, F., {Panem}, C., {Pourbaix}, D., {Randich}, S., {Sartoretti},
  P., {Siddiqui}, H.~I., {Soubiran}, C., {Walton}, N.~A., {Arenou}, F.,
  {Bastian}, U., {Cropper}, M., {Drimmel}, R., {Katz}, D., {Lattanzi}, M.~G.,
  {Bakker}, J., {Cacciari}, C., {Casta{\~n}eda}, J., {Chaoul}, L., {Cheek}, N.,
  {De Angeli}, F., {Fabricius}, C., {Guerra}, R., {Holl}, B., {Masana}, E.,
  {Messineo}, R., {Mowlavi}, N., {Nienartowicz}, K., {Panuzzo}, P., {Portell},
  J., {Riello}, M., {Seabroke}, G.~M., {Tanga}, P., {Th{\'e}venin}, F.,
  {Gracia-Abril}, G., {Comoretto}, G., {Garcia-Reinaldos}, M., {Teyssier}, D.,
  {Altmann}, M., {Andrae}, R., {Audard}, M., {Bellas-Velidis}, I., {Benson},
  K., {Berthier}, J., {Blomme}, R., {Burgess}, P., {Busso}, G., {Carry}, B.,
  {Cellino}, A., {Clementini}, G., {Clotet}, M., {Creevey}, O., {Davidson}, M.,
  {De Ridder}, J., {Delchambre}, L., {Dell'Oro}, A., {Ducourant}, C.,
  {Fern{\'a}ndez-Hern{\'a}ndez}, J., {Fouesneau}, M., {Fr{\'e}mat}, Y.,
  {Galluccio}, L., {Garc{\'\i}a-Torres}, M., {Gonz{\'a}lez-N{\'u}{\~n}ez}, J.,
  {Gonz{\'a}lez-Vidal}, J.~J., {Gosset}, E., {Guy}, L.~P., {Halbwachs}, J.~L.,
  {Hambly}, N.~C., {Harrison}, D.~L., {Hern{\'a}ndez}, J., {Hestroffer}, D.,
  {Hodgkin}, S.~T., {Hutton}, A., {Jasniewicz}, G., {Jean-Antoine-Piccolo}, A.,
  {Jordan}, S., {Korn}, A.~J., {Krone-Martins}, A., {Lanzafame}, A.~C.,
  {Lebzelter}, T., {L{\"o}ffler}, W., {Manteiga}, M., {Marrese}, P.~M.,
  {Mart{\'\i}n-Fleitas}, J.~M., {Moitinho}, A., {Mora}, A., {Muinonen}, K.,
  {Osinde}, J., {Pancino}, E., {Pauwels}, T., {Petit}, J.~M., {Recio-Blanco},
  A., {Richards}, P.~J., {Rimoldini}, L., {Robin}, A.~C., {Sarro}, L.~M.,
  {Siopis}, C., {Smith}, M., {Sozzetti}, A., {S{\"u}veges}, M., {Torra}, J.,
  {van Reeven}, W., {Abbas}, U., {Abreu Aramburu}, A., {Accart}, S., {Aerts},
  C., {Altavilla}, G., {{\'A}lvarez}, M.~A., {Alvarez}, R., {Alves}, J.,
  {Anderson}, R.~I., {Andrei}, A.~H., {Anglada Varela}, E., {Antiche}, E.,
  {Antoja}, T., {Arcay}, B., {Astraatmadja}, T.~L., {Bach}, N., {Baker}, S.~G.,
  {Balaguer-N{\'u}{\~n}ez}, L., {Balm}, P., {Barache}, C., {Barata}, C.,
  {Barbato}, D., {Barblan}, F., {Barklem}, P.~S., {Barrado}, D., {Barros}, M.,
  {Bartholom{\'e} Mu{\~n}oz}, L., {Bassilana}, J.~L., {Becciani}, U.,
  {Bellazzini}, M., {Berihuete}, A., {Bertone}, S., {Bianchi}, L.,
  {Bienaym{\'e}}, O., {Blanco-Cuaresma}, S., {Boch}, T., {Boeche}, C.,
  {Bombrun}, A., {Borrachero}, R., {Bouquillon}, S., {Bourda}, G., {Bragaglia},
  A., {Bramante}, L., {Breddels}, M.~A., {Brouillet}, N., {Br{\"u}semeister},
  T., {Brugaletta}, E., {Bucciarelli}, B., {Burlacu}, A., {Busonero}, D.,
  {Butkevich}, A.~G., {Buzzi}, R., {Caffau}, E., {Cancelliere}, R.,
  {Cannizzaro}, G., {Carballo}, R., {Carlucci}, T., {Carrasco}, J.~M.,
  {Casamiquela}, L., {Castellani}, M., {Castro-Ginard}, A., {Charlot}, P.,
  {Chemin}, L., {Chiavassa}, A., {Cocozza}, G., {Costigan}, G., {Cowell}, S.,
  {Crifo}, F., {Crosta}, M., {Crowley}, C., {Cuypers}, J., {Dafonte}, C.,
  {Damerdji}, Y., {Dapergolas}, A., {David}, P., {David}, M., {de Laverny}, P.,
  {De Luise}, F., {De March}, R., {de Martino}, D., {de Souza}, R., {de
  Torres}, A., {Debosscher}, J., {del Pozo}, E., {Delbo}, M., {Delgado}, A.,
  {Delgado}, H.~E., {Diakite}, S., {Diener}, C., {Distefano}, E., {Dolding},
  C., {Drazinos}, P., {Dur{\'a}n}, J., {Edvardsson}, B., {Enke}, H.,
  {Eriksson}, K., {Esquej}, P., {Eynard Bontemps}, G., {Fabre}, C., {Fabrizio},
  M., {Faigler}, S., {Falc{\~a}o}, A.~J., {Farr{\`a}s Casas}, M., {Federici},
  L., {Fedorets}, G., {Fernique}, P., {Figueras}, F., {Filippi}, F.,
  {Findeisen}, K., {Fonti}, A., {Fraile}, E., {Fraser}, M., {Fr{\'e}zouls}, B.,
  {Gai}, M., {Galleti}, S., {Garabato}, D., {Garc{\'\i}a-Sedano}, F.,
  {Garofalo}, A., {Garralda}, N., {Gavel}, A., {Gavras}, P., {Gerssen}, J.,
  {Geyer}, R., {Giacobbe}, P., {Gilmore}, G., {Girona}, S., {Giuffrida}, G.,
  {Glass}, F., {Gomes}, M., {Granvik}, M., {Gueguen}, A., {Guerrier}, A.,
  {Guiraud}, J., {Guti{\'e}}, R., {Haigron}, R., {Hatzidimitriou}, D.,
  {Hauser}, M., {Haywood}, M., {Heiter}, U., {Helmi}, A., {Heu}, J., {Hilger},
  T., {Hobbs}, D., {Hofmann}, W., {Holland}, G., {Huckle}, H.~E., {Hypki}, A.,
  {Icardi}, V., {Jan{\ss}en}, K., {Jevardat de Fombelle}, G., {Jonker}, P.~G.,
  {Juh{\'a}sz}, {\'A}.~L., {Julbe}, F., {Karampelas}, A., {Kewley}, A., {Klar},
  J., {Kochoska}, A., {Kohley}, R., {Kolenberg}, K., {Kontizas}, M.,
  {Kontizas}, E., {Koposov}, S.~E., {Kordopatis}, G., {Kostrzewa-Rutkowska},
  Z., {Koubsky}, P., {Lambert}, S., {Lanza}, A.~F., {Lasne}, Y., {Lavigne},
  J.~B., {Le Fustec}, Y., {Le Poncin-Lafitte}, C., {Lebreton}, Y., {Leccia},
  S., {Leclerc}, N., {Lecoeur-Taibi}, I., {Lenhardt}, H., {Leroux}, F., {Liao},
  S., {Licata}, E., {Lindstr{\o}m}, H.~E.~P., {Lister}, T.~A., {Livanou}, E.,
  {Lobel}, A., {L{\'o}pez}, M., {Managau}, S., {Mann}, R.~G., {Mantelet}, G.,
  {Marchal}, O., {Marchant}, J.~M., {Marconi}, M., {Marinoni}, S.,
  {Marschalk{\'o}}, G., {Marshall}, D.~J., {Martino}, M., {Marton}, G., {Mary},
  N., {Massari}, D., {Matijevi{\v{c}}}, G., {Mazeh}, T., {McMillan}, P.~J.,
  {Messina}, S., {Michalik}, D., {Millar}, N.~R., {Molina}, D., {Molinaro}, R.,
  {Moln{\'a}r}, L., {Montegriffo}, P., {Mor}, R., {Morbidelli}, R., {Morel},
  T., {Morris}, D., {Mulone}, A.~F., {Muraveva}, T., {Musella}, I., {Nelemans},
  G., {Nicastro}, L., {Noval}, L., {O'Mullane}, W., {Ord{\'e}novic}, C.,
  {Ord{\'o}{\~n}ez-Blanco}, D., {Osborne}, P., {Pagani}, C., {Pagano}, I.,
  {Pailler}, F., {Palacin}, H., {Palaversa}, L., {Panahi}, A., {Pawlak}, M.,
  {Piersimoni}, A.~M., {Pineau}, F.~X., {Plachy}, E., {Plum}, G., {Poggio}, E.,
  {Poujoulet}, E., {Pr{\v{s}}a}, A., {Pulone}, L., {Racero}, E., {Ragaini}, S.,
  {Rambaux}, N., {Ramos-Lerate}, M., {Regibo}, S., {Reyl{\'e}}, C., {Riclet},
  F., {Ripepi}, V., {Riva}, A., {Rivard}, A., {Rixon}, G., {Roegiers}, T.,
  {Roelens}, M., {Romero-G{\'o}mez}, M., {Rowell}, N., {Royer}, F.,
  {Ruiz-Dern}, L., {Sadowski}, G., {Sagrist{\`a} Sell{\'e}s}, T., {Sahlmann},
  J., {Salgado}, J., {Salguero}, E., {Sanna}, N., {Santana-Ros}, T., {Sarasso},
  M., {Savietto}, H., {Schultheis}, M., {Sciacca}, E., {Segol}, M., {Segovia},
  J.~C., {S{\'e}gransan}, D., {Shih}, I.~C., {Siltala}, L., {Silva}, A.~F.,
  {Smart}, R.~L., {Smith}, K.~W., {Solano}, E., {Solitro}, F., {Sordo}, R.,
  {Soria Nieto}, S., {Souchay}, J., {Spagna}, A., {Spoto}, F., {Stampa}, U.,
  {Steele}, I.~A., {Steidelm{\"u}ller}, H., {Stephenson}, C.~A., {Stoev}, H.,
  {Suess}, F.~F., {Surdej}, J., {Szabados}, L., {Szegedi-Elek}, E., {Tapiador},
  D., {Taris}, F., {Tauran}, G., {Taylor}, M.~B., {Teixeira}, R., {Terrett},
  D., {Teyssand ier}, P., {Thuillot}, W., {Titarenko}, A., {Torra Clotet}, F.,
  {Turon}, C., {Ulla}, A., {Utrilla}, E., {Uzzi}, S., {Vaillant}, M.,
  {Valentini}, G., {Valette}, V., {van Elteren}, A., {Van Hemelryck}, E.,
  {Vaschetto}, M., {Vecchiato}, A., {Veljanoski}, J., {Viala}, Y., {Vicente},
  D., {Vogt}, S., {von Essen}, C., {Voss}, H., {Votruba}, V., {Voutsinas}, S.,
  {Walmsley}, G., {Weiler}, M., {Wertz}, O., {Wevers}, T., {Wyrzykowski},
  {\L}., {Yoldas}, A., {{\v{Z}}erjal}, M., {Ziaeepour}, H., {Zorec}, J.,
  {Zschocke}, S., {Zucker}, S., {Zurbach}, C., \& {Zwitter}, T.
  2018{\natexlab{a}}, \aap, 616, A10

\bibitem[{{Gaia Collaboration} {et~al.}(2018{\natexlab{b}}){Gaia
  Collaboration}, {Brown}, {Vallenari}, {Prusti}, {de Bruijne}, {Babusiaux},
  {Bailer-Jones}, {Biermann}, {Evans}, {Eyer}, {Jansen}, {Jordi}, {Klioner},
  {Lammers}, {Lindegren}, {Luri}, {Mignard}, {Panem}, {Pourbaix}, {Randich},
  {Sartoretti}, {Siddiqui}, {Soubiran}, {van Leeuwen}, {Walton}, {Arenou},
  {Bastian}, {Cropper}, {Drimmel}, {Katz}, {Lattanzi}, {Bakker}, {Cacciari},
  {Casta{\~n}eda}, {Chaoul}, {Cheek}, {De Angeli}, {Fabricius}, {Guerra},
  {Holl}, {Masana}, {Messineo}, {Mowlavi}, {Nienartowicz}, {Panuzzo},
  {Portell}, {Riello}, {Seabroke}, {Tanga}, {Th{\'e}venin}, {Gracia-Abril},
  {Comoretto}, {Garcia-Reinaldos}, {Teyssier}, {Altmann}, {Andrae}, {Audard},
  {Bellas-Velidis}, {Benson}, {Berthier}, {Blomme}, {Burgess}, {Busso},
  {Carry}, {Cellino}, {Clementini}, {Clotet}, {Creevey}, {Davidson}, {De
  Ridder}, {Delchambre}, {Dell'Oro}, {Ducourant},
  {Fern{\'a}ndez-Hern{\'a}ndez}, {Fouesneau}, {Fr{\'e}mat}, {Galluccio},
  {Garc{\'\i}a-Torres}, {Gonz{\'a}lez-N{\'u}{\~n}ez}, {Gonz{\'a}lez-Vidal},
  {Gosset}, {Guy}, {Halbwachs}, {Hambly}, {Harrison}, {Hern{\'a}ndez},
  {Hestroffer}, {Hodgkin}, {Hutton}, {Jasniewicz}, {Jean-Antoine-Piccolo},
  {Jordan}, {Korn}, {Krone-Martins}, {Lanzafame}, {Lebzelter}, {L{\"o}ffler},
  {Manteiga}, {Marrese}, {Mart{\'\i}n-Fleitas}, {Moitinho}, {Mora}, {Muinonen},
  {Osinde}, {Pancino}, {Pauwels}, {Petit}, {Recio-Blanco}, {Richards},
  {Rimoldini}, {Robin}, {Sarro}, {Siopis}, {Smith}, {Sozzetti}, {S{\"u}veges},
  {Torra}, {van Reeven}, {Abbas}, {Abreu Aramburu}, {Accart}, {Aerts},
  {Altavilla}, {{\'A}lvarez}, {Alvarez}, {Alves}, {Anderson}, {Andrei},
  {Anglada Varela}, {Antiche}, {Antoja}, {Arcay}, {Astraatmadja}, {Bach},
  {Baker}, {Balaguer-N{\'u}{\~n}ez}, {Balm}, {Barache}, {Barata}, {Barbato},
  {Barblan}, {Barklem}, {Barrado}, {Barros}, {Barstow}, {Bartholom{\'e}
  Mu{\~n}oz}, {Bassilana}, {Becciani}, {Bellazzini}, {Berihuete}, {Bertone},
  {Bianchi}, {Bienaym{\'e}}, {Blanco-Cuaresma}, {Boch}, {Boeche}, {Bombrun},
  {Borrachero}, {Bossini}, {Bouquillon}, {Bourda}, {Bragaglia}, {Bramante},
  {Breddels}, {Bressan}, {Brouillet}, {Br{\"u}semeister}, {Brugaletta},
  {Bucciarelli}, {Burlacu}, {Busonero}, {Butkevich}, {Buzzi}, {Caffau},
  {Cancelliere}, {Cannizzaro}, {Cantat-Gaudin}, {Carballo}, {Carlucci},
  {Carrasco}, {Casamiquela}, {Castellani}, {Castro-Ginard}, {Charlot},
  {Chemin}, {Chiavassa}, {Cocozza}, {Costigan}, {Cowell}, {Crifo}, {Crosta},
  {Crowley}, {Cuypers}, {Dafonte}, {Damerdji}, {Dapergolas}, {David}, {David},
  {de Laverny}, {De Luise}, {De March}, {de Martino}, {de Souza}, {de Torres},
  {Debosscher}, {del Pozo}, {Delbo}, {Delgado}, {Delgado}, {Di Matteo},
  {Diakite}, {Diener}, {Distefano}, {Dolding}, {Drazinos}, {Dur{\'a}n},
  {Edvardsson}, {Enke}, {Eriksson}, {Esquej}, {Eynard Bontemps}, {Fabre},
  {Fabrizio}, {Faigler}, {Falc{\~a}o}, {Farr{\`a}s Casas}, {Federici},
  {Fedorets}, {Fernique}, {Figueras}, {Filippi}, {Findeisen}, {Fonti},
  {Fraile}, {Fraser}, {Fr{\'e}zouls}, {Gai}, {Galleti}, {Garabato},
  {Garc{\'\i}a-Sedano}, {Garofalo}, {Garralda}, {Gavel}, {Gavras}, {Gerssen},
  {Geyer}, {Giacobbe}, {Gilmore}, {Girona}, {Giuffrida}, {Glass}, {Gomes},
  {Granvik}, {Gueguen}, {Guerrier}, {Guiraud}, {Guti{\'e}rrez-S{\'a}nchez},
  {Haigron}, {Hatzidimitriou}, {Hauser}, {Haywood}, {Heiter}, {Helmi}, {Heu},
  {Hilger}, {Hobbs}, {Hofmann}, {Holland}, {Huckle}, {Hypki}, {Icardi},
  {Jan{\ss}en}, {Jevardat de Fombelle}, {Jonker}, {Juh{\'a}sz}, {Julbe},
  {Karampelas}, {Kewley}, {Klar}, {Kochoska}, {Kohley}, {Kolenberg},
  {Kontizas}, {Kontizas}, {Koposov}, {Kordopatis}, {Kostrzewa-Rutkowska},
  {Koubsky}, {Lambert}, {Lanza}, {Lasne}, {Lavigne}, {Le Fustec}, {Le
  Poncin-Lafitte}, {Lebreton}, {Leccia}, {Leclerc}, {Lecoeur-Taibi},
  {Lenhardt}, {Leroux}, {Liao}, {Licata}, {Lindstr{\o}m}, {Lister}, {Livanou},
  {Lobel}, {L{\'o}pez}, {Managau}, {Mann}, {Mantelet}, {Marchal}, {Marchant},
  {Marconi}, {Marinoni}, {Marschalk{\'o}}, {Marshall}, {Martino}, {Marton},
  {Mary}, {Massari}, {Matijevi{\v{c}}}, {Mazeh}, {McMillan}, {Messina},
  {Michalik}, {Millar}, {Molina}, {Molinaro}, {Moln{\'a}r}, {Montegriffo},
  {Mor}, {Morbidelli}, {Morel}, {Morris}, {Mulone}, {Muraveva}, {Musella},
  {Nelemans}, {Nicastro}, {Noval}, {O'Mullane}, {Ord{\'e}novic},
  {Ord{\'o}{\~n}ez-Blanco}, {Osborne}, {Pagani}, {Pagano}, {Pailler},
  {Palacin}, {Palaversa}, {Panahi}, {Pawlak}, {Piersimoni}, {Pineau}, {Plachy},
  {Plum}, {Poggio}, {Poujoulet}, {Pr{\v{s}}a}, {Pulone}, {Racero}, {Ragaini},
  {Rambaux}, {Ramos-Lerate}, {Regibo}, {Reyl{\'e}}, {Riclet}, {Ripepi}, {Riva},
  {Rivard}, {Rixon}, {Roegiers}, {Roelens}, {Romero-G{\'o}mez}, {Rowell},
  {Royer}, {Ruiz-Dern}, {Sadowski}, {Sagrist{\`a} Sell{\'e}s}, {Sahlmann},
  {Salgado}, {Salguero}, {Sanna}, {Santana-Ros}, {Sarasso}, {Savietto},
  {Schultheis}, {Sciacca}, {Segol}, {Segovia}, {S{\'e}gransan}, {Shih},
  {Siltala}, {Silva}, {Smart}, {Smith}, {Solano}, {Solitro}, {Sordo}, {Soria
  Nieto}, {Souchay}, {Spagna}, {Spoto}, {Stampa}, {Steele},
  {Steidelm{\"u}ller}, {Stephenson}, {Stoev}, {Suess}, {Surdej}, {Szabados},
  {Szegedi-Elek}, {Tapiador}, {Taris}, {Tauran}, {Taylor}, {Teixeira},
  {Terrett}, {Teyssand ier}, {Thuillot}, {Titarenko}, {Torra Clotet}, {Turon},
  {Ulla}, {Utrilla}, {Uzzi}, {Vaillant}, {Valentini}, {Valette}, {van Elteren},
  {Van Hemelryck}, {van Leeuwen}, {Vaschetto}, {Vecchiato}, {Veljanoski},
  {Viala}, {Vicente}, {Vogt}, {von Essen}, {Voss}, {Votruba}, {Voutsinas},
  {Walmsley}, {Weiler}, {Wertz}, {Wevers}, {Wyrzykowski}, {Yoldas},
  {{\v{Z}}erjal}, {Ziaeepour}, {Zorec}, {Zschocke}, {Zucker}, {Zurbach}, \&
  {Zwitter}}]{gaia2018b}
{Gaia Collaboration}, {Brown}, A.~G.~A., {Vallenari}, A., {Prusti}, T., {de
  Bruijne}, J.~H.~J., {Babusiaux}, C., {Bailer-Jones}, C.~A.~L., {Biermann},
  M., {Evans}, D.~W., {Eyer}, L., {Jansen}, F., {Jordi}, C., {Klioner}, S.~A.,
  {Lammers}, U., {Lindegren}, L., {Luri}, X., {Mignard}, F., {Panem}, C.,
  {Pourbaix}, D., {Randich}, S., {Sartoretti}, P., {Siddiqui}, H.~I.,
  {Soubiran}, C., {van Leeuwen}, F., {Walton}, N.~A., {Arenou}, F., {Bastian},
  U., {Cropper}, M., {Drimmel}, R., {Katz}, D., {Lattanzi}, M.~G., {Bakker},
  J., {Cacciari}, C., {Casta{\~n}eda}, J., {Chaoul}, L., {Cheek}, N., {De
  Angeli}, F., {Fabricius}, C., {Guerra}, R., {Holl}, B., {Masana}, E.,
  {Messineo}, R., {Mowlavi}, N., {Nienartowicz}, K., {Panuzzo}, P., {Portell},
  J., {Riello}, M., {Seabroke}, G.~M., {Tanga}, P., {Th{\'e}venin}, F.,
  {Gracia-Abril}, G., {Comoretto}, G., {Garcia-Reinaldos}, M., {Teyssier}, D.,
  {Altmann}, M., {Andrae}, R., {Audard}, M., {Bellas-Velidis}, I., {Benson},
  K., {Berthier}, J., {Blomme}, R., {Burgess}, P., {Busso}, G., {Carry}, B.,
  {Cellino}, A., {Clementini}, G., {Clotet}, M., {Creevey}, O., {Davidson}, M.,
  {De Ridder}, J., {Delchambre}, L., {Dell'Oro}, A., {Ducourant}, C.,
  {Fern{\'a}ndez-Hern{\'a}ndez}, J., {Fouesneau}, M., {Fr{\'e}mat}, Y.,
  {Galluccio}, L., {Garc{\'\i}a-Torres}, M., {Gonz{\'a}lez-N{\'u}{\~n}ez}, J.,
  {Gonz{\'a}lez-Vidal}, J.~J., {Gosset}, E., {Guy}, L.~P., {Halbwachs}, J.~L.,
  {Hambly}, N.~C., {Harrison}, D.~L., {Hern{\'a}ndez}, J., {Hestroffer}, D.,
  {Hodgkin}, S.~T., {Hutton}, A., {Jasniewicz}, G., {Jean-Antoine-Piccolo}, A.,
  {Jordan}, S., {Korn}, A.~J., {Krone-Martins}, A., {Lanzafame}, A.~C.,
  {Lebzelter}, T., {L{\"o}ffler}, W., {Manteiga}, M., {Marrese}, P.~M.,
  {Mart{\'\i}n-Fleitas}, J.~M., {Moitinho}, A., {Mora}, A., {Muinonen}, K.,
  {Osinde}, J., {Pancino}, E., {Pauwels}, T., {Petit}, J.~M., {Recio-Blanco},
  A., {Richards}, P.~J., {Rimoldini}, L., {Robin}, A.~C., {Sarro}, L.~M.,
  {Siopis}, C., {Smith}, M., {Sozzetti}, A., {S{\"u}veges}, M., {Torra}, J.,
  {van Reeven}, W., {Abbas}, U., {Abreu Aramburu}, A., {Accart}, S., {Aerts},
  C., {Altavilla}, G., {{\'A}lvarez}, M.~A., {Alvarez}, R., {Alves}, J.,
  {Anderson}, R.~I., {Andrei}, A.~H., {Anglada Varela}, E., {Antiche}, E.,
  {Antoja}, T., {Arcay}, B., {Astraatmadja}, T.~L., {Bach}, N., {Baker}, S.~G.,
  {Balaguer-N{\'u}{\~n}ez}, L., {Balm}, P., {Barache}, C., {Barata}, C.,
  {Barbato}, D., {Barblan}, F., {Barklem}, P.~S., {Barrado}, D., {Barros}, M.,
  {Barstow}, M.~A., {Bartholom{\'e} Mu{\~n}oz}, S., {Bassilana}, J.~L.,
  {Becciani}, U., {Bellazzini}, M., {Berihuete}, A., {Bertone}, S., {Bianchi},
  L., {Bienaym{\'e}}, O., {Blanco-Cuaresma}, S., {Boch}, T., {Boeche}, C.,
  {Bombrun}, A., {Borrachero}, R., {Bossini}, D., {Bouquillon}, S., {Bourda},
  G., {Bragaglia}, A., {Bramante}, L., {Breddels}, M.~A., {Bressan}, A.,
  {Brouillet}, N., {Br{\"u}semeister}, T., {Brugaletta}, E., {Bucciarelli}, B.,
  {Burlacu}, A., {Busonero}, D., {Butkevich}, A.~G., {Buzzi}, R., {Caffau}, E.,
  {Cancelliere}, R., {Cannizzaro}, G., {Cantat-Gaudin}, T., {Carballo}, R.,
  {Carlucci}, T., {Carrasco}, J.~M., {Casamiquela}, L., {Castellani}, M.,
  {Castro-Ginard}, A., {Charlot}, P., {Chemin}, L., {Chiavassa}, A., {Cocozza},
  G., {Costigan}, G., {Cowell}, S., {Crifo}, F., {Crosta}, M., {Crowley}, C.,
  {Cuypers}, J., {Dafonte}, C., {Damerdji}, Y., {Dapergolas}, A., {David}, P.,
  {David}, M., {de Laverny}, P., {De Luise}, F., {De March}, R., {de Martino},
  D., {de Souza}, R., {de Torres}, A., {Debosscher}, J., {del Pozo}, E.,
  {Delbo}, M., {Delgado}, A., {Delgado}, H.~E., {Di Matteo}, P., {Diakite}, S.,
  {Diener}, C., {Distefano}, E., {Dolding}, C., {Drazinos}, P., {Dur{\'a}n},
  J., {Edvardsson}, B., {Enke}, H., {Eriksson}, K., {Esquej}, P., {Eynard
  Bontemps}, G., {Fabre}, C., {Fabrizio}, M., {Faigler}, S., {Falc{\~a}o},
  A.~J., {Farr{\`a}s Casas}, M., {Federici}, L., {Fedorets}, G., {Fernique},
  P., {Figueras}, F., {Filippi}, F., {Findeisen}, K., {Fonti}, A., {Fraile},
  E., {Fraser}, M., {Fr{\'e}zouls}, B., {Gai}, M., {Galleti}, S., {Garabato},
  D., {Garc{\'\i}a-Sedano}, F., {Garofalo}, A., {Garralda}, N., {Gavel}, A.,
  {Gavras}, P., {Gerssen}, J., {Geyer}, R., {Giacobbe}, P., {Gilmore}, G.,
  {Girona}, S., {Giuffrida}, G., {Glass}, F., {Gomes}, M., {Granvik}, M.,
  {Gueguen}, A., {Guerrier}, A., {Guiraud}, J., {Guti{\'e}rrez-S{\'a}nchez},
  R., {Haigron}, R., {Hatzidimitriou}, D., {Hauser}, M., {Haywood}, M.,
  {Heiter}, U., {Helmi}, A., {Heu}, J., {Hilger}, T., {Hobbs}, D., {Hofmann},
  W., {Holland}, G., {Huckle}, H.~E., {Hypki}, A., {Icardi}, V., {Jan{\ss}en},
  K., {Jevardat de Fombelle}, G., {Jonker}, P.~G., {Juh{\'a}sz}, {\'A}.~L.,
  {Julbe}, F., {Karampelas}, A., {Kewley}, A., {Klar}, J., {Kochoska}, A.,
  {Kohley}, R., {Kolenberg}, K., {Kontizas}, M., {Kontizas}, E., {Koposov},
  S.~E., {Kordopatis}, G., {Kostrzewa-Rutkowska}, Z., {Koubsky}, P., {Lambert},
  S., {Lanza}, A.~F., {Lasne}, Y., {Lavigne}, J.~B., {Le Fustec}, Y., {Le
  Poncin-Lafitte}, C., {Lebreton}, Y., {Leccia}, S., {Leclerc}, N.,
  {Lecoeur-Taibi}, I., {Lenhardt}, H., {Leroux}, F., {Liao}, S., {Licata}, E.,
  {Lindstr{\o}m}, H.~E.~P., {Lister}, T.~A., {Livanou}, E., {Lobel}, A.,
  {L{\'o}pez}, M., {Managau}, S., {Mann}, R.~G., {Mantelet}, G., {Marchal}, O.,
  {Marchant}, J.~M., {Marconi}, M., {Marinoni}, S., {Marschalk{\'o}}, G.,
  {Marshall}, D.~J., {Martino}, M., {Marton}, G., {Mary}, N., {Massari}, D.,
  {Matijevi{\v{c}}}, G., {Mazeh}, T., {McMillan}, P.~J., {Messina}, S.,
  {Michalik}, D., {Millar}, N.~R., {Molina}, D., {Molinaro}, R., {Moln{\'a}r},
  L., {Montegriffo}, P., {Mor}, R., {Morbidelli}, R., {Morel}, T., {Morris},
  D., {Mulone}, A.~F., {Muraveva}, T., {Musella}, I., {Nelemans}, G.,
  {Nicastro}, L., {Noval}, L., {O'Mullane}, W., {Ord{\'e}novic}, C.,
  {Ord{\'o}{\~n}ez-Blanco}, D., {Osborne}, P., {Pagani}, C., {Pagano}, I.,
  {Pailler}, F., {Palacin}, H., {Palaversa}, L., {Panahi}, A., {Pawlak}, M.,
  {Piersimoni}, A.~M., {Pineau}, F.~X., {Plachy}, E., {Plum}, G., {Poggio}, E.,
  {Poujoulet}, E., {Pr{\v{s}}a}, A., {Pulone}, L., {Racero}, E., {Ragaini}, S.,
  {Rambaux}, N., {Ramos-Lerate}, M., {Regibo}, S., {Reyl{\'e}}, C., {Riclet},
  F., {Ripepi}, V., {Riva}, A., {Rivard}, A., {Rixon}, G., {Roegiers}, T.,
  {Roelens}, M., {Romero-G{\'o}mez}, M., {Rowell}, N., {Royer}, F.,
  {Ruiz-Dern}, L., {Sadowski}, G., {Sagrist{\`a} Sell{\'e}s}, T., {Sahlmann},
  J., {Salgado}, J., {Salguero}, E., {Sanna}, N., {Santana-Ros}, T., {Sarasso},
  M., {Savietto}, H., {Schultheis}, M., {Sciacca}, E., {Segol}, M., {Segovia},
  J.~C., {S{\'e}gransan}, D., {Shih}, I.~C., {Siltala}, L., {Silva}, A.~F.,
  {Smart}, R.~L., {Smith}, K.~W., {Solano}, E., {Solitro}, F., {Sordo}, R.,
  {Soria Nieto}, S., {Souchay}, J., {Spagna}, A., {Spoto}, F., {Stampa}, U.,
  {Steele}, I.~A., {Steidelm{\"u}ller}, H., {Stephenson}, C.~A., {Stoev}, H.,
  {Suess}, F.~F., {Surdej}, J., {Szabados}, L., {Szegedi-Elek}, E., {Tapiador},
  D., {Taris}, F., {Tauran}, G., {Taylor}, M.~B., {Teixeira}, R., {Terrett},
  D., {Teyssand ier}, P., {Thuillot}, W., {Titarenko}, A., {Torra Clotet}, F.,
  {Turon}, C., {Ulla}, A., {Utrilla}, E., {Uzzi}, S., {Vaillant}, M.,
  {Valentini}, G., {Valette}, V., {van Elteren}, A., {Van Hemelryck}, E., {van
  Leeuwen}, M., {Vaschetto}, M., {Vecchiato}, A., {Veljanoski}, J., {Viala},
  Y., {Vicente}, D., {Vogt}, S., {von Essen}, C., {Voss}, H., {Votruba}, V.,
  {Voutsinas}, S., {Walmsley}, G., {Weiler}, M., {Wertz}, O., {Wevers}, T.,
  {Wyrzykowski}, {\L}., {Yoldas}, A., {{\v{Z}}erjal}, M., {Ziaeepour}, H.,
  {Zorec}, J., {Zschocke}, S., {Zucker}, S., {Zurbach}, C., \& {Zwitter}, T.
  2018{\natexlab{b}}, \aap, 616, A1

\bibitem[{{Gaia Collaboration} {et~al.}(2018{\natexlab{c}}){Gaia
  Collaboration}, {Helmi}, {van Leeuwen}, {McMillan}, {Massari}, {Antoja},
  {Robin}, {Lindegren}, {Bastian}, {Arenou}, {Babusiaux}, {Biermann},
  {Breddels}, {Hobbs}, {Jordi}, {Pancino}, {Reyl{\'e}}, {Veljanoski}, {Brown},
  {Vallenari}, {Prusti}, {de Bruijne}, {Bailer-Jones}, {Evans}, {Eyer},
  {Jansen}, {Klioner}, {Lammers}, {Luri}, {Mignard}, {Panem}, {Pourbaix},
  {Randich}, {Sartoretti}, {Siddiqui}, {Soubiran}, {Walton}, {Cropper},
  {Drimmel}, {Katz}, {Lattanzi}, {Bakker}, {Cacciari}, {Casta{\~n}eda},
  {Chaoul}, {Cheek}, {De Angeli}, {Fabricius}, {Guerra}, {Holl}, {Masana},
  {Messineo}, {Mowlavi}, {Nienartowicz}, {Panuzzo}, {Portell}, {Riello},
  {Seabroke}, {Tanga}, {Th{\'e}venin}, {Gracia-Abril}, {Comoretto},
  {Garcia-Reinaldos}, {Teyssier}, {Altmann}, {Andrae}, {Audard},
  {Bellas-Velidis}, {Benson}, {Berthier}, {Blomme}, {Burgess}, {Busso},
  {Carry}, {Cellino}, {Clementini}, {Clotet}, {Creevey}, {Davidson}, {De
  Ridder}, {Delchambre}, {Dell'Oro}, {Ducourant},
  {Fern{\'a}ndez-Hern{\'a}ndez}, {Fouesneau}, {Fr{\'e}mat}, {Galluccio},
  {Garc{\'\i}a-Torres}, {Gonz{\'a}lez-N{\'u}{\~n}ez}, {Gonz{\'a}lez-Vidal},
  {Gosset}, {Guy}, {Halbwachs}, {Hambly}, {Harrison}, {Hern{\'a}ndez},
  {Hestroffer}, {Hodgkin}, {Hutton}, {Jasniewicz}, {Jean-Antoine-Piccolo},
  {Jordan}, {Korn}, {Krone-Martins}, {Lanzafame}, {Lebzelter}, {L{\"o}ffler},
  {Manteiga}, {Marrese}, {Mart{\'\i}n-Fleitas}, {Moitinho}, {Mora}, {Muinonen},
  {Osinde}, {Pauwels}, {Petit}, {Recio-Blanco}, {Richards}, {Rimoldini},
  {Sarro}, {Siopis}, {Smith}, {Sozzetti}, {S{\"u}veges}, {Torra}, {van Reeven},
  {Abbas}, {Abreu Aramburu}, {Accart}, {Aerts}, {Altavilla}, {{\'A}lvarez},
  {Alvarez}, {Alves}, {Anderson}, {Andrei}, {Anglada Varela}, {Antiche},
  {Arcay}, {Astraatmadja}, {Bach}, {Baker}, {Balaguer-N{\'u}{\~n}ez}, {Balm},
  {Barache}, {Barata}, {Barbato}, {Barblan}, {Barklem}, {Barrado}, {Barros},
  {Barstow}, {Bartholom{\'e} Mu{\~n}oz}, {Bassilana}, {Becciani}, {Bellazzini},
  {Berihuete}, {Bertone}, {Bianchi}, {Bienaym{\'e}}, {Blanco-Cuaresma}, {Boch},
  {Boeche}, {Bombrun}, {Borrachero}, {Bossini}, {Bouquillon}, {Bourda},
  {Bragaglia}, {Bramante}, {Bressan}, {Brouillet}, {Br{\"u}semeister},
  {Brugaletta}, {Bucciarelli}, {Burlacu}, {Busonero}, {Butkevich}, {Buzzi},
  {Caffau}, {Cancelliere}, {Cannizzaro}, {Cantat-Gaudin}, {Carballo},
  {Carlucci}, {Carrasco}, {Casamiquela}, {Castellani}, {Castro-Ginard},
  {Charlot}, {Chemin}, {Chiavassa}, {Cocozza}, {Costigan}, {Cowell}, {Crifo},
  {Crosta}, {Crowley}, {Cuypers}, {Dafonte}, {Damerdji}, {Dapergolas}, {David},
  {David}, {de Laverny}, {De Luise}, {De March}, {de Martino}, {de Souza}, {de
  Torres}, {Debosscher}, {del Pozo}, {Delbo}, {Delgado}, {Delgado}, {Di
  Matteo}, {Diakite}, {Diener}, {Distefano}, {Dolding}, {Drazinos},
  {Dur{\'a}n}, {Edvardsson}, {Enke}, {Eriksson}, {Esquej}, {Eynard Bontemps},
  {Fabre}, {Fabrizio}, {Faigler}, {Falc{\~a}o}, {Farr{\`a}s Casas}, {Federici},
  {Fedorets}, {Fernique}, {Figueras}, {Filippi}, {Findeisen}, {Fonti},
  {Fraile}, {Fraser}, {Fr{\'e}zouls}, {Gai}, {Galleti}, {Garabato},
  {Garc{\'\i}a-Sedano}, {Garofalo}, {Garralda}, {Gavel}, {Gavras}, {Gerssen},
  {Geyer}, {Giacobbe}, {Gilmore}, {Girona}, {Giuffrida}, {Glass}, {Gomes},
  {Granvik}, {Gueguen}, {Guerrier}, {Guiraud}, {Guti{\'e}rrez-S{\'a}nchez},
  {Hofmann}, {Holland}, {Huckle}, {Hypki}, {Icardi}, {Jan{\ss}en}, {Jevardat de
  Fombelle}, {Jonker}, {Juh{\'a}sz}, {Julbe}, {Karampelas}, {Kewley}, {Klar},
  {Kochoska}, {Kohley}, {Kolenberg}, {Kontizas}, {Kontizas}, {Koposov},
  {Kordopatis}, {Kostrzewa-Rutkowska}, {Koubsky}, {Lambert}, {Lanza}, {Lasne},
  {Lavigne}, {Le Fustec}, {Le Poncin-Lafitte}, {Lebreton}, {Leccia}, {Leclerc},
  {Lecoeur-Taibi}, {Lenhardt}, {Leroux}, {Liao}, {Licata}, {Lindstr{\o}m},
  {Lister}, {Livanou}, {Lobel}, {L{\'o}pez}, {Managau}, {Mann}, {Mantelet},
  {Marchal}, {Marchant}, {Marconi}, {Marinoni}, {Marschalk{\'o}}, {Marshall},
  {Martino}, {Marton}, {Mary}, {Matijevi{\v{c}}}, {Mazeh}, {Messina},
  {Michalik}, {Millar}, {Molina}, {Molinaro}, {Moln{\'a}r}, {Montegriffo},
  {Mor}, {Morbidelli}, {Morel}, {Morris}, {Mulone}, {Muraveva}, {Musella},
  {Nelemans}, {Nicastro}, {Noval}, {O'Mullane}, {Ord{\'e}novic},
  {Ord{\'o}{\~n}ez-Blanco}, {Osborne}, {Pagani}, {Pagano}, {Pailler},
  {Palacin}, {Palaversa}, {Panahi}, {Pawlak}, {Piersimoni}, {Pineau}, {Plachy},
  {Plum}, {Poggio}, {Poujoulet}, {Pr{\v{s}}a}, {Pulone}, {Racero}, {Ragaini},
  {Rambaux}, {Ramos-Lerate}, {Regibo}, {Riclet}, {Ripepi}, {Riva}, {Rivard},
  {Rixon}, {Roegiers}, {Roelens}, {Romero-G{\'o}mez}, {Rowell}, {Royer},
  {Ruiz-Dern}, {Sadowski}, {Sagrist{\`a} Sell{\'e}s}, {Sahlmann}, {Salgado},
  {Salguero}, {Sanna}, {Santana-Ros}, {Sarasso}, {Savietto}, {Schultheis},
  {Sciacca}, {Segol}, {Segovia}, {S{\'e}gransan}, {Shih}, {Siltala}, {Silva},
  {Smart}, {Smith}, {Solano}, {Solitro}, {Sordo}, {Soria Nieto}, {Souchay},
  {Spagna}, {Spoto}, {Stampa}, {Steele}, {Steidelm{\"u}ller}, {Stephenson},
  {Stoev}, {Suess}, {Surdej}, {Szabados}, {Szegedi-Elek}, {Tapiador}, {Taris},
  {Tauran}, {Taylor}, {Teixeira}, {Terrett}, {Teyssand ier}, {Thuillot},
  {Titarenko}, {Torra Clotet}, {Turon}, {Ulla}, {Utrilla}, {Uzzi}, {Vaillant},
  {Valentini}, {Valette}, {van Elteren}, {Van Hemelryck}, {van Leeuwen},
  {Vaschetto}, {Vecchiato}, {Viala}, {Vicente}, {Vogt}, {von Essen}, {Voss},
  {Votruba}, {Voutsinas}, {Walmsley}, {Weiler}, {Wertz}, {Wevems},
  {Wyrzykowski}, {Yoldas}, {{\v{Z}}erjal}, {Ziaeepour}, {Zorec}, {Zschocke},
  {Zucker}, {Zurbach}, \& {Zwitter}}]{helmi2018b}
{Gaia Collaboration}, {Helmi}, A., {van Leeuwen}, F., {McMillan}, P.~J.,
  {Massari}, D., {Antoja}, T., {Robin}, A.~C., {Lindegren}, L., {Bastian}, U.,
  {Arenou}, F., {Babusiaux}, C., {Biermann}, M., {Breddels}, M.~A., {Hobbs},
  D., {Jordi}, C., {Pancino}, E., {Reyl{\'e}}, C., {Veljanoski}, J., {Brown},
  A.~G.~A., {Vallenari}, A., {Prusti}, T., {de Bruijne}, J.~H.~J.,
  {Bailer-Jones}, C.~A.~L., {Evans}, D.~W., {Eyer}, L., {Jansen}, F.,
  {Klioner}, S.~A., {Lammers}, U., {Luri}, X., {Mignard}, F., {Panem}, C.,
  {Pourbaix}, D., {Randich}, S., {Sartoretti}, P., {Siddiqui}, H.~I.,
  {Soubiran}, C., {Walton}, N.~A., {Cropper}, M., {Drimmel}, R., {Katz}, D.,
  {Lattanzi}, M.~G., {Bakker}, J., {Cacciari}, C., {Casta{\~n}eda}, J.,
  {Chaoul}, L., {Cheek}, N., {De Angeli}, F., {Fabricius}, C., {Guerra}, R.,
  {Holl}, B., {Masana}, E., {Messineo}, R., {Mowlavi}, N., {Nienartowicz}, K.,
  {Panuzzo}, P., {Portell}, J., {Riello}, M., {Seabroke}, G.~M., {Tanga}, P.,
  {Th{\'e}venin}, F., {Gracia-Abril}, G., {Comoretto}, G., {Garcia-Reinaldos},
  M., {Teyssier}, D., {Altmann}, M., {Andrae}, R., {Audard}, M.,
  {Bellas-Velidis}, I., {Benson}, K., {Berthier}, J., {Blomme}, R., {Burgess},
  P., {Busso}, G., {Carry}, B., {Cellino}, A., {Clementini}, G., {Clotet}, M.,
  {Creevey}, O., {Davidson}, M., {De Ridder}, J., {Delchambre}, L., {Dell'Oro},
  A., {Ducourant}, C., {Fern{\'a}ndez-Hern{\'a}ndez}, J., {Fouesneau}, M.,
  {Fr{\'e}mat}, Y., {Galluccio}, L., {Garc{\'\i}a-Torres}, M.,
  {Gonz{\'a}lez-N{\'u}{\~n}ez}, J., {Gonz{\'a}lez-Vidal}, J.~J., {Gosset}, E.,
  {Guy}, L.~P., {Halbwachs}, J.~L., {Hambly}, N.~C., {Harrison}, D.~L.,
  {Hern{\'a}ndez}, J., {Hestroffer}, D., {Hodgkin}, S.~T., {Hutton}, A.,
  {Jasniewicz}, G., {Jean-Antoine-Piccolo}, A., {Jordan}, S., {Korn}, A.~J.,
  {Krone-Martins}, A., {Lanzafame}, A.~C., {Lebzelter}, T., {L{\"o}ffler}, W.,
  {Manteiga}, M., {Marrese}, P.~M., {Mart{\'\i}n-Fleitas}, J.~M., {Moitinho},
  A., {Mora}, A., {Muinonen}, K., {Osinde}, J., {Pauwels}, T., {Petit}, J.~M.,
  {Recio-Blanco}, A., {Richards}, P.~J., {Rimoldini}, L., {Sarro}, L.~M.,
  {Siopis}, C., {Smith}, M., {Sozzetti}, A., {S{\"u}veges}, M., {Torra}, J.,
  {van Reeven}, W., {Abbas}, U., {Abreu Aramburu}, A., {Accart}, S., {Aerts},
  C., {Altavilla}, G., {{\'A}lvarez}, M.~A., {Alvarez}, R., {Alves}, J.,
  {Anderson}, R.~I., {Andrei}, A.~H., {Anglada Varela}, E., {Antiche}, E.,
  {Arcay}, B., {Astraatmadja}, T.~L., {Bach}, N., {Baker}, S.~G.,
  {Balaguer-N{\'u}{\~n}ez}, L., {Balm}, P., {Barache}, C., {Barata}, C.,
  {Barbato}, D., {Barblan}, F., {Barklem}, P.~S., {Barrado}, D., {Barros}, M.,
  {Barstow}, M.~A., {Bartholom{\'e} Mu{\~n}oz}, S., {Bassilana}, J.~L.,
  {Becciani}, U., {Bellazzini}, M., {Berihuete}, A., {Bertone}, S., {Bianchi},
  L., {Bienaym{\'e}}, O., {Blanco-Cuaresma}, S., {Boch}, T., {Boeche}, C.,
  {Bombrun}, A., {Borrachero}, R., {Bossini}, D., {Bouquillon}, S., {Bourda},
  G., {Bragaglia}, A., {Bramante}, L., {Bressan}, A., {Brouillet}, N.,
  {Br{\"u}semeister}, T., {Brugaletta}, E., {Bucciarelli}, B., {Burlacu}, A.,
  {Busonero}, D., {Butkevich}, A.~G., {Buzzi}, R., {Caffau}, E., {Cancelliere},
  R., {Cannizzaro}, G., {Cantat-Gaudin}, T., {Carballo}, R., {Carlucci}, T.,
  {Carrasco}, J.~M., {Casamiquela}, L., {Castellani}, M., {Castro-Ginard}, A.,
  {Charlot}, P., {Chemin}, L., {Chiavassa}, A., {Cocozza}, G., {Costigan}, G.,
  {Cowell}, S., {Crifo}, F., {Crosta}, M., {Crowley}, C., {Cuypers}, J.,
  {Dafonte}, C., {Damerdji}, Y., {Dapergolas}, A., {David}, P., {David}, M.,
  {de Laverny}, P., {De Luise}, F., {De March}, R., {de Martino}, D., {de
  Souza}, R., {de Torres}, A., {Debosscher}, J., {del Pozo}, E., {Delbo}, M.,
  {Delgado}, A., {Delgado}, H.~E., {Di Matteo}, P., {Diakite}, S., {Diener},
  C., {Distefano}, E., {Dolding}, C., {Drazinos}, P., {Dur{\'a}n}, J.,
  {Edvardsson}, B., {Enke}, H., {Eriksson}, K., {Esquej}, P., {Eynard
  Bontemps}, G., {Fabre}, C., {Fabrizio}, M., {Faigler}, S., {Falc{\~a}o},
  A.~J., {Farr{\`a}s Casas}, M., {Federici}, L., {Fedorets}, G., {Fernique},
  P., {Figueras}, F., {Filippi}, F., {Findeisen}, K., {Fonti}, A., {Fraile},
  E., {Fraser}, M., {Fr{\'e}zouls}, B., {Gai}, M., {Galleti}, S., {Garabato},
  D., {Garc{\'\i}a-Sedano}, F., {Garofalo}, A., {Garralda}, N., {Gavel}, A.,
  {Gavras}, P., {Gerssen}, J., {Geyer}, R., {Giacobbe}, P., {Gilmore}, G.,
  {Girona}, S., {Giuffrida}, G., {Glass}, F., {Gomes}, M., {Granvik}, M.,
  {Gueguen}, A., {Guerrier}, A., {Guiraud}, J., {Guti{\'e}rrez-S{\'a}nchez},
  R., {Hofmann}, W., {Holland}, G., {Huckle}, H.~E., {Hypki}, A., {Icardi}, V.,
  {Jan{\ss}en}, K., {Jevardat de Fombelle}, G., {Jonker}, P.~G., {Juh{\'a}sz},
  {\'A}.~L., {Julbe}, F., {Karampelas}, A., {Kewley}, A., {Klar}, J.,
  {Kochoska}, A., {Kohley}, R., {Kolenberg}, K., {Kontizas}, M., {Kontizas},
  E., {Koposov}, S.~E., {Kordopatis}, G., {Kostrzewa-Rutkowska}, Z., {Koubsky},
  P., {Lambert}, S., {Lanza}, A.~F., {Lasne}, Y., {Lavigne}, J.~B., {Le
  Fustec}, Y., {Le Poncin-Lafitte}, C., {Lebreton}, Y., {Leccia}, S.,
  {Leclerc}, N., {Lecoeur-Taibi}, I., {Lenhardt}, H., {Leroux}, F., {Liao}, S.,
  {Licata}, E., {Lindstr{\o}m}, H.~E.~P., {Lister}, T.~A., {Livanou}, E.,
  {Lobel}, A., {L{\'o}pez}, M., {Managau}, S., {Mann}, R.~G., {Mantelet}, G.,
  {Marchal}, O., {Marchant}, J.~M., {Marconi}, M., {Marinoni}, S.,
  {Marschalk{\'o}}, G., {Marshall}, D.~J., {Martino}, M., {Marton}, G., {Mary},
  N., {Matijevi{\v{c}}}, G., {Mazeh}, T., {Messina}, S., {Michalik}, D.,
  {Millar}, N.~R., {Molina}, D., {Molinaro}, R., {Moln{\'a}r}, L.,
  {Montegriffo}, P., {Mor}, R., {Morbidelli}, R., {Morel}, T., {Morris}, D.,
  {Mulone}, A.~F., {Muraveva}, T., {Musella}, I., {Nelemans}, G., {Nicastro},
  L., {Noval}, L., {O'Mullane}, W., {Ord{\'e}novic}, C.,
  {Ord{\'o}{\~n}ez-Blanco}, D., {Osborne}, P., {Pagani}, C., {Pagano}, I.,
  {Pailler}, F., {Palacin}, H., {Palaversa}, L., {Panahi}, A., {Pawlak}, M.,
  {Piersimoni}, A.~M., {Pineau}, F.~X., {Plachy}, E., {Plum}, G., {Poggio}, E.,
  {Poujoulet}, E., {Pr{\v{s}}a}, A., {Pulone}, L., {Racero}, E., {Ragaini}, S.,
  {Rambaux}, N., {Ramos-Lerate}, M., {Regibo}, S., {Riclet}, F., {Ripepi}, V.,
  {Riva}, A., {Rivard}, A., {Rixon}, G., {Roegiers}, T., {Roelens}, M.,
  {Romero-G{\'o}mez}, M., {Rowell}, N., {Royer}, F., {Ruiz-Dern}, L.,
  {Sadowski}, G., {Sagrist{\`a} Sell{\'e}s}, T., {Sahlmann}, J., {Salgado}, J.,
  {Salguero}, E., {Sanna}, N., {Santana-Ros}, T., {Sarasso}, M., {Savietto},
  H., {Schultheis}, M., {Sciacca}, E., {Segol}, M., {Segovia}, J.~C.,
  {S{\'e}gransan}, D., {Shih}, I.~C., {Siltala}, L., {Silva}, A.~F., {Smart},
  R.~L., {Smith}, K.~W., {Solano}, E., {Solitro}, F., {Sordo}, R., {Soria
  Nieto}, S., {Souchay}, J., {Spagna}, A., {Spoto}, F., {Stampa}, U., {Steele},
  I.~A., {Steidelm{\"u}ller}, H., {Stephenson}, C.~A., {Stoev}, H., {Suess},
  F.~F., {Surdej}, J., {Szabados}, L., {Szegedi-Elek}, E., {Tapiador}, D.,
  {Taris}, F., {Tauran}, G., {Taylor}, M.~B., {Teixeira}, R., {Terrett}, D.,
  {Teyssand ier}, P., {Thuillot}, W., {Titarenko}, A., {Torra Clotet}, F.,
  {Turon}, C., {Ulla}, A., {Utrilla}, E., {Uzzi}, S., {Vaillant}, M.,
  {Valentini}, G., {Valette}, V., {van Elteren}, A., {Van Hemelryck}, E., {van
  Leeuwen}, M., {Vaschetto}, M., {Vecchiato}, A., {Viala}, Y., {Vicente}, D.,
  {Vogt}, S., {von Essen}, C., {Voss}, H., {Votruba}, V., {Voutsinas}, S.,
  {Walmsley}, G., {Weiler}, M., {Wertz}, O., {Wevems}, T., {Wyrzykowski},
  {\L}., {Yoldas}, A., {{\v{Z}}erjal}, M., {Ziaeepour}, H., {Zorec}, J.,
  {Zschocke}, S., {Zucker}, S., {Zurbach}, C., \& {Zwitter}, T.
  2018{\natexlab{c}}, \aap, 616, A12

\bibitem[{{Gaia Collaboration} {et~al.}(2016){Gaia Collaboration}, {Prusti},
  {de Bruijne}, {Brown}, {Vallenari}, {Babusiaux}, {Bailer-Jones}, {Bastian},
  {Biermann}, {Evans}, \& et~al.}]{gaia2016}
{Gaia Collaboration}, {Prusti}, T., {de Bruijne}, J.~H.~J., {Brown}, A.~G.~A.,
  {Vallenari}, A., {Babusiaux}, C., {Bailer-Jones}, C.~A.~L., {Bastian}, U.,
  {Biermann}, M., {Evans}, D.~W., \& et~al. 2016, \aap, 595, A1

\bibitem[{{Geha} {et~al.}(2009){Geha}, {Willman}, {Simon}, {Strigari}, {Kirby},
  {Law}, \& {Strader}}]{geha2009}
{Geha}, M., {Willman}, B., {Simon}, J.~D., {Strigari}, L.~E., {Kirby}, E.~N.,
  {Law}, D.~R., \& {Strader}, J. 2009, \apj, 692, 1464

\bibitem[{{Girardi} {et~al.}(2002){Girardi}, {Bertelli}, {Bressan}, {Chiosi},
  {Groenewegen}, {Marigo}, {Salasnich}, \& {Weiss}}]{girardi2002}
{Girardi}, L., {Bertelli}, G., {Bressan}, A., {Chiosi}, C., {Groenewegen},
  M.~A.~T., {Marigo}, P., {Salasnich}, B., \& {Weiss}, A. 2002, \aap, 391, 195

\bibitem[{{Gravity Collaboration} {et~al.}(2018){Gravity Collaboration},
  {Abuter}, {Amorim}, {Anugu}, {Baub{\"o}ck}, {Benisty}, {Berger}, {Blind},
  {Bonnet}, {Brandner}, {Buron}, {Collin}, {Chapron}, {Cl{\'e}net}, {Coud{\'e}
  Du Foresto}, {de Zeeuw}, {Deen}, {Delplancke-Str{\"o}bele}, {Dembet},
  {Dexter}, {Duvert}, {Eckart}, {Eisenhauer}, {Finger}, {F{\"o}rster
  Schreiber}, {F{\'e}dou}, {Garcia}, {Garcia Lopez}, {Gao}, {Gendron},
  {Genzel}, {Gillessen}, {Gordo}, {Habibi}, {Haubois}, {Haug}, {Hau{\ss}mann},
  {Henning}, {Hippler}, {Horrobin}, {Hubert}, {Hubin}, {Jimenez Rosales},
  {Jochum}, {Jocou}, {Kaufer}, {Kellner}, {Kendrew}, {Kervella}, {Kok},
  {Kulas}, {Lacour}, {Lapeyr{\`e}re}, {Lazareff}, {Le Bouquin}, {L{\'e}na},
  {Lippa}, {Lenzen}, {M{\'e}rand}, {M{\"u}ler}, {Neumann}, {Ott}, {Palanca},
  {Paumard}, {Pasquini}, {Perraut}, {Perrin}, {Pfuhl}, {Plewa}, {Rabien},
  {Ram{\'\i}rez}, {Ramos}, {Rau}, {Rodr{\'\i}guez-Coira}, {Rohloff}, {Rousset},
  {Sanchez-Bermudez}, {Scheithauer}, {Sch{\"o}ller}, {Schuler}, {Spyromilio},
  {Straub}, {Straubmeier}, {Sturm}, {Tacconi}, {Tristram}, {Vincent}, {von
  Fellenberg}, {Wank}, {Waisberg}, {Widmann}, {Wieprecht}, {Wiest},
  {Wiezorrek}, {Woillez}, {Yazici}, {Ziegler}, \& {Zins}}]{gravity2018}
{Gravity Collaboration}, {Abuter}, R., {Amorim}, A., {Anugu}, N.,
  {Baub{\"o}ck}, M., {Benisty}, M., {Berger}, J.~P., {Blind}, N., {Bonnet}, H.,
  {Brandner}, W., {Buron}, A., {Collin}, C., {Chapron}, F., {Cl{\'e}net}, Y.,
  {Coud{\'e} Du Foresto}, V., {de Zeeuw}, P.~T., {Deen}, C.,
  {Delplancke-Str{\"o}bele}, F., {Dembet}, R., {Dexter}, J., {Duvert}, G.,
  {Eckart}, A., {Eisenhauer}, F., {Finger}, G., {F{\"o}rster Schreiber}, N.~M.,
  {F{\'e}dou}, P., {Garcia}, P., {Garcia Lopez}, R., {Gao}, F., {Gendron}, E.,
  {Genzel}, R., {Gillessen}, S., {Gordo}, P., {Habibi}, M., {Haubois}, X.,
  {Haug}, M., {Hau{\ss}mann}, F., {Henning}, T., {Hippler}, S., {Horrobin}, M.,
  {Hubert}, Z., {Hubin}, N., {Jimenez Rosales}, A., {Jochum}, L., {Jocou}, K.,
  {Kaufer}, A., {Kellner}, S., {Kendrew}, S., {Kervella}, P., {Kok}, Y.,
  {Kulas}, M., {Lacour}, S., {Lapeyr{\`e}re}, V., {Lazareff}, B., {Le Bouquin},
  J.~B., {L{\'e}na}, P., {Lippa}, M., {Lenzen}, R., {M{\'e}rand}, A.,
  {M{\"u}ler}, E., {Neumann}, U., {Ott}, T., {Palanca}, L., {Paumard}, T.,
  {Pasquini}, L., {Perraut}, K., {Perrin}, G., {Pfuhl}, O., {Plewa}, P.~M.,
  {Rabien}, S., {Ram{\'\i}rez}, A., {Ramos}, J., {Rau}, C.,
  {Rodr{\'\i}guez-Coira}, G., {Rohloff}, R.~R., {Rousset}, G.,
  {Sanchez-Bermudez}, J., {Scheithauer}, S., {Sch{\"o}ller}, M., {Schuler}, N.,
  {Spyromilio}, J., {Straub}, O., {Straubmeier}, C., {Sturm}, E., {Tacconi},
  L.~J., {Tristram}, K.~R.~W., {Vincent}, F., {von Fellenberg}, S., {Wank}, I.,
  {Waisberg}, I., {Widmann}, F., {Wieprecht}, E., {Wiest}, M., {Wiezorrek}, E.,
  {Woillez}, J., {Yazici}, S., {Ziegler}, D., \& {Zins}, G. 2018, \aap, 615,
  L15

\bibitem[{{Gregory} {et~al.}(2020){Gregory}, {Collins}, {Erkal}, {Tollerud},
  {Delorme}, {Hill}, {Sand}, {Strader}, \& {Willman}}]{gregory2020}
{Gregory}, A.~L., {Collins}, M. L.~M., {Erkal}, D., {Tollerud}, E., {Delorme},
  M., {Hill}, L., {Sand}, D.~J., {Strader}, J., \& {Willman}, B. 2020, \mnras,
  496, 1092

\bibitem[{{Grillmair}(2009)}]{grillmair2009}
{Grillmair}, C.~J. 2009, \apj, 693, 1118

\bibitem[{{Grillmair}(2011)}]{grillmair2011}
---. 2011, \apj, 738, 98

\bibitem[{{Hansen} {et~al.}(2017){Hansen}, {Simon}, {Marshall}, {Li},
  {Carollo}, {DePoy}, {Nagasawa}, {Bernstein}, {Drlica-Wagner}, {Abdalla},
  {Allam}, {Annis}, {Bechtol}, {Benoit-L{\'e}vy}, {Brooks}, {Buckley-Geer},
  {Carnero Rosell}, {Carrasco Kind}, {Carretero}, {Cunha}, {da Costa}, {Desai},
  {Eifler}, {Fausti Neto}, {Flaugher}, {Frieman}, {Garc{\'\i}a-Bellido},
  {Gaztanaga}, {Gerdes}, {Gruen}, {Gruendl}, {Gschwend}, {Gutierrez}, {James},
  {Krause}, {Kuehn}, {Kuropatkin}, {Lahav}, {Miquel}, {Plazas}, {Romer},
  {Sanchez}, {Santiago}, {Scarpine}, {Smith}, {Soares-Santos}, {Sobreira},
  {Suchyta}, {Swanson}, {Tarle}, {Walker}, \& {DES Collaboration}}]{hansen2017}
{Hansen}, T.~T., {Simon}, J.~D., {Marshall}, J.~L., {Li}, T.~S., {Carollo}, D.,
  {DePoy}, D.~L., {Nagasawa}, D.~Q., {Bernstein}, R.~A., {Drlica-Wagner}, A.,
  {Abdalla}, F.~B., {Allam}, S., {Annis}, J., {Bechtol}, K., {Benoit-L{\'e}vy},
  A., {Brooks}, D., {Buckley-Geer}, E., {Carnero Rosell}, A., {Carrasco Kind},
  M., {Carretero}, J., {Cunha}, C.~E., {da Costa}, L.~N., {Desai}, S.,
  {Eifler}, T.~F., {Fausti Neto}, A., {Flaugher}, B., {Frieman}, J.,
  {Garc{\'\i}a-Bellido}, J., {Gaztanaga}, E., {Gerdes}, D.~W., {Gruen}, D.,
  {Gruendl}, R.~A., {Gschwend}, J., {Gutierrez}, G., {James}, D.~J., {Krause},
  E., {Kuehn}, K., {Kuropatkin}, N., {Lahav}, O., {Miquel}, R., {Plazas},
  A.~A., {Romer}, A.~K., {Sanchez}, E., {Santiago}, B., {Scarpine}, V.,
  {Smith}, R.~C., {Soares-Santos}, M., {Sobreira}, F., {Suchyta}, E.,
  {Swanson}, M.~E.~C., {Tarle}, G., {Walker}, A.~R., \& {DES Collaboration}.
  2017, \apj, 838, 44

\bibitem[{{Hargis} {et~al.}(2016){Hargis}, {Kimmig}, {Willman}, {Caldwell},
  {Walker}, {Strader}, {Sand }, {Grillmair}, \& {Yoon}}]{hargis2016}
{Hargis}, J.~R., {Kimmig}, B., {Willman}, B., {Caldwell}, N., {Walker}, M.~G.,
  {Strader}, J., {Sand }, D.~J., {Grillmair}, C.~J., \& {Yoon}, J.~H. 2016,
  \apj, 818, 39

\bibitem[{{Hargis} {et~al.}(2014){Hargis}, {Willman}, \& {Peter}}]{hargis2014}
{Hargis}, J.~R., {Willman}, B., \& {Peter}, A. H.~G. 2014, \apjl, 795, L13

\bibitem[{{Ibata} {et~al.}(2020){Ibata}, {Bellazzini}, {Thomas}, {Malhan},
  {Martin}, {Famaey}, \& {Siebert}}]{ibata2020}
{Ibata}, R., {Bellazzini}, M., {Thomas}, G., {Malhan}, K., {Martin}, N.,
  {Famaey}, B., \& {Siebert}, A. 2020, \apjl, 891, L19

\bibitem[{{Ji} {et~al.}(2016){Ji}, {Frebel}, {Simon}, \& {Chiti}}]{ji2016}
{Ji}, A.~P., {Frebel}, A., {Simon}, J.~D., \& {Chiti}, A. 2016, \apj, 830, 93

\bibitem[{{Ji} {et~al.}(2020){Ji}, {Li}, {Simon}, {Marshall}, {Vivas}, {Pace},
  {Bechtol}, {Drlica-Wagner}, {Koposov}, {Hansen}, {Allam}, {Gruendl},
  {Johnson}, {McNanna}, {No{\"e}l}, {Tucker}, \& {Walker}}]{ji2020}
{Ji}, A.~P., {Li}, T.~S., {Simon}, J.~D., {Marshall}, J., {Vivas}, A.~K.,
  {Pace}, A.~B., {Bechtol}, K., {Drlica-Wagner}, A., {Koposov}, S.~E.,
  {Hansen}, T.~T., {Allam}, S., {Gruendl}, R.~A., {Johnson}, M.~D., {McNanna},
  M., {No{\"e}l}, N.~E.~D., {Tucker}, D.~L., \& {Walker}, A.~R. 2020, \apj,
  889, 27

\bibitem[{{Ji} {et~al.}(2019){Ji}, {Simon}, {Frebel}, {Venn}, \&
  {Hansen}}]{ji2019}
{Ji}, A.~P., {Simon}, J.~D., {Frebel}, A., {Venn}, K.~A., \& {Hansen}, T.~T.
  2019, \apj, 870, 83

\bibitem[{{Kacharov} {et~al.}(2017){Kacharov}, {Battaglia}, {Rejkuba}, {Cole},
  {Carrera}, {Fraternali}, {Wilkinson}, {Gallart}, {Irwin}, \&
  {Tolstoy}}]{kacharov2017}
{Kacharov}, N., {Battaglia}, G., {Rejkuba}, M., {Cole}, A.~A., {Carrera}, R.,
  {Fraternali}, F., {Wilkinson}, M.~I., {Gallart}, C.~G., {Irwin}, M., \&
  {Tolstoy}, E. 2017, \mnras, 466, 2006

\bibitem[{{Kallivayalil} {et~al.}(2018){Kallivayalil}, {Sales}, {Zivick},
  {Fritz}, {Del Pino}, {Sohn}, {Besla}, {van der Marel}, {Navarro}, \&
  {Sacchi}}]{kallivayalil2018}
{Kallivayalil}, N., {Sales}, L.~V., {Zivick}, P., {Fritz}, T.~K., {Del Pino},
  A., {Sohn}, S.~T., {Besla}, G., {van der Marel}, R.~P., {Navarro}, J.~F., \&
  {Sacchi}, E. 2018, \apj, 867, 19

\bibitem[{{Kallivayalil} {et~al.}(2013){Kallivayalil}, {van der Marel},
  {Besla}, {Anderson}, \& {Alcock}}]{kallivayalil2013}
{Kallivayalil}, N., {van der Marel}, R.~P., {Besla}, G., {Anderson}, J., \&
  {Alcock}, C. 2013, \apj, 764, 161

\bibitem[{{Kirby} {et~al.}(2013){Kirby}, {Boylan-Kolchin}, {Cohen}, {Geha},
  {Bullock}, \& {Kaplinghat}}]{kirby2013}
{Kirby}, E.~N., {Boylan-Kolchin}, M., {Cohen}, J.~G., {Geha}, M., {Bullock},
  J.~S., \& {Kaplinghat}, M. 2013, \apj, 770, 16

\bibitem[Kim et al.(2016)]{kim2016} Kim, D., Jerjen, H., Geha, M., et al.\ 2016, \apj, 833, 16
  
\bibitem[{{Kirby} {et~al.}(2015){Kirby}, {Cohen}, {Simon}, \&
  {Guhathakurta}}]{kirby2015}
{Kirby}, E.~N., {Cohen}, J.~G., {Simon}, J.~D., \& {Guhathakurta}, P. 2015,
  \apjl, 814, L7

\bibitem[{{Kirby} {et~al.}(2017){Kirby}, {Cohen}, {Simon}, {Guhathakurta},
  {Thygesen}, \& {Duggan}}]{kirby2017}
{Kirby}, E.~N., {Cohen}, J.~G., {Simon}, J.~D., {Guhathakurta}, P., {Thygesen},
  A.~O., \& {Duggan}, G.~E. 2017, \apj, 838, 83

\bibitem[{{Koch} {et~al.}(2013){Koch}, {Feltzing}, {Ad{\'e}n}, \&
  {Matteucci}}]{koch2013}
{Koch}, A., {Feltzing}, S., {Ad{\'e}n}, D., \& {Matteucci}, F. 2013, \aap, 554,
  A5

\bibitem[{{Koch} {et~al.}(2014){Koch}, {Hansen}, {Feltzing}, \&
  {Wilkinson}}]{koch2014}
{Koch}, A., {Hansen}, T., {Feltzing}, S., \& {Wilkinson}, M.~I. 2014, \apj,
  780, 91

\bibitem[{{Koch} {et~al.}(2008){Koch}, {McWilliam}, {Grebel}, {Zucker}, \&
  {Belokurov}}]{koch2008b}
{Koch}, A., {McWilliam}, A., {Grebel}, E.~K., {Zucker}, D.~B., \& {Belokurov},
  V. 2008, \apjl, 688, L13

\bibitem[{{Koch} {et~al.}(2009){Koch}, {Wilkinson}, {Kleyna}, {Irwin},
  {Zucker}, {Belokurov}, {Gilmore}, {Fellhauer}, \& {Evans}}]{koch2009}
{Koch}, A., {Wilkinson}, M.~I., {Kleyna}, J.~T., {Irwin}, M., {Zucker}, D.~B.,
  {Belokurov}, V., {Gilmore}, G.~F., {Fellhauer}, M., \& {Evans}, N.~W. 2009,
  \apj, 690, 453

\bibitem[{{Koposov} {et~al.}(2015){Koposov}, {Belokurov}, {Torrealba}, \&
  {Evans}}]{koposov2015}
{Koposov}, S.~E., {Belokurov}, V., {Torrealba}, G., \& {Evans}, N.~W. 2015,
  \apj, 805, 130

\bibitem[{{Koposov} {et~al.}(2011){Koposov}, {Gilmore}, {Walker}, {Belokurov},
  {Wyn Evans}, {Fellhauer}, {Gieren}, {Geisler}, {Monaco}, {Norris}, {Okamoto},
  {Pe{\~n}arrubia}, {Wilkinson}, {Wyse}, \& {Zucker}}]{koposov2011}
{Koposov}, S.~E., {Gilmore}, G., {Walker}, M.~G., {Belokurov}, V., {Wyn Evans},
  N., {Fellhauer}, M., {Gieren}, W., {Geisler}, D., {Monaco}, L., {Norris},
  J.~E., {Okamoto}, S., {Pe{\~n}arrubia}, J., {Wilkinson}, M., {Wyse},
  R.~F.~G., \& {Zucker}, D.~B. 2011, \apj, 736, 146

\bibitem[{{Koposov} {et~al.}(2018){Koposov}, {Walker}, {Belokurov}, {Casey},
  {Geringer-Sameth}, {Mackey}, {Da Costa}, {Erkal}, {Jethwa}, {Mateo},
  {Olszewski}, \& {Bailey}}]{koposov2018}
{Koposov}, S.~E., {Walker}, M.~G., {Belokurov}, V., {Casey}, A.~R.,
  {Geringer-Sameth}, A., {Mackey}, D., {Da Costa}, G., {Erkal}, D., {Jethwa},
  P., {Mateo}, M., {Olszewski}, E.~W., \& {Bailey}, III, J.~I. 2018, ArXiv
  e-prints

\bibitem[{{Li} {et~al.}(2017){Li}, {Simon}, {Drlica-Wagner}, {Bechtol}, {Wang},
  {Garc{\'\i}a-Bellido}, {Frieman}, {Marshall}, {James}, {Strigari}, {Pace},
  {Balbinot}, {Zhang}, {Abbott}, {Allam}, {Benoit-L{\'e}vy}, {Bernstein},
  {Bertin}, {Brooks}, {Burke}, {Carnero Rosell}, {Carrasco Kind}, {Carretero},
  {Cunha}, {D'Andrea}, {da Costa}, {DePoy}, {Desai}, {Diehl}, {Eifler},
  {Flaugher}, {Goldstein}, {Gruen}, {Gruendl}, {Gschwend}, {Gutierrez},
  {Krause}, {Kuehn}, {Lin}, {Maia}, {March}, {Menanteau}, {Miquel}, {Plazas},
  {Romer}, {Sanchez}, {Santiago}, {Schubnell}, {Sevilla-Noarbe}, {Smith},
  {Sobreira}, {Suchyta}, {Tarle}, {Thomas}, {Tucker}, {Walker}, {Wechsler},
  {Wester}, {Yanny}, \& {DES Collaboration}}]{li2017}
{Li}, T.~S., {Simon}, J.~D., {Drlica-Wagner}, A., {Bechtol}, K., {Wang}, M.~Y.,
  {Garc{\'\i}a-Bellido}, J., {Frieman}, J., {Marshall}, J.~L., {James}, D.~J.,
  {Strigari}, L., {Pace}, A.~B., {Balbinot}, E., {Zhang}, Y., {Abbott},
  T.~M.~C., {Allam}, S., {Benoit-L{\'e}vy}, A., {Bernstein}, G.~M., {Bertin},
  E., {Brooks}, D., {Burke}, D.~L., {Carnero Rosell}, A., {Carrasco Kind}, M.,
  {Carretero}, J., {Cunha}, C.~E., {D'Andrea}, C.~B., {da Costa}, L.~N.,
  {DePoy}, D.~L., {Desai}, S., {Diehl}, H.~T., {Eifler}, T.~F., {Flaugher}, B.,
  {Goldstein}, D.~A., {Gruen}, D., {Gruendl}, R.~A., {Gschwend}, J.,
  {Gutierrez}, G., {Krause}, E., {Kuehn}, K., {Lin}, H., {Maia}, M.~A.~G.,
  {March}, M., {Menanteau}, F., {Miquel}, R., {Plazas}, A.~A., {Romer}, A.~K.,
  {Sanchez}, E., {Santiago}, B., {Schubnell}, M., {Sevilla-Noarbe}, I.,
  {Smith}, R.~C., {Sobreira}, F., {Suchyta}, E., {Tarle}, G., {Thomas}, D.,
  {Tucker}, D.~L., {Walker}, A.~R., {Wechsler}, R.~H., {Wester}, W., {Yanny},
  B., \& {DES Collaboration}. 2017, \apj, 838, 8

\bibitem[{{Li} {et~al.}(2018{\natexlab{a}}){Li}, {Simon}, {Kuehn}, {Pace},
  {Erkal}, {Bechtol}, {Yanny}, {Drlica-Wagner}, {Marshall}, {Lidman},
  {Balbinot}, {Carollo}, {Jenkins}, {Mart{\'\i}nez-V{\'a}zquez}, {Shipp},
  {Stringer}, {Vivas}, {Walker}, {Wechsler}, {Abdalla}, {Allam}, {Annis},
  {Avila}, {Bertin}, {Brooks}, {Buckley-Geer}, {Burke}, {Carnero Rosell},
  {Carrasco Kind}, {Carretero}, {Cunha}, {D'Andrea}, {da Costa}, {Davis}, {De
  Vicente}, {Doel}, {Eifler}, {Evrard}, {Flaugher}, {Frieman},
  {Garc{\'\i}a-Bellido}, {Gaztanaga}, {Gerdes}, {Gruen}, {Gruendl}, {Gschwend},
  {Gutierrez}, {Hartley}, {Hollowood}, {Honscheid}, {James}, {Krause}, {Maia},
  {March}, {Menanteau}, {Miquel}, {Plazas}, {Sanchez}, {Santiago}, {Scarpine},
  {Schindler}, {Schubnell}, {Sevilla-Noarbe}, {Smith}, {Smith},
  {Soares-Santos}, {Sobreira}, {Suchyta}, {Swanson}, {Tarle}, {Tucker}, \& {DES
  Collaboration}}]{li2018a}
{Li}, T.~S., {Simon}, J.~D., {Kuehn}, K., {Pace}, A.~B., {Erkal}, D.,
  {Bechtol}, K., {Yanny}, B., {Drlica-Wagner}, A., {Marshall}, J.~L., {Lidman},
  C., {Balbinot}, E., {Carollo}, D., {Jenkins}, S.,
  {Mart{\'\i}nez-V{\'a}zquez}, C.~E., {Shipp}, N., {Stringer}, K.~M., {Vivas},
  A.~K., {Walker}, A.~R., {Wechsler}, R.~H., {Abdalla}, F.~B., {Allam}, S.,
  {Annis}, J., {Avila}, S., {Bertin}, E., {Brooks}, D., {Buckley-Geer}, E.,
  {Burke}, D.~L., {Carnero Rosell}, A., {Carrasco Kind}, M., {Carretero}, J.,
  {Cunha}, C.~E., {D'Andrea}, C.~B., {da Costa}, L.~N., {Davis}, C., {De
  Vicente}, J., {Doel}, P., {Eifler}, T.~F., {Evrard}, A.~E., {Flaugher}, B.,
  {Frieman}, J., {Garc{\'\i}a-Bellido}, J., {Gaztanaga}, E., {Gerdes}, D.~W.,
  {Gruen}, D., {Gruendl}, R.~A., {Gschwend}, J., {Gutierrez}, G., {Hartley},
  W.~G., {Hollowood}, D.~L., {Honscheid}, K., {James}, D.~J., {Krause}, E.,
  {Maia}, M.~A.~G., {March}, M., {Menanteau}, F., {Miquel}, R., {Plazas},
  A.~A., {Sanchez}, E., {Santiago}, B., {Scarpine}, V., {Schindler}, R.,
  {Schubnell}, M., {Sevilla-Noarbe}, I., {Smith}, M., {Smith}, R.~C.,
  {Soares-Santos}, M., {Sobreira}, F., {Suchyta}, E., {Swanson}, M.~E.~C.,
  {Tarle}, G., {Tucker}, D.~L., \& {DES Collaboration}. 2018{\natexlab{a}},
  \apj, 866, 22

\bibitem[{{Li} {et~al.}(2018{\natexlab{b}}){Li}, {Simon}, {Pace}, {Torrealba},
  {Kuehn}, {Drlica-Wagner}, {Bechtol}, {Vivas}, {van der Marel}, {Wood},
  {Yanny}, {Belokurov}, {Jethwa}, {Zucker}, {Lewis}, {Kron}, {Nidever},
  {S{\'a}nchez-Conde}, {Ji}, {Conn}, {James}, {Martin}, {Martinez-Delgado},
  {No{\"e}l}, \& {MagLiteS Collaboration}}]{li2018b}
{Li}, T.~S., {Simon}, J.~D., {Pace}, A.~B., {Torrealba}, G., {Kuehn}, K.,
  {Drlica-Wagner}, A., {Bechtol}, K., {Vivas}, A.~K., {van der Marel}, R.~P.,
  {Wood}, M., {Yanny}, B., {Belokurov}, V., {Jethwa}, P., {Zucker}, D.~B.,
  {Lewis}, G., {Kron}, R., {Nidever}, D.~L., {S{\'a}nchez-Conde}, M.~A., {Ji},
  A.~P., {Conn}, B.~C., {James}, D.~J., {Martin}, N.~F., {Martinez-Delgado},
  D., {No{\"e}l}, N.~E.~D., \& {MagLiteS Collaboration}. 2018{\natexlab{b}},
  \apj, 857, 145

\bibitem[{{Li} {et~al.}(2020){Li}, {Qian}, {Han}, {Li}, {Wang}, \&
  {Jing}}]{li2020}
{Li}, Z.-Z., {Qian}, Y.-Z., {Han}, J., {Li}, T.~S., {Wang}, W., \& {Jing},
  Y.~P. 2020, \apj, 894, 10

\bibitem[{{Lindegren} {et~al.}(2018){Lindegren}, {Hern{\'a}ndez}, {Bombrun},
  {Klioner}, {Bastian}, {Ramos-Lerate}, {de Torres}, {Steidelm{\"u}ller},
  {Stephenson}, {Hobbs}, {Lammers}, {Biermann}, {Geyer}, {Hilger}, {Michalik},
  {Stampa}, {McMillan}, {Casta{\~n}eda}, {Clotet}, {Comoretto}, {Davidson},
  {Fabricius}, {Gracia}, {Hambly}, {Hutton}, {Mora}, {Portell}, {van Leeuwen},
  {Abbas}, {Abreu}, {Altmann}, {Andrei}, {Anglada}, {Balaguer-N{\'u}{\~n}ez},
  {Barache}, {Becciani}, {Bertone}, {Bianchi}, {Bouquillon}, {Bourda},
  {Br{\"u}semeister}, {Bucciarelli}, {Busonero}, {Buzzi}, {Cancelliere},
  {Carlucci}, {Charlot}, {Cheek}, {Crosta}, {Crowley}, {de Bruijne}, {de
  Felice}, {Drimmel}, {Esquej}, {Fienga}, {Fraile}, {Gai}, {Garralda},
  {Gonz{\'a}lez-Vidal}, {Guerra}, {Hauser}, {Hofmann}, {Holl}, {Jordan},
  {Lattanzi}, {Lenhardt}, {Liao}, {Licata}, {Lister}, {L{\"o}ffler},
  {Marchant}, {Martin-Fleitas}, {Messineo}, {Mignard}, {Morbidelli}, {Poggio},
  {Riva}, {Rowell}, {Salguero}, {Sarasso}, {Sciacca}, {Siddiqui}, {Smart},
  {Spagna}, {Steele}, {Taris}, {Torra}, {van Elteren}, {van Reeven}, \&
  {Vecchiato}}]{lindegren2018}
{Lindegren}, L., {Hern{\'a}ndez}, J., {Bombrun}, A., {Klioner}, S., {Bastian},
  U., {Ramos-Lerate}, M., {de Torres}, A., {Steidelm{\"u}ller}, H.,
  {Stephenson}, C., {Hobbs}, D., {Lammers}, U., {Biermann}, M., {Geyer}, R.,
  {Hilger}, T., {Michalik}, D., {Stampa}, U., {McMillan}, P.~J.,
  {Casta{\~n}eda}, J., {Clotet}, M., {Comoretto}, G., {Davidson}, M.,
  {Fabricius}, C., {Gracia}, G., {Hambly}, N.~C., {Hutton}, A., {Mora}, A.,
  {Portell}, J., {van Leeuwen}, F., {Abbas}, U., {Abreu}, A., {Altmann}, M.,
  {Andrei}, A., {Anglada}, E., {Balaguer-N{\'u}{\~n}ez}, L., {Barache}, C.,
  {Becciani}, U., {Bertone}, S., {Bianchi}, L., {Bouquillon}, S., {Bourda}, G.,
  {Br{\"u}semeister}, T., {Bucciarelli}, B., {Busonero}, D., {Buzzi}, R.,
  {Cancelliere}, R., {Carlucci}, T., {Charlot}, P., {Cheek}, N., {Crosta}, M.,
  {Crowley}, C., {de Bruijne}, J., {de Felice}, F., {Drimmel}, R., {Esquej},
  P., {Fienga}, A., {Fraile}, E., {Gai}, M., {Garralda}, N.,
  {Gonz{\'a}lez-Vidal}, J.~J., {Guerra}, R., {Hauser}, M., {Hofmann}, W.,
  {Holl}, B., {Jordan}, S., {Lattanzi}, M.~G., {Lenhardt}, H., {Liao}, S.,
  {Licata}, E., {Lister}, T., {L{\"o}ffler}, W., {Marchant}, J.,
  {Martin-Fleitas}, J.~M., {Messineo}, R., {Mignard}, F., {Morbidelli}, R.,
  {Poggio}, E., {Riva}, A., {Rowell}, N., {Salguero}, E., {Sarasso}, M.,
  {Sciacca}, E., {Siddiqui}, H., {Smart}, R.~L., {Spagna}, A., {Steele}, I.,
  {Taris}, F., {Torra}, J., {van Elteren}, A., {van Reeven}, W., \&
  {Vecchiato}, A. 2018, \aap, 616, A2

\bibitem[{{Longeard} {et~al.}(2020{\natexlab{a}}){Longeard}, {Martin}, {Ibata},
  {Starkenburg}, {Jablonka}, {Aguado}, {Carlberg}, {C{\^o}t{\'e}},
  {Gonz{\'a}lez Hern{\'a}ndez}, {Lucchesi}, {Malhan}, {Navarro},
  {S{\'a}nchez-Janssen}, {Thomas}, {Venn}, \& {McConnachie}}]{longeard2020b}
{Longeard}, N., {Martin}, N., {Ibata}, R., {Starkenburg}, E., {Jablonka}, P.,
  {Aguado}, D.~S., {Carlberg}, R.~G., {C{\^o}t{\'e}}, P., {Gonz{\'a}lez
  Hern{\'a}ndez}, J.~I., {Lucchesi}, R., {Malhan}, K., {Navarro}, J.~F.,
  {S{\'a}nchez-Janssen}, R., {Thomas}, G.~F., {Venn}, K., \& {McConnachie},
  A.~W. 2020{\natexlab{a}}, arXiv e-prints, arXiv:2005.05976

\bibitem[{{Longeard} {et~al.}(2018){Longeard}, {Martin}, {Starkenburg},
  {Ibata}, {Collins}, {Geha}, {Laevens}, {Rich}, {Aguado}, {Arentsen},
  {Carlberg}, {C{\^o}t{\'e}}, {Hill}, {Jablonka}, {Gonz{\'a}lez Hern{\'a}ndez},
  {Navarro}, {S{\'a}nchez-Janssen}, {Tolstoy}, {Venn}, \&
  {Youakim}}]{longeard2018}
{Longeard}, N., {Martin}, N., {Starkenburg}, E., {Ibata}, R.~A., {Collins}, M.
  L.~M., {Geha}, M., {Laevens}, B. P.~M., {Rich}, R.~M., {Aguado}, D.~S.,
  {Arentsen}, A., {Carlberg}, R.~G., {C{\^o}t{\'e}}, P., {Hill}, V.,
  {Jablonka}, P., {Gonz{\'a}lez Hern{\'a}ndez}, J.~I., {Navarro}, J.~F.,
  {S{\'a}nchez-Janssen}, R., {Tolstoy}, E., {Venn}, K.~A., \& {Youakim}, K.
  2018, \mnras, 480, 2609

\bibitem[{{Longeard} {et~al.}(2020{\natexlab{b}}){Longeard}, {Martin},
  {Starkenburg}, {Ibata}, {Collins}, {Laevens}, {Mackey}, {Rich}, {Aguado},
  {Arentsen}, {Jablonka}, {Gonz{\'a}lez Hern{\'a}ndez}, {Navarro}, \&
  {S{\'a}nchez-Janssen}}]{longeard2020a}
{Longeard}, N., {Martin}, N., {Starkenburg}, E., {Ibata}, R.~A., {Collins}, M.
  L.~M., {Laevens}, B. P.~M., {Mackey}, D., {Rich}, R.~M., {Aguado}, D.~S.,
  {Arentsen}, A., {Jablonka}, P., {Gonz{\'a}lez Hern{\'a}ndez}, J.~I.,
  {Navarro}, J.~F., \& {S{\'a}nchez-Janssen}, R. 2020{\natexlab{b}}, \mnras,
  491, 356

\bibitem[{{Marshall} {et~al.}(2019){Marshall}, {Hansen}, {Simon}, {Li},
  {Bernstein}, {Kuehn}, {Pace}, {DePoy}, {Palmese}, {Pieres}, {Strigari},
  {Drlica-Wagner}, {Bechtol}, {Lidman}, {Nagasawa}, {Bertin}, {Brooks},
  {Buckley-Geer}, {Burke}, {Carnero Rosell}, {Carrasco Kind}, {Carretero},
  {Cunha}, {D'Andrea}, {da Costa}, {De Vicente}, {Desai}, {Doel}, {Eifler},
  {Flaugher}, {Fosalba}, {Frieman}, {Garc{\'\i}a-Bellido}, {Gaztanaga},
  {Gerdes}, {Gruendl}, {Gschwend}, {Gutierrez}, {Hartley}, {Hollowood},
  {Honscheid}, {Hoyle}, {James}, {Kuropatkin}, {Maia}, {Menanteau}, {Miller},
  {Miquel}, {Plazas}, {Sanchez}, {Santiago}, {Scarpine}, {Schubnell},
  {Serrano}, {Sevilla-Noarbe}, {Smith}, {Soares-Santos}, {Suchyta}, {Swanson},
  {Tarle}, {Wester}, \& {DES Collaboration}}]{marshall2019}
{Marshall}, J.~L., {Hansen}, T., {Simon}, J.~D., {Li}, T.~S., {Bernstein},
  R.~A., {Kuehn}, K., {Pace}, A.~B., {DePoy}, D.~L., {Palmese}, A., {Pieres},
  A., {Strigari}, L., {Drlica-Wagner}, A., {Bechtol}, K., {Lidman}, C.,
  {Nagasawa}, D.~Q., {Bertin}, E., {Brooks}, D., {Buckley-Geer}, E., {Burke},
  D.~L., {Carnero Rosell}, A., {Carrasco Kind}, M., {Carretero}, J., {Cunha},
  C.~E., {D'Andrea}, C.~B., {da Costa}, L.~N., {De Vicente}, J., {Desai}, S.,
  {Doel}, P., {Eifler}, T.~F., {Flaugher}, B., {Fosalba}, P., {Frieman}, J.,
  {Garc{\'\i}a-Bellido}, J., {Gaztanaga}, E., {Gerdes}, D.~W., {Gruendl},
  R.~A., {Gschwend}, J., {Gutierrez}, G., {Hartley}, W.~G., {Hollowood}, D.~L.,
  {Honscheid}, K., {Hoyle}, B., {James}, D.~J., {Kuropatkin}, N., {Maia},
  M.~A.~G., {Menanteau}, F., {Miller}, C.~J., {Miquel}, R., {Plazas}, A.~A.,
  {Sanchez}, E., {Santiago}, B., {Scarpine}, V., {Schubnell}, M., {Serrano},
  S., {Sevilla-Noarbe}, I., {Smith}, M., {Soares-Santos}, M., {Suchyta}, E.,
  {Swanson}, M.~E.~C., {Tarle}, G., {Wester}, W., \& {DES Collaboration}. 2019,
  \apj, 882, 177

\bibitem[{{Martin} {et~al.}(2016{\natexlab{a}}){Martin}, {Geha}, {Ibata},
  {Collins}, {Laevens}, {Bell}, {Rix}, {Ferguson}, {Chambers}, {Wainscoat}, \&
  {Waters}}]{martin2016c}
{Martin}, N.~F., {Geha}, M., {Ibata}, R.~A., {Collins}, M. L.~M., {Laevens}, B.
  P.~M., {Bell}, E.~F., {Rix}, H.-W., {Ferguson}, A. M.~N., {Chambers}, K.~C.,
  {Wainscoat}, R.~J., \& {Waters}, C. 2016{\natexlab{a}}, \mnras, 458, L59

\bibitem[{{Martin} {et~al.}(2004){Martin}, {Ibata}, {Bellazzini}, {Irwin},
  {Lewis}, \& {Dehnen}}]{martin2004a}
{Martin}, N.~F., {Ibata}, R.~A., {Bellazzini}, M., {Irwin}, M.~J., {Lewis},
  G.~F., \& {Dehnen}, W. 2004, \mnras, 348, 12

\bibitem[{{Martin} {et~al.}(2007){Martin}, {Ibata}, {Chapman}, {Irwin}, \&
  {Lewis}}]{martin2007b}
{Martin}, N.~F., {Ibata}, R.~A., {Chapman}, S.~C., {Irwin}, M., \& {Lewis},
  G.~F. 2007, \mnras, 380, 281

\bibitem[{{Martin} {et~al.}(2016{\natexlab{b}}){Martin}, {Ibata}, {Collins},
  {Rich}, {Bell}, {Ferguson}, {Laevens}, {Rix}, {Chapman}, \&
  {Koch}}]{martin2016b}
{Martin}, N.~F., {Ibata}, R.~A., {Collins}, M. L.~M., {Rich}, R.~M., {Bell},
  E.~F., {Ferguson}, A. M.~N., {Laevens}, B. P.~M., {Rix}, H.-W., {Chapman},
  S.~C., \& {Koch}, A. 2016{\natexlab{b}}, \apj, 818, 40

\bibitem[{{Massari} \& {Helmi}(2018)}]{massari2018}
{Massari}, D. \& {Helmi}, A. 2018, \aap, 620, A155

\bibitem[{{Mateo} {et~al.}(2008){Mateo}, {Olszewski}, \& {Walker}}]{mateo2008}
{Mateo}, M., {Olszewski}, E.~W., \& {Walker}, M.~G. 2008, \apj, 675, 201

\bibitem[{{Mau} {et~al.}(2020){Mau}, {Cerny}, {Pace}, {Choi}, {Drlica-Wagner},
  {Santana-Silva}, {Riley}, {Erkal}, {Stringfellow}, {Adam{\'o}w}, {Carlin},
  {Gruendl}, {Hernandez-Lang}, {Kuropatkin}, {Li}, {Mart{\'\i}nez-V{\'a}zquez},
  {Morganson}, {Mutlu-Pakdil}, {Neilsen}, {Nidever}, {Olsen}, {Sand},
  {Tollerud}, {Tucker}, {Yanny}, {Zenteno}, {Allam}, {Barkhouse}, {Bechtol},
  {Bell}, {Balaji}, {Crnojevi{\'c}}, {Esteves}, {Ferguson}, {Gallart},
  {Hughes}, {James}, {Jethwa}, {Johnson}, {Kuehn}, {Majewski}, {Mao},
  {Massana}, {McNanna}, {Monachesi}, {Nadler}, {No{\"e}l}, {Palmese},
  {Paz-Chinchon}, {Pieres}, {Sanchez}, {Shipp}, {Simon}, {Soares-Santos},
  {Tavangar}, {van der Marel}, {Vivas}, {Walker}, \& {Wechsler}}]{mau2020}
{Mau}, S., {Cerny}, W., {Pace}, A.~B., {Choi}, Y., {Drlica-Wagner}, A.,
  {Santana-Silva}, L., {Riley}, A.~H., {Erkal}, D., {Stringfellow}, G.~S.,
  {Adam{\'o}w}, M., {Carlin}, J.~L., {Gruendl}, R.~A., {Hernandez-Lang}, D.,
  {Kuropatkin}, N., {Li}, T.~S., {Mart{\'\i}nez-V{\'a}zquez}, C.~E.,
  {Morganson}, E., {Mutlu-Pakdil}, B., {Neilsen}, E.~H., {Nidever}, D.~L.,
  {Olsen}, K.~A.~G., {Sand}, D.~J., {Tollerud}, E.~J., {Tucker}, D.~L.,
  {Yanny}, B., {Zenteno}, A., {Allam}, S., {Barkhouse}, W.~A., {Bechtol}, K.,
  {Bell}, E.~F., {Balaji}, P., {Crnojevi{\'c}}, D., {Esteves}, J., {Ferguson},
  P.~S., {Gallart}, C., {Hughes}, A.~K., {James}, D.~J., {Jethwa}, P.,
  {Johnson}, L.~C., {Kuehn}, K., {Majewski}, S., {Mao}, Y.~Y., {Massana}, P.,
  {McNanna}, M., {Monachesi}, A., {Nadler}, E.~O., {No{\"e}l}, N.~E.~D.,
  {Palmese}, A., {Paz-Chinchon}, F., {Pieres}, A., {Sanchez}, J., {Shipp}, N.,
  {Simon}, J.~D., {Soares-Santos}, M., {Tavangar}, K., {van der Marel}, R.~P.,
  {Vivas}, A.~K., {Walker}, A.~R., \& {Wechsler}, R.~H. 2020, \apj, 890, 136

\bibitem[{{Mau} {et~al.}(2019){Mau}, {Drlica-Wagner}, {Bechtol}, {Pace}, {Li},
  {Soares-Santos}, {Kuropatkin}, {Allam}, {Tucker}, {Santana-Silva}, {Yanny},
  {Jethwa}, {Palmese}, {Vivas}, {Burgad}, {Chen}, \& {BLISS
  Collaboration}}]{mau2019}
{Mau}, S., {Drlica-Wagner}, A., {Bechtol}, K., {Pace}, A.~B., {Li}, T.,
  {Soares-Santos}, M., {Kuropatkin}, N., {Allam}, S., {Tucker}, D.,
  {Santana-Silva}, L., {Yanny}, B., {Jethwa}, P., {Palmese}, A., {Vivas}, K.,
  {Burgad}, C., {Chen}, H.~Y., \& {BLISS Collaboration}. 2019, \apj, 875, 154

\bibitem[{{McConnachie}(2012)}]{mcconnachie2012}
{McConnachie}, A.~W. 2012, \aj, 144, 4

\bibitem[{{Mu{\~n}oz} {et~al.}(2006){Mu{\~n}oz}, {Carlin}, {Frinchaboy},
  {Nidever}, {Majewski}, \& {Patterson}}]{munoz2006b}
{Mu{\~n}oz}, R.~R., {Carlin}, J.~L., {Frinchaboy}, P.~M., {Nidever}, D.~L.,
  {Majewski}, S.~R., \& {Patterson}, R.~J. 2006, \apjl, 650, L51

\bibitem[{{Nagasawa} {et~al.}(2018){Nagasawa}, {Marshall}, {Li}, {Hansen},
  {Simon}, {Bernstein}, {Balbinot}, {Drlica-Wagner}, {Pace}, {Strigari},
  {Pellegrino}, {DePoy}, {Suntzeff}, {Bechtol}, {Walker}, {Abbott}, {Abdalla},
  {Allam}, {Annis}, {Benoit-L{\'e}vy}, {Bertin}, {Brooks}, {Carnero Rosell},
  {Carrasco Kind}, {Carretero}, {Cunha}, {D'Andrea}, {da Costa}, {Davis},
  {Desai}, {Doel}, {Eifler}, {Flaugher}, {Fosalba}, {Frieman},
  {Garc{\'\i}a-Bellido}, {Gaztanaga}, {Gerdes}, {Gruen}, {Gruendl}, {Gschwend},
  {Gutierrez}, {Hartley}, {Honscheid}, {James}, {Jeltema}, {Krause}, {Kuehn},
  {Kuhlmann}, {Kuropatkin}, {March}, {Miquel}, {Nord}, {Roodman}, {Sanchez},
  {Santiago}, {Scarpine}, {Schindler}, {Schubnell}, {Sevilla-Noarbe}, {Smith},
  {Smith}, {Soares-Santos}, {Sobreira}, {Suchyta}, {Tarle}, {Thomas}, {Tucker},
  {Wechsler}, {Wolf}, \& {Yanny}}]{nagasawa2018}
{Nagasawa}, D.~Q., {Marshall}, J.~L., {Li}, T.~S., {Hansen}, T.~T., {Simon},
  J.~D., {Bernstein}, R.~A., {Balbinot}, E., {Drlica-Wagner}, A., {Pace},
  A.~B., {Strigari}, L.~E., {Pellegrino}, C.~M., {DePoy}, D.~L., {Suntzeff},
  N.~B., {Bechtol}, K., {Walker}, A.~R., {Abbott}, T.~M.~C., {Abdalla}, F.~B.,
  {Allam}, S., {Annis}, J., {Benoit-L{\'e}vy}, A., {Bertin}, E., {Brooks}, D.,
  {Carnero Rosell}, A., {Carrasco Kind}, M., {Carretero}, J., {Cunha}, C.~E.,
  {D'Andrea}, C.~B., {da Costa}, L.~N., {Davis}, C., {Desai}, S., {Doel}, P.,
  {Eifler}, T.~F., {Flaugher}, B., {Fosalba}, P., {Frieman}, J.,
  {Garc{\'\i}a-Bellido}, J., {Gaztanaga}, E., {Gerdes}, D.~W., {Gruen}, D.,
  {Gruendl}, R.~A., {Gschwend}, J., {Gutierrez}, G., {Hartley}, W.~G.,
  {Honscheid}, K., {James}, D.~J., {Jeltema}, T., {Krause}, E., {Kuehn}, K.,
  {Kuhlmann}, S., {Kuropatkin}, N., {March}, M., {Miquel}, R., {Nord}, B.,
  {Roodman}, A., {Sanchez}, E., {Santiago}, B., {Scarpine}, V., {Schindler},
  R., {Schubnell}, M., {Sevilla-Noarbe}, I., {Smith}, M., {Smith}, R.~C.,
  {Soares-Santos}, M., {Sobreira}, F., {Suchyta}, E., {Tarle}, G., {Thomas},
  D., {Tucker}, D.~L., {Wechsler}, R.~H., {Wolf}, R.~C., \& {Yanny}, B. 2018,
  \apj, 852, 99

\bibitem[{{Norris} {et~al.}(2008){Norris}, {Gilmore}, {Wyse}, {Wilkinson},
  {Belokurov}, {Evans}, \& {Zucker}}]{norris2008}
{Norris}, J.~E., {Gilmore}, G., {Wyse}, R. F.~G., {Wilkinson}, M.~I.,
  {Belokurov}, V., {Evans}, N.~W., \& {Zucker}, D.~B. 2008, \apjl, 689, L113

\bibitem[{{Norris} {et~al.}(2010{\natexlab{a}}){Norris}, {Gilmore}, {Wyse},
  {Yong}, \& {Frebel}}]{norris2010b}
{Norris}, J.~E., {Gilmore}, G., {Wyse}, R. F.~G., {Yong}, D., \& {Frebel}, A.
  2010{\natexlab{a}}, \apjl, 722, L104

\bibitem[{{Norris} {et~al.}(2010{\natexlab{b}}){Norris}, {Wyse}, {Gilmore},
  {Yong}, {Frebel}, {Wilkinson}, {Belokurov}, \& {Zucker}}]{norris2010a}
{Norris}, J.~E., {Wyse}, R.~F.~G., {Gilmore}, G., {Yong}, D., {Frebel}, A.,
  {Wilkinson}, M.~I., {Belokurov}, V., \& {Zucker}, D.~B. 2010{\natexlab{b}},
  \apj, 723, 1632

\bibitem[{{Pace} {et~al.}(2020){Pace}, {Kaplinghat}, {Kirby}, {Simon},
  {Tollerud}, {Mu{\~n}oz}, {C{\^o}t{\'e}}, {Djorgovski}, \& {Geha}}]{pace2020}
{Pace}, A.~B., {Kaplinghat}, M., {Kirby}, E., {Simon}, J.~D., {Tollerud}, E.,
  {Mu{\~n}oz}, R.~R., {C{\^o}t{\'e}}, P., {Djorgovski}, S.~G., \& {Geha}, M.
  2020, \mnras, 495, 3022

\bibitem[{{Pace} \& {Li}(2019)}]{pace2019}
{Pace}, A.~B. \& {Li}, T.~S. 2019, \apj, 875, 77

\bibitem[{{Patel} {et~al.}(2020){Patel}, {Kallivayalil}, {Garavito-Camargo},
  {Besla}, {Weisz}, {van der Marel}, {Boylan-Kolchin}, {Pawlowski}, \&
  {G{\'o}mez}}]{patel2020}
{Patel}, E., {Kallivayalil}, N., {Garavito-Camargo}, N., {Besla}, G., {Weisz},
  D.~R., {van der Marel}, R.~P., {Boylan-Kolchin}, M., {Pawlowski}, M.~S., \&
  {G{\'o}mez}, F.~A. 2020, \apj, 893, 121

\bibitem[{{Posti} \& {Helmi}(2019)}]{posti2019}
{Posti}, L. \& {Helmi}, A. 2019, \aap, 621, A56

\bibitem[{{Roederer} \& {Kirby}(2014)}]{roederer2014}
{Roederer}, I.~U. \& {Kirby}, E.~N. 2014, \mnras, 440, 2665

\bibitem[{{Schlegel} {et~al.}(1998){Schlegel}, {Finkbeiner}, \&
  {Davis}}]{schlegel1998}
{Schlegel}, D.~J., {Finkbeiner}, D.~P., \& {Davis}, M. 1998, \apj, 500, 525

\bibitem[{{Sch{\"o}nrich} {et~al.}(2010){Sch{\"o}nrich}, {Binney}, \&
  {Dehnen}}]{schonrich2010}
{Sch{\"o}nrich}, R., {Binney}, J., \& {Dehnen}, W. 2010, \mnras, 403, 1829

\bibitem[{{Simon}(2018)}]{simon2018}
{Simon}, J.~D. 2018, \apj, 863, 89

\bibitem[{{Simon} {et~al.}(2015){Simon}, {Drlica-Wagner}, {Li}, {Nord}, {Geha},
  {Bechtol}, {Balbinot}, {Buckley-Geer}, {Lin}, {Marshall}, {Santiago},
  {Strigari}, {Wang}, {Wechsler}, {Yanny}, {Abbott}, {Bauer}, {Bernstein},
  {Bertin}, {Brooks}, {Burke}, {Capozzi}, {Carnero Rosell}, {Carrasco Kind},
  {D'Andrea}, {da Costa}, {DePoy}, {Desai}, {Diehl}, {Dodelson}, {Cunha},
  {Estrada}, {Evrard}, {Fausti Neto}, {Fernandez}, {Finley}, {Flaugher},
  {Frieman}, {Gaztanaga}, {Gerdes}, {Gruen}, {Gruendl}, {Honscheid}, {James},
  {Kent}, {Kuehn}, {Kuropatkin}, {Lahav}, {Maia}, {March}, {Martini}, {Miller},
  {Miquel}, {Ogando}, {Romer}, {Roodman}, {Rykoff}, {Sako}, {Sanchez},
  {Schubnell}, {Sevilla}, {Smith}, {Soares-Santos}, {Sobreira}, {Suchyta},
  {Swanson}, {Tarle}, {Thaler}, {Tucker}, {Vikram}, {Walker}, {Wester}, \& {DES
  Collaboration}}]{simon2015}
{Simon}, J.~D., {Drlica-Wagner}, A., {Li}, T.~S., {Nord}, B., {Geha}, M.,
  {Bechtol}, K., {Balbinot}, E., {Buckley-Geer}, E., {Lin}, H., {Marshall}, J.,
  {Santiago}, B., {Strigari}, L., {Wang}, M., {Wechsler}, R.~H., {Yanny}, B.,
  {Abbott}, T., {Bauer}, A.~H., {Bernstein}, G.~M., {Bertin}, E., {Brooks}, D.,
  {Burke}, D.~L., {Capozzi}, D., {Carnero Rosell}, A., {Carrasco Kind}, M.,
  {D'Andrea}, C.~B., {da Costa}, L.~N., {DePoy}, D.~L., {Desai}, S., {Diehl},
  H.~T., {Dodelson}, S., {Cunha}, C.~E., {Estrada}, J., {Evrard}, A.~E.,
  {Fausti Neto}, A., {Fernandez}, E., {Finley}, D.~A., {Flaugher}, B.,
  {Frieman}, J., {Gaztanaga}, E., {Gerdes}, D., {Gruen}, D., {Gruendl}, R.~A.,
  {Honscheid}, K., {James}, D., {Kent}, S., {Kuehn}, K., {Kuropatkin}, N.,
  {Lahav}, O., {Maia}, M.~A.~G., {March}, M., {Martini}, P., {Miller}, C.~J.,
  {Miquel}, R., {Ogando}, R., {Romer}, A.~K., {Roodman}, A., {Rykoff}, E.~S.,
  {Sako}, M., {Sanchez}, E., {Schubnell}, M., {Sevilla}, I., {Smith}, R.~C.,
  {Soares-Santos}, M., {Sobreira}, F., {Suchyta}, E., {Swanson}, M.~E.~C.,
  {Tarle}, G., {Thaler}, J., {Tucker}, D., {Vikram}, V., {Walker}, A.~R.,
  {Wester}, W., \& {DES Collaboration}. 2015, \apj, 808, 95

\bibitem[{{Simon} {et~al.}(2010){Simon}, {Frebel}, {McWilliam}, {Kirby}, \&
  {Thompson}}]{simon2010}
{Simon}, J.~D., {Frebel}, A., {McWilliam}, A., {Kirby}, E.~N., \& {Thompson},
  I.~B. 2010, \apj, 716, 446

\bibitem[{{Simon} \& {Geha}(2007)}]{simon2007}
{Simon}, J.~D. \& {Geha}, M. 2007, \apj, 670, 313

\bibitem[{{Simon} {et~al.}(2011){Simon}, {Geha}, {Minor}, {Martinez}, {Kirby},
  {Bullock}, {Kaplinghat}, {Strigari}, {Willman}, {Choi}, {Tollerud}, \&
  {Wolf}}]{simon2011}
{Simon}, J.~D., {Geha}, M., {Minor}, Q.~E., {Martinez}, G.~D., {Kirby}, E.~N.,
  {Bullock}, J.~S., {Kaplinghat}, M., {Strigari}, L.~E., {Willman}, B., {Choi},
  P.~I., {Tollerud}, E.~J., \& {Wolf}, J. 2011, \apj, 733, 46

\bibitem[{{Simon} {et~al.}(2017){Simon}, {Li}, {Drlica-Wagner}, {Bechtol},
  {Marshall}, {James}, {Wang}, {Strigari}, {Balbinot}, {Kuehn}, {Walker},
  {Abbott}, {Allam}, {Annis}, {Benoit-L{\'e}vy}, {Brooks}, {Buckley-Geer},
  {Burke}, {Carnero Rosell}, {Carrasco Kind}, {Carretero}, {Cunha}, {D'Andrea},
  {da Costa}, {DePoy}, {Desai}, {Doel}, {Fernandez}, {Flaugher}, {Frieman},
  {Garc{\'\i}a-Bellido}, {Gaztanaga}, {Goldstein}, {Gruen}, {Gutierrez},
  {Kuropatkin}, {Maia}, {Martini}, {Menanteau}, {Miller}, {Miquel}, {Neilsen},
  {Nord}, {Ogando}, {Plazas}, {Romer}, {Rykoff}, {Sanchez}, {Santiago},
  {Scarpine}, {Schubnell}, {Sevilla-Noarbe}, {Smith}, {Sobreira}, {Suchyta},
  {Swanson}, {Tarle}, {Whiteway}, {Yanny}, \& {DES Collaboration}}]{simon2017}
{Simon}, J.~D., {Li}, T.~S., {Drlica-Wagner}, A., {Bechtol}, K., {Marshall},
  J.~L., {James}, D.~J., {Wang}, M.~Y., {Strigari}, L., {Balbinot}, E.,
  {Kuehn}, K., {Walker}, A.~R., {Abbott}, T.~M.~C., {Allam}, S., {Annis}, J.,
  {Benoit-L{\'e}vy}, A., {Brooks}, D., {Buckley-Geer}, E., {Burke}, D.~L.,
  {Carnero Rosell}, A., {Carrasco Kind}, M., {Carretero}, J., {Cunha}, C.~E.,
  {D'Andrea}, C.~B., {da Costa}, L.~N., {DePoy}, D.~L., {Desai}, S., {Doel},
  P., {Fernandez}, E., {Flaugher}, B., {Frieman}, J., {Garc{\'\i}a-Bellido},
  J., {Gaztanaga}, E., {Goldstein}, D.~A., {Gruen}, D., {Gutierrez}, G.,
  {Kuropatkin}, N., {Maia}, M.~A.~G., {Martini}, P., {Menanteau}, F., {Miller},
  C.~J., {Miquel}, R., {Neilsen}, E., {Nord}, B., {Ogando}, R., {Plazas},
  A.~A., {Romer}, A.~K., {Rykoff}, E.~S., {Sanchez}, E., {Santiago}, B.,
  {Scarpine}, V., {Schubnell}, M., {Sevilla-Noarbe}, I., {Smith}, R.~C.,
  {Sobreira}, F., {Suchyta}, E., {Swanson}, M.~E.~C., {Tarle}, G., {Whiteway},
  L., {Yanny}, B., \& {DES Collaboration}. 2017, \apj, 838, 11

\bibitem[{{Simon} {et~al.}(2019){Simon}, {Li}, {Erkal}, {Pace},
  {Drlica-Wagner}, {James}, {Marshall}, {Bechtol}, {Hansen}, {Kuehn}, {Lidman},
  {Allam}, {Annis}, {Avila}, {Bertin}, {Brooks}, {Burke}, {Carnero Rosell},
  {Carrasco Kind}, {Carretero}, {da Costa}, {De Vicente}, {Desai}, {Doel},
  {Eifler}, {Everett}, {Fosalba}, {Frieman}, {Garcia-Bellido}, {Gaztanaga},
  {Gerdes}, {Gruen}, {Gruendl}, {Gschwend}, {Gutierrez}, {Hollowood},
  {Honscheid}, {Krause}, {Kuropatkin}, {MacCrann}, {Maia}, {March}, {Miquel},
  {Palmese}, {Paz-Chinchon}, {Plazas}, {Reil}, {Roodman}, {Sanchez},
  {Santiago}, {Scarpine}, {Schubnell}, {Serrano}, {Smith}, {Suchyta}, {Tarle},
  \& {Walker}}]{simon2019}
{Simon}, J.~D., {Li}, T.~S., {Erkal}, D., {Pace}, A.~B., {Drlica-Wagner}, A.,
  {James}, D.~J., {Marshall}, J.~L., {Bechtol}, K., {Hansen}, T., {Kuehn}, K.,
  {Lidman}, C., {Allam}, S., {Annis}, J., {Avila}, S., {Bertin}, E., {Brooks},
  D., {Burke}, D.~L., {Carnero Rosell}, A., {Carrasco Kind}, M., {Carretero},
  J., {da Costa}, L.~N., {De Vicente}, J., {Desai}, S., {Doel}, P., {Eifler},
  T.~F., {Everett}, S., {Fosalba}, P., {Frieman}, J., {Garcia-Bellido}, J.,
  {Gaztanaga}, E., {Gerdes}, D.~W., {Gruen}, D., {Gruendl}, R.~A., {Gschwend},
  J., {Gutierrez}, G., {Hollowood}, D.~L., {Honscheid}, K., {Krause}, E.,
  {Kuropatkin}, N., {MacCrann}, N., {Maia}, M.~A.~G., {March}, M., {Miquel},
  R., {Palmese}, A., {Paz-Chinchon}, F., {Plazas}, A.~A., {Reil}, K.,
  {Roodman}, A., {Sanchez}, E., {Santiago}, B., {Scarpine}, V., {Schubnell},
  M., {Serrano}, S., {Smith}, M., {Suchyta}, E., {Tarle}, G., \& {Walker},
  A.~R. 2019, arXiv e-prints, arXiv:1911.08493

\bibitem[{{Spencer} {et~al.}(2018){Spencer}, {Mateo}, {Olszewski}, {Walker},
  {McConnachie}, \& {Kirby}}]{spencer2018}
{Spencer}, M.~E., {Mateo}, M., {Olszewski}, E.~W., {Walker}, M.~G.,
  {McConnachie}, A.~W., \& {Kirby}, E.~N. 2018, \aj, 156, 257

\bibitem[{{Spencer} {et~al.}(2017){Spencer}, {Mateo}, {Walker}, \&
  {Olszewski}}]{spencer2017}
{Spencer}, M.~E., {Mateo}, M., {Walker}, M.~G., \& {Olszewski}, E.~W. 2017,
  \apj, 836, 202

\bibitem[{{Spite} {et~al.}(2018){Spite}, {Spite}, {Fran{\c{c}}ois},
  {Bonifacio}, {Caffau}, \& {Salvadori}}]{spite2018}
{Spite}, M., {Spite}, F., {Fran{\c{c}}ois}, P., {Bonifacio}, P., {Caffau}, E.,
  \& {Salvadori}, S. 2018, \aap, 617, A56

\bibitem[{{Torrealba} {et~al.}(2019){Torrealba}, {Belokurov}, {Koposov}, {Li},
  {Walker}, {Sanders}, {Geringer-Sameth}, {Zucker}, {Kuehn}, {Evans}, \&
  {Dehnen}}]{torrealba2019}
{Torrealba}, G., {Belokurov}, V., {Koposov}, S.~E., {Li}, T.~S., {Walker},
  M.~G., {Sanders}, J.~L., {Geringer-Sameth}, A., {Zucker}, D.~B., {Kuehn}, K.,
  {Evans}, N.~W., \& {Dehnen}, W. 2019, \mnras, 488, 2743

\bibitem[{{Torrealba} {et~al.}(2016){Torrealba}, {Koposov}, {Belokurov},
  {Irwin}, {Collins}, {Spencer}, {Ibata}, {Mateo}, {Bonaca}, \&
  {Jethwa}}]{torrealba2016b}
{Torrealba}, G., {Koposov}, S.~E., {Belokurov}, V., {Irwin}, M., {Collins}, M.,
  {Spencer}, M., {Ibata}, R., {Mateo}, M., {Bonaca}, A., \& {Jethwa}, P. 2016,
  \mnras, 463, 712

\bibitem[{{Venn} {et~al.}(2017){Venn}, {Starkenburg}, {Malo}, {Martin}, \&
  {Laevens}}]{venn2017}
{Venn}, K.~A., {Starkenburg}, E., {Malo}, L., {Martin}, N., \& {Laevens},
  B.~P.~M. 2017, \mnras, 466, 3741

\bibitem[{{Walker} {et~al.}(2009{\natexlab{a}}){Walker}, {Belokurov}, {Evans},
  {Irwin}, {Mateo}, {Olszewski}, \& {Gilmore}}]{walker2009a}
{Walker}, M.~G., {Belokurov}, V., {Evans}, N.~W., {Irwin}, M.~J., {Mateo}, M.,
  {Olszewski}, E.~W., \& {Gilmore}, G. 2009{\natexlab{a}}, \apjl, 694, L144

\bibitem[{{Walker} {et~al.}(2009{\natexlab{b}}){Walker}, {Mateo}, \&
  {Olszewski}}]{walker2009b}
{Walker}, M.~G., {Mateo}, M., \& {Olszewski}, E.~W. 2009{\natexlab{b}}, \aj,
  137, 3100

\bibitem[{{Walker} {et~al.}(2015{\natexlab{a}}){Walker}, {Mateo}, {Olszewski},
  {Bailey}, {Koposov}, {Belokurov}, \& {Evans}}]{walker2015a}
{Walker}, M.~G., {Mateo}, M., {Olszewski}, E.~W., {Bailey}, John~I., I.,
  {Koposov}, S.~E., {Belokurov}, V., \& {Evans}, N.~W. 2015{\natexlab{a}},
  \apj, 808, 108

\bibitem[{{Walker} {et~al.}(2016){Walker}, {Mateo}, {Olszewski}, {Koposov},
  {Belokurov}, {Jethwa}, {Nidever}, {Bonnivard}, {Bailey}, {Bell}, \&
  {Loebman}}]{walker2016}
{Walker}, M.~G., {Mateo}, M., {Olszewski}, E.~W., {Koposov}, S., {Belokurov},
  V., {Jethwa}, P., {Nidever}, D.~L., {Bonnivard}, V., {Bailey}, John~I., I.,
  {Bell}, E.~F., \& {Loebman}, S.~R. 2016, \apj, 819, 53

\bibitem[{{Walker} {et~al.}(2015{\natexlab{b}}){Walker}, {Olszewski}, \&
  {Mateo}}]{walker2015b}
{Walker}, M.~G., {Olszewski}, E.~W., \& {Mateo}, M. 2015{\natexlab{b}}, \mnras,
  448, 2717

\bibitem[{{Watkins} {et~al.}(2019){Watkins}, {van der Marel}, {Sohn}, \&
  {Evans}}]{watkins2019}
{Watkins}, L.~L., {van der Marel}, R.~P., {Sohn}, S.~T., \& {Evans}, N.~W.
  2019, \apj, 873, 118

\bibitem[{{Weiler}(2018)}]{weiler2018}
{Weiler}, M. 2018, \aap, 617, A138

\bibitem[{{Willman} {et~al.}(2011){Willman}, {Geha}, {Strader}, {Strigari},
  {Simon}, {Kirby}, {Ho}, \& {Warres}}]{willman2011}
{Willman}, B., {Geha}, M., {Strader}, J., {Strigari}, L.~E., {Simon}, J.~D.,
  {Kirby}, E., {Ho}, N., \& {Warres}, A. 2011, \aj, 142, 128

\end{thebibliography}
\end{document}